\documentclass[]{aa}
\usepackage{epsfig}
\def\lsim{\vcenter{\hbox{$<$}\offinterlineskip\hbox{$\sim$}}}
\def\gsim{\vcenter{\hbox{$>$}\offinterlineskip\hbox{$\sim$}}}
\begin{document}
\newcommand{\Aliii}{Al~{\sc iii}}
\newcommand{\Feiii}{Fe~{\sc iii}}
\newcommand{\Feiv}{Fe~{\sc iv}}
\newcommand{\Fev}{Fe~{\sc v}}
\newcommand{\Fevi}{Fe~{\sc vi}}
\newcommand{\Hei}{He~{\sc i}}
\newcommand{\Heii}{He~{\sc ii}}
\newcommand{\Heiii}{He~{\sc iii}}
\newcommand{\Niii}{N~{\sc iii}}
\newcommand{\Niv}{N~{\sc iv}}
\newcommand{\Nv}{N~{\sc v}}
\newcommand{\Ovi}{O~{\sc vi}}
\newcommand{\Siii}{Si~{\sc ii}}
\newcommand{\Siiii}{Si~{\sc iii}}
\newcommand{\Siiv}{Si~{\sc iv}}
\newcommand{\Siv}{Si~{\sc v}}
\newcommand{\Cii}{C~{\sc ii}}
\newcommand{\Ciii}{C~{\sc iii}}
\newcommand{\Civ}{C~{\sc iv}}
\newcommand{\Cv}{C~{\sc v}}
\newcommand{\pis}{$^1$P$^0$-$^1$S}
\newcommand{\piis}{$^2$P$^0$-$^2$S}
\newcommand{\piid}{$^2$P$^0$-$^2$D}
\newcommand{\piiip}{$^3$P$^0$-$^3$P}
\newcommand{\sip}{$^1$S-$^1$P$^0$}
\newcommand{\siip}{$^2$S-$^2$P$^0$}
\newcommand{\sviip}{a$^7$S-z$^7$P$^0$}
\newcommand{\gvh}{a$^5$G-z$^5$H$^0$}
\renewcommand{\topfraction}{1.}
\renewcommand{\bottomfraction}{1.}
\renewcommand{\textfraction}{0.}
\title{Modelling the orbital modulation of ultraviolet resonance lines in 
       high-mass X-ray binaries\thanks{Based
       on observations obtained with the International Ultraviolet Explorer at
       Villafranca Tracking Station (ESA) and at Goddard Space Flight Center
       (NASA)}}
\author{Jacco Th. van Loon\inst{1}, Lex Kaper\inst{2} \and Godelieve
        Hammerschlag-Hensberge\inst{2}}
\institute{Astrophysics Group, School of Chemistry \& Physics, Keele
           University, Staffordshire ST5 5BG, United Kingdom
\and       Astronomical Institute, University of Amsterdam, Kruislaan 403,
           NL-1098 SJ Amsterdam, The Netherlands}
\offprints{Jacco van Loon: jacco@astro.keele.ac.uk}
\date{Received date; accepted date}
\titlerunning{Modelling UV resonance lines in HMXBs}
\authorrunning{van Loon et al.}
\abstract{
The stellar-wind structure in high-mass X-ray binaries (HMXBs) is investigated
through modelling of their ultraviolet (UV) resonance lines. For the OB
supergiants in two systems, Vela X-1 and 4U1700$-$37, high-resolution UV
spectra are available; for Cyg X-1, SMC X-1, and LMC X-4 low-resolution
spectra are used. In order to account for the non-monotonic velocity structure
of the stellar wind, a modified version of the Sobolev Exact Integration (SEI)
method by Lamers et al. (1987) is applied.\\
The orbital modulation of the UV resonance lines provides information on the
size of the Str\"{o}mgren zone surrounding the X-ray source. The amplitude of
the observed orbital modulation (known as the Hatchett-McCray effect),
however, also depends on the density- and velocity structure of the ambient
wind. Model profiles are presented that illustrate the effect on the
appearance of the HM effect by varying stellar-wind parameters. The $q$
parameter of Hatchett \& McCray (1977), as well as other parameters describing
the supergiant's wind structure, are derived for the 5 systems. The X-ray
luminosity needed to create the observed size of the Str\"{o}mgren zone is
consistent with the observed X-ray flux. The derived wind parameters are
compared to those determined in single OB supergiants of similar spectral
type.\\
Our models naturally explain the observed absence of the HM effect in
4U1700$-$37. The orbital modulation in Vela X-1 indicates that besides the
Str\"{o}mgren zone other structures are present in the stellar wind (such as a
photo-ionization wake). The ratio of the wind velocity and the escape velocity
is found to be lower in OB supergiants in HMXBs than in single OB supergiants
of the same effective temperature.
}
\maketitle

\section{Introduction}

In a high-mass X-ray binary (HMXB) a massive, early-type star transfers mass
onto a compact companion, a neutron star or a black hole. The potential energy
associated with the accreted matter is efficiently converted into X-rays
(Davidson \& Ostriker 1973). The mass transfer can take place in two different
ways: (i) via the massive star's dense stellar wind which is partly
intercepted by the strong gravitational field of the compact companion; or
(ii) via a flow of matter towards the compact star through the inner
Lagrangian point (Roche-lobe overflow). In the latter case an accretion disk
is expected to be present in the system, as well as a rapidly spinning X-ray
source. The massive star is either an OB supergiant with a dense
radiation-driven wind or a Be-type star characterised by strong H$\alpha$
emission arising from a slowly outflowing, dense equatorial disk. For reviews
on this subject we refer to Lewin et al.\ (1995), van Paradijs (1998) and
Kaper (2001).

The accretion-induced X-ray luminosity gives a direct measure of the wind
density and velocity at the orbit of the X-ray source. In this sense, the
compact object acts as a probe in the stellar wind. In OB-supergiant systems,
the X-ray source strongly affects the supergiant's radiation-driven wind. The
X-ray source ionizes the surrounding wind regions creating an extended
(Str\"{o}mgren) zone of strong ionization that trails the X-ray source in its
orbit. This causes the orbital modulation of ultraviolet (UV) resonance lines
(Hatchett \& McCray 1977; Kaper et al.\ 1993). Hatchett \& McCray predicted
this effect to be observable in UV resonance lines of the HMXB
HD153919/4U1700$-$37. The HM-effect has, however, not been detected in spectra
obtained with the {\it International Ultraviolet Explorer} (IUE), as shown by
Dupree et al.\ (1978). Although Hammerschlag-Hensberge et al.\ (1990) reported
variations with orbital phase in some subordinate lines, Kaper et al.\ (1990)
reported that these variations are caused by variable Raman-scattered emission
lines and are not due to the HM-effect (see also Kaper et al.\ 1993).

%
%
\begin{table*}[]
\caption[]{System parameters of the HMXBs in our sample. References are: a)
Walborn (1973); b) Chevalier \& Ilovaisky (1977); c) Webster et al.\ (1972);
d) Conti (1978); e) Bregman et al.\ (1973); f) Osmer (1973); g) Sadakane et
al.\ (1985); h) Heap \& Corcoran (1992); i) Bolton (1975); j) Levine et al.\
(1991); k) Rappaport \& Joss (1983); l) Nagase (1989); m) Kemp (1977); n) van
Paradijs \& Kuiper (1984); o) Deeter et al.\ (1987); p) Haberl et al.\ (1989);
q) Bolton \& Herbst (1976); r) Kaper (1998).}
\begin{tabular}{lll|lllll|lll}
\hline\hline
\multicolumn{3}{c|}{X-ray companion} &
\multicolumn{5}{c|}{Optical primary} &
\multicolumn{3}{c}{} \\
Name                         &
$M_{\rm X}/M$\rlap{$_\odot$} &
$L_{\rm X}$(erg/s)$^{\rm r}$ &
Name                         &
spec.type                    &
$m_{\rm V}$                  &
$M_{\rm \star}/M_\odot$      &
$R_{\rm \star}/R_\odot$      &
$a/R_\odot$                  &
$P_{\rm orb}$(d)             &
$d$(kpc)                     \\
\hline
Cyg X-1                      &
13$^{\rm i}$                 &
$1.3\times10^{37}$           &
HDE226868                    &
O9.7 Iab$^{\rm a}$           &
8.87$^{\rm e}$               &
21$^{\rm i}$                 &
\llap{1}8$^{\rm i}$          &
43$^{\rm i}$                 &
5.60000$^{\rm m}$            &
1.8$^{\rm d}$                \\
LMC X-4                      &
1.38$^{\rm j}$               &
$2.2\times10^{38}$           &
Sk-Ph                        &
O8 V-III$^{\rm b}$           &
\llap{1}4$^{\rm b}$          &
14.7$^{\rm j}$               &
7.6$^{\rm j}$                &
12$^{\rm j}$                 &
1.40839$^{\rm j}$            &
\llap{5}0                    \\
SMC X-1                      &
1.05$^{\rm k}$               &
$2.6\times10^{38}$           &
Sk 160                       &
B0 Ib$^{\rm c}$              &
\llap{1}3.2$^{\rm f}$        &
17.0$^{\rm k}$               &
\llap{1}6.5$^{\rm k}$        &
25 $^{\rm k}$                &
3.89239$^{\rm n}$            &
\llap{6}0                    \\
Vela X-1                     &
1.77$^{\rm l}$               &
$4.6\times10^{36}$           &
HD77581                      &
B0.5 Iab$^{\rm d}$           &
6.88$^{\rm g}$               &
23.0$^{\rm l}$               &
\llap{3}4.0$^{\rm l}$        &
52.9$^{\rm l}$               &
8.964416$^{\rm o}$           &
1.9$^{\rm g}$                \\
4U1700$-$37                  &
1.8$^{\rm h}$                &
$9.8\times10^{35}$           &
HD153919                     &
O6.5 Iaf$^{+{\rm a}}$        &
6.51$^{\rm h}$               &
52$^{\rm h}$                 &
\llap{1}8$^{\rm h}$          &
36$^{\rm h}$                 &
3.411652$^{\rm p}$           &
1.7$^{\rm q}$                \\
\hline
\end{tabular}
\end{table*}

Dupree et al.\ (1980) did detect the HM-effect in IUE spectra of HD77581/Vela
X-1. The B-supergiant's wind profiles are less saturated while the
Str\"{o}mgren zone is expected to be larger than in the case of
HD153919/4U1700$-$37. A detailed study of the orbital modulation of the UV
resonance lines of HD77581/Vela X-1 indicated that the velocity and density
structure of the stellar wind cannot be a monotonically rising function with
distance from the star (Kaper et al.\ 1993), a conclusion that has been
derived from observations of single early-type stars as well (Lucy 1982).
Additional absorption (e.g.\ due to material trailing the X-ray source in its
orbit) appears in the line profiles at late orbital phases (Sadakane et al.\
1985; Kaper et al.\ 1994).

HD153919 and HD77581 are the only OB-supergiants in HMXBs bright enough in the
UV to have been observed with sufficient signal-to-noise in the
high-resolution mode of IUE. Kallman et al.\ (1987) and Payne \& Coe (1987)
studied the HD77581/Vela X-1 system in low-resolution mode and trailing mode,
respectively, to search for the signature of rapid modulation of the
Str\"{o}mgren zone as a reaction to the periodically varying X-ray emission
from the pulsar. Although these early experiments with IUE were unsuccessful,
the periodic modulation was recently detected using the Faint Object
Spectrograph onboard the {\it Hubble Space Telescope} (HST) by Boroson et al.\
(1996). Treves et al.\ (1980) detected the orbital modulation of UV resonance
lines in low-resolution IUE spectra of HD226868/Cyg X-1. This source is too
faint to obtain useful high-resolution IUE spectra (Davis \& Hartmann 1983).
Bonnet-Bidaud et al.\ (1981) and van der Klis et al.\ (1982) discussed the
appearance of the HM-effect in low-resolution IUE spectra of Sk 160/SMC X-1
and Sk-Ph/LMC X-4. Two high-resolution UV spectra of Sk 160/SMC X-1 confirmed
the orbital modulation of the \Siiv\ resonance doublet (Hammerschlag-Hensberge
et al.\ 1984). Vrtilek et al.\ (1997) and Boroson et al.\ (1999) discussed
observations of LMC X-4, including high-resolution HST/GHRS spectra, in terms
of a shadow wind. In this system the X-ray flux is so strong that only in the
X-ray shadow behind the supergiant companion a normal stellar wind can develop
(Blondin 1994). This represents the most extreme case of the HM effect; in
practice, only Roche-lobe overflow systems will include a shadow wind. Recent
HST/STIS observations by Kaper et al.\ (in preparation) confirm the presence
of a shadow wind in LMC X-4, as well as a photo-ionization wake at the
(leading) interface between the shadow wind and the X-ray ionization zone. We
have not included the peculiar O7 III/black-hole candidate system \#32/LMC X-1
(Hutchings et al.\ 1983, 1987; Cowley et al.\ 1995) because the system is
embedded in a nebula with strong (UV) emission lines.

Although detailed observational studies of the HM effect in individual systems
exist, a quantitative analysis based on state-of-the-art stellar-wind models
is lacking. In the past two decades significant improvement has been made in
modelling the stellar-wind profiles of (single) OB-type stars (Groenewegen et
al.\ 1989; Groenewegen \& Lamers 1989, 1991; Haser et al.\ 1998). We present a
quantitative analysis of the UV spectral variability in the five HMXBs with
OB-supergiant companion observed with IUE, comparing the complete set of IUE
spectra. The model spectra are obtained using a modified version of the
Sobolev Exact Integration (SEI) method introduced by Lamers et al.\ (1987).
Our new method allows to take into account the non-monotonic wind structure
observed in many (single) OB-type stars by adding ``turbulence'', as well as
an extended Str\"{o}mgren zone around the X-ray source. With this analysis,
fundamental parameters of the system are derived.

An overview of the data and spectral line variability is given in Sect.\ 2.
X-ray eclipse spectra, UV continuum lightcurves and additional photometry are
described in Appendices A--C, and the applied variability and error analysis
method is described in Appendix D. Sect.\ 3 introduces the radiation transfer
code ``SEI'' (Lamers et al.\ 1987) and the modifications that we implemented
to be able to compute line profiles for HMXBs. Sect.\ 4 describes our attempts
to model the UV line variability observed in HD77581/Vela X-1 and
HD153919/4U1700$-$37. The derived terminal velocities, ionization fractions
and sizes of the ionization zones are discussed in Sect.\ 5.

\section{IUE observations of High-Mass X-ray Binaries}

%
%
\begin{table*}[]
\caption[]{SWP number, Reduced Julian Day (RJD) = JD$-$2\,440\,000 and
mid-exposure orbital phase $\phi$ of the IUE spectra. Spectra used to
construct $\phi\sim0$ and $\phi\sim0.5$ averages are indicated by a suffix
$\circ$ and $+$, respectively. Orbital parameters:}
\vspace{-3mm}
\begin{tabular}{lll|lll|lll|lll|lll}
\multicolumn{3}{l}{HDE226868/Cyg X-1}                 &
\multicolumn{4}{l}{$\phi_0 \equiv$ RJD 2804.245}      &
\multicolumn{3}{l}{P=$5{\fd}60000(24)$}               &
\multicolumn{5}{l}{(Kemp 1977)}                       \\
\multicolumn{3}{l}{Sk 160/SMC X-1}                    &
\multicolumn{4}{l}{$\phi_0 \equiv$ RJD 3000.1626(8)}  &
\multicolumn{3}{l}{P=$3{\fd}89239(2)$}                &
\multicolumn{5}{l}{(van Paradijs \& Kuiper 1984)}     \\
\multicolumn{3}{l}{Sk-Ph/LMC X-4}                     &
\multicolumn{4}{l}{$\phi_0 \equiv$ RJD 7742.4904(2)}  &
\multicolumn{3}{l}{P=$1{\fd}40839(1)$}                &
\multicolumn{5}{l}{(Levine et al.\ 1991)}             \\
\multicolumn{3}{l}{HD77581/Vela X-1}                  &
\multicolumn{4}{l}{$\phi_0 \equiv$ RJD 4279.0466(37)} &
\multicolumn{3}{l}{P=$8{\fd}964416(49)$}              &
\multicolumn{5}{l}{(Deeter et al.\ 1987)}             \\
\multicolumn{3}{l}{HD153919/4U1700$-$37}              &
\multicolumn{4}{l}{$\phi_0 \equiv$ RJD 6161.3400(30)} &
\multicolumn{3}{l}{P=$3{\fd}411652(26)$}              &
\multicolumn{5}{l}{(Haberl et al.\ 1989)}             \\
\hline\hline
SWP                  & RJD  & $\phi$ &
 1968                & 3699 & 0.775  &
21569                & 5656 & 0.728  &
22324                & 5752 & 0.407  &
 4753                & 3958 & 0.2124 \\
\multicolumn{3}{c|}{\it Cyg X-1}     &
 2020\rlap{$^\circ$} & 3705 & 0.221  &
21607                & 5660 & 0.888  &
32961\rlap{$^+$}     & 7214 & 0.432  &
 5180                & 4003 & 0.3758 \\
 1273\rlap{$^\circ$} & 3599 & 0.929  &
 2044\rlap{$^\circ$} & 3708 & 0.906  &
21608\rlap{$^\circ$} & 5660 & 0.019  &
32967\rlap{$^+$}     & 7215 & 0.543  &
25586                & 6160 & 0.7770 \\
 1445                & 3629 & 0.317  &
 6203\rlap{$^\circ$} & 4102 & 0.185  &
21609\rlap{$^\circ$} & 5660 & 0.066  &
33085\rlap{$^+$}     & 7233 & 0.550  &
25587                & 6160 & 0.7884 \\
 1451\rlap{$^+$}     & 3629 & 0.441  &
 6207\rlap{$^+$}     & 4103 & 0.406  &
21611                & 5661 & 0.166  &
46144                & 8933 & 0.165  &
25588                & 6160 & 0.8020 \\
 1478\rlap{$^\circ$} & 3632 & 0.971  &
 6219\rlap{$^+$}     & 4104 & 0.705  &
21612                & 5661 & 0.217  &
46151                & 8934 & 0.278  &
25589                & 6160 & 0.8169 \\
 1500\rlap{$^+$}     & 3636 & 0.540  &
 6224\rlap{$^\circ$} & 4105 & 0.949  &
21613                & 5661 & 0.267  &
46167                & 8935 & 0.406  &
25590                & 6160 & 0.8299 \\
 1515\rlap{$^\circ$} & 3638 & 0.047  &
 7092                & 4182 & 0.826  &
21614                & 5661 & 0.315  &
\multicolumn{3}{c}{\it 4U1700$-$37}  &
25591                & 6160 & 0.8423 \\
 1979                & 3701 & 0.270  &
 7094                & 4182 & 0.878  &
21615                & 5661 & 0.360  &
 1476                & 3632 & 0.7960 &
25592                & 6160 & 0.8545 \\
 2049\rlap{$^+$}     & 3708 & 0.538  &
 8662\rlap{$^\circ$} & 4334 & 0.850  &
21616                & 5661 & 0.408  &
 1714\rlap{$^\circ$} & 3664 & 0.0093 &
25596\rlap{$^\circ$} & 6161 & 0.0976 \\
 3012                & 3799 & 0.714  &
 8673                & 4335 & 0.130  &
21617                & 5661 & 0.454  &
 1960                & 3700 & 0.6277 &
25597                & 6161 & 0.1074 \\
 3079                & 3802 & 0.257  &
 8687                & 4336 & 0.396  &
21618\rlap{$^+$}     & 5661 & 0.494  &
 1961                & 3700 & 0.6393 &
25598                & 6161 & 0.1228 \\
 3107                & 3804 & 0.640  &
 8701                & 4337 & 0.617  &
21619\rlap{$^+$}     & 5661 & 0.534  &
 1969                & 3701 & 0.8508 &
25599                & 6161 & 0.1371 \\
 3518\rlap{$^\circ$} & 3845 & 0.024  &
\multicolumn{3}{c|}{\it LMC X-4}     &
21625\rlap{$^\circ$} & 5662 & 0.003  &
 1970                & 3701 & 0.8669 &
25600                & 6161 & 0.1462 \\
 3535                & 3848 & 0.388  &
 1477\rlap{$^\circ$} & 3632 & 0.950  &
\multicolumn{3}{c|}{\it Vela X-1}    &
 1972                & 3701 & 0.9098 &
25607                & 6162 & 0.3640 \\
 3940                & 3892 & 0.281  &
 2045                & 3708 & 0.587  &
 1442\rlap{$^+$}     & 3628 & 0.453  &
 1973\rlap{$^\circ$} & 3701 & 0.9241 &
25608                & 6162 & 0.3746 \\
 3965                & 3894 & 0.732  &
 6202                & 4102 & 0.368  &
 1488\rlap{$^\circ$} & 3634 & 0.057  &
 1975\rlap{$^\circ$} & 3701 & 0.9619 &
25609                & 6162 & 0.3844 \\
 3966                & 3894 & 0.744  &
 6204\rlap{$^+$}     & 4102 & 0.505  &
 2087                & 3712 & 0.856  &
 1986                & 3702 & 0.2536 &
25610                & 6162 & 0.3954 \\
 5178\rlap{$^\circ$} & 4002 & 0.049  &
 6208                & 4103 & 0.115  &
 3510                & 3845 & 0.604  &
 1987                & 3702 & 0.2740 &
25611                & 6162 & 0.4068 \\
 5181                & 4003 & 0.084  &
 6220\rlap{$^\circ$} & 4104 & 0.945  &
 3519                & 3846 & 0.714  &
 1991                & 3703 & 0.4572 &
25612                & 6162 & 0.4174 \\
 5183                & 4003 & 0.110  &
 6223\rlap{$^+$}     & 4105 & 0.482  &
 3550                & 3850 & 0.140  &
 1992\rlap{$^+$}     & 3703 & 0.4716 &
25613                & 6162 & 0.4279 \\
 5475                & 4035 & 0.780  &
 6225                & 4105 & 0.609  &
 3649\rlap{$^+$}     & 3862 & 0.537  &
 1994\rlap{$^+$}     & 3703 & 0.5038 &
25614                & 6162 & 0.4377 \\
 7784                & 4265 & 0.953  &
 6226                & 4105 & 0.660  &
 4718                & 3954 & 0.770  &
 1995                & 3703 & 0.5186 &
25615                & 6162 & 0.4550 \\
 9340                & 4412 & 0.169  &
 7066                & 4179 & 0.508  &
18823\rlap{$^+$}     & 5323 & 0.457  &
 2002                & 3703 & 0.6092 &
25616                & 6162 & 0.4658 \\
 9364                & 4416 & 0.834  &
 7091                & 4182 & 0.454  &
18958\rlap{$^+$}     & 5341 & 0.529  &
 2003                & 3703 & 0.6357 &
25617\rlap{$^+$}     & 6162 & 0.4778 \\
 9394                & 4419 & 0.395  &
 8663                & 4334 & 0.396  &
18970                & 5343 & 0.743  &
 2004                & 3703 & 0.6540 &
25618\rlap{$^+$}     & 6163 & 0.4880 \\
 9397                & 4419 & 0.491  &
 8664\rlap{$^+$}     & 4334 & 0.446  &
18983\rlap{$^\circ$} & 5345 & 0.976  &
 2006                & 3704 & 0.7411 &
25619\rlap{$^+$}     & 6163 & 0.4996 \\
 9413                & 4422 & 0.940  &
 8674                & 4335 & 0.175  &
19012                & 5351 & 0.609  &
 2008                & 3704 & 0.7672 &
25620\rlap{$^+$}     & 6163 & 0.5295 \\
 9421                & 4423 & 0.117  &
 8675                & 4335 & 0.221  &
19061                & 5357 & 0.282  &
 2009                & 3704 & 0.7818 &
25621\rlap{$^+$}     & 6163 & 0.5413 \\
 9439                & 4425 & 0.474  &
 8686                & 4336 & 0.802  &
22278                & 5746 & 0.733  &
 2106\rlap{$^\circ$} & 3715 & 0.0384 &
28730                & 6632 & 0.1508 \\
 9459                & 4426 & 0.767  &
 8688                & 4336 & 0.896  &
22287                & 5747 & 0.845  &
 2153                & 3720 & 0.4570 &
28731                & 6632 & 0.1638 \\
\multicolumn{3}{c|}{\it SMC X-1}     &
 8689                & 4336 & 0.937  &
22297\rlap{$^\circ$} & 5748 & 0.975  &
 4742\rlap{$^\circ$} & 3957 & 0.9378 &
28732                & 6632 & 0.1775 \\
 1520\rlap{$^+$}     & 3639 & 0.361  &
 9366                & 4416 & 0.372  &
22301\rlap{$^\circ$} & 5749 & 0.068  &
 4751                & 3958 & 0.1839 &
                     &      &        \\
 1533\rlap{$^\circ$} & 3642 & 0.930  &
21472                & 5646 & 0.585  &
22309                & 5751 & 0.290  &
 4752                & 3958 & 0.1978 &
                     &      &        \\
\hline
\end{tabular}
\end{table*}

Our sample comprises five HMXBs; only for HD77581/Vela X-1 and
HD153919/4U1700$-$37 high-resolution IUE spectra will be discussed. The system
parameters (Table 1) are collected from the literature. Vela X-1 is an X-ray
pulsar, identifying the compact object as a neutron star. For 4U1700$-$37 no
X-ray pulsations have been detected, but the X-ray spectrum suggests that the
X-ray source is a neutron star (White \& Marshall 1983; Reynolds et al.\
1999). Both HMXBs are wind-fed systems (Kaper 1998) and have relatively modest
X-ray luminosities. Cyg X-1 is a well-known black-hole candidate. The other
two sources, SMC X-1 and LMC X-4 are located in the Magellanic Clouds. The low
metallicity in the Magellanic Clouds and the occcurence of Roche-lobe overflow
due to their tight orbits account for their high X-ray luminosities. Sk-Ph/LMC
X-4 contains a (sub)giant rather than a supergiant. The short pulse period of
both SMC X-1 and LMC X-4 suggests that these neutron stars are surrounded by
an accretion disk. Periods of 60 and 30 days in the lightcurves of
respectively SMC X-1 (Wojdowski et al.\ 1998) and LMC X-4 (Heemskerk \& van
Paradijs 1989) are interpreted as the precession period of this disk.

The observation logs of the IUE spectra are listed in Table 2. Some spectra
were included in previous studies (Dupree et al.\ 1978, 1980; Treves et al.\
1980; Bonnet-Bidaud et al.\ 1981; van der Klis et al.\ 1982; Sadakane et al.\
1985; Hammerschlag-Hensberge et al.\ 1990; Kaper et al.\ 1993), but many are
used here for the first time. The low-resolution spectra (HDE226868/Cyg X-1,
Sk 160/SMC X-1 and Sk-Ph/LMC X-4) are sampled on a grid of 1 \AA\ per point.
The IUEDR software package (Giddings 1983) provided by STARLINK was used for
data reduction. We refer to Kaper et al.\ (1993) for details on the data
reduction of the high-resolution spectra of HD77581/Vela X-1 and
HD153919/4U1700$-$37, sampled on a grid of 0.1 \AA\ per point. The
determination and subtraction of the inter-order background light has been
improved for the high-resolution IUE final archive spectra in the INES system
(http://ines.vilspa.esa.es) using NEWSIPS (Gonz\'{a}lez-Riestra et al.\ 2000),
which may be important for the \Nv\ resonance line around 1240 \AA.

The exposure times are typically 45, 37, 45, 150 and 30 minutes for
HDE226868/Cyg X-1, Sk 160/SMC X-1, Sk-Ph/LMC X-4, HD77581/Vela X-1 and
HD153919/4U1700$-$37, respectively, corresponding to a spread in orbital phase
of 0.006, 0.007, 0.022, 0.012 and 0.006 per spectrum, respectively. Average
X-ray eclipse ($\phi=0$) spectra, which should be representative of the
``undisturbed'' stellar wind of the OB companion, are presented in Appendix A.
The continuum variability (which becomes apparent when normalising the
spectra), known as ellipsoidal variations, is due to the ``pear-like'' shape
of the supergiant filling its Roche lobe. These variations are discussed in
Appendices B \& C. The method we used for our error and variability analysis
is based on the calculation of covariances and is described in Appendix D.

The strongest variability is detected in the ultraviolet \siip\ resonance
doublets that are formed in the stellar wind; e.g.\ \Civ\ ($\lambda_0=1548.20$
\& 1550.774 \AA, $\Delta v=498$ km s$^{-1}$), \Siiv\ ($\lambda_0=1393.755$ \&
1402.77 \AA, $\Delta v=1927$ km s$^{-1}$), and \Nv\ ($\lambda_0=1238.821$ \&
1242.804 \AA, $\Delta v=961$ km s$^{-1}$). The orbital modulation of the mean
flux in these lines is shown in Figs.\ 1 (HDE226868/Cyg X-1, Sk 160/SMC X-1
and Sk-Ph/LMC X-4) and 2 (HD77581/Vela X-1 and HD153919/4U1700$-$37). The
error bars follow from the covariance-based procedure. In the luminous
(Roche-Lobe overflow) X-ray sources Cyg X-1, SMC X-1 and LMC X-4, the X-rays
ionize most of the stellar wind, leaving only a shadow wind unaffected. In the
fainter (wind-fed) X-ray sources Vela X-1 and 4U1700$-$37 the stellar wind is
much less disturbed.

%
%
\begin{figure*}[]
\centerline{\vbox{
\hbox{\psfig{figure=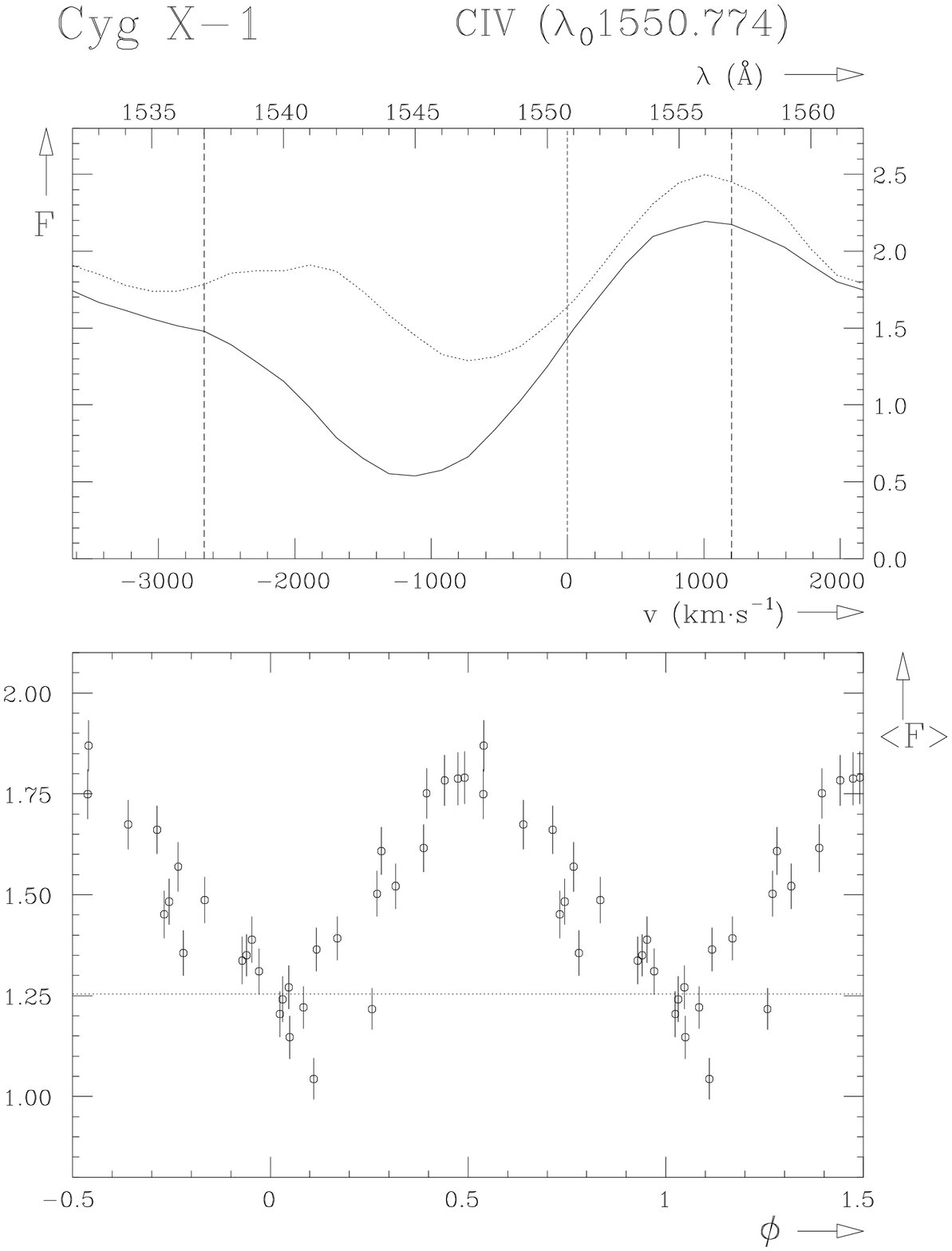,width=57mm}
      \psfig{figure=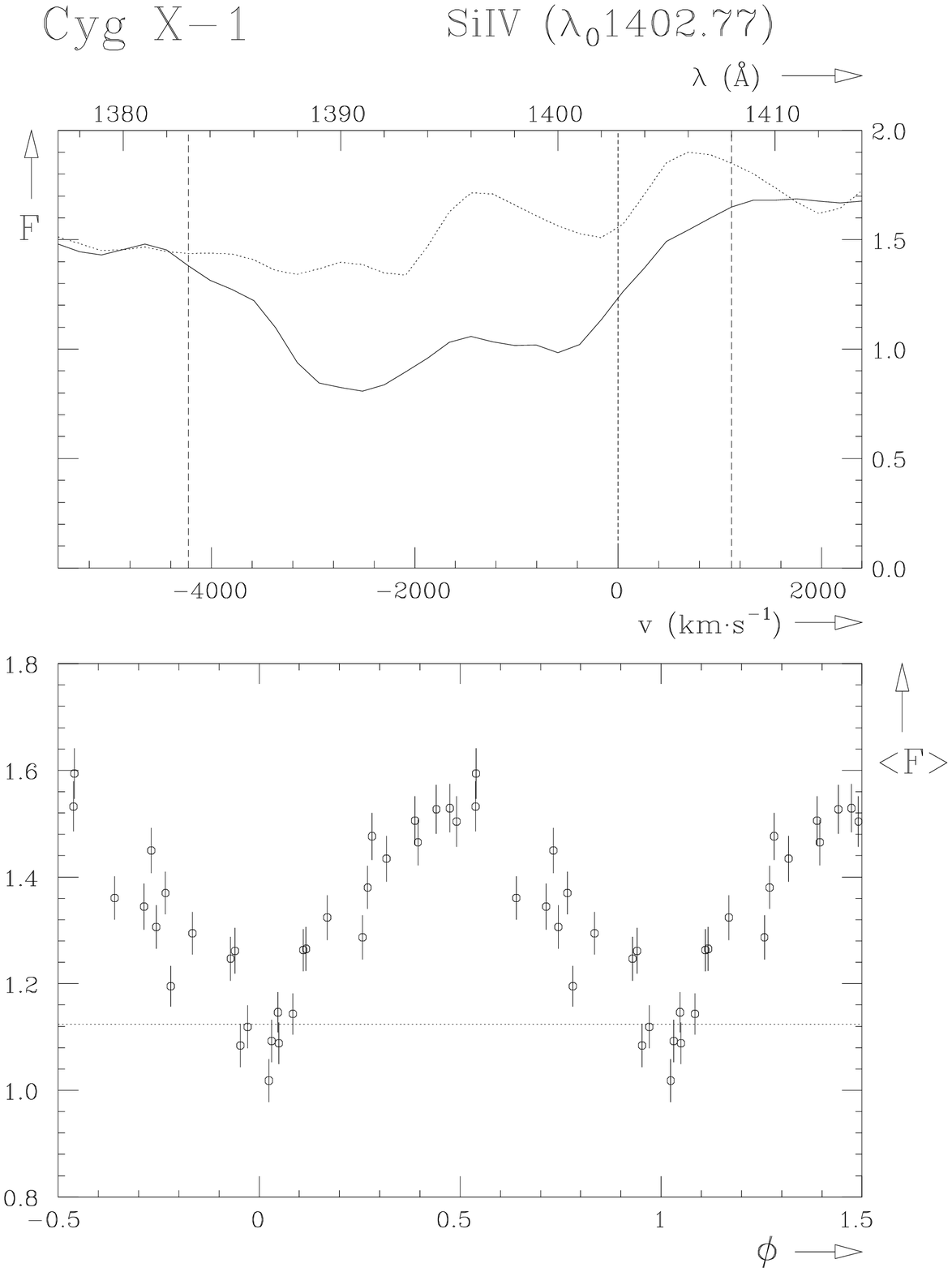,width=57mm}
      \psfig{figure=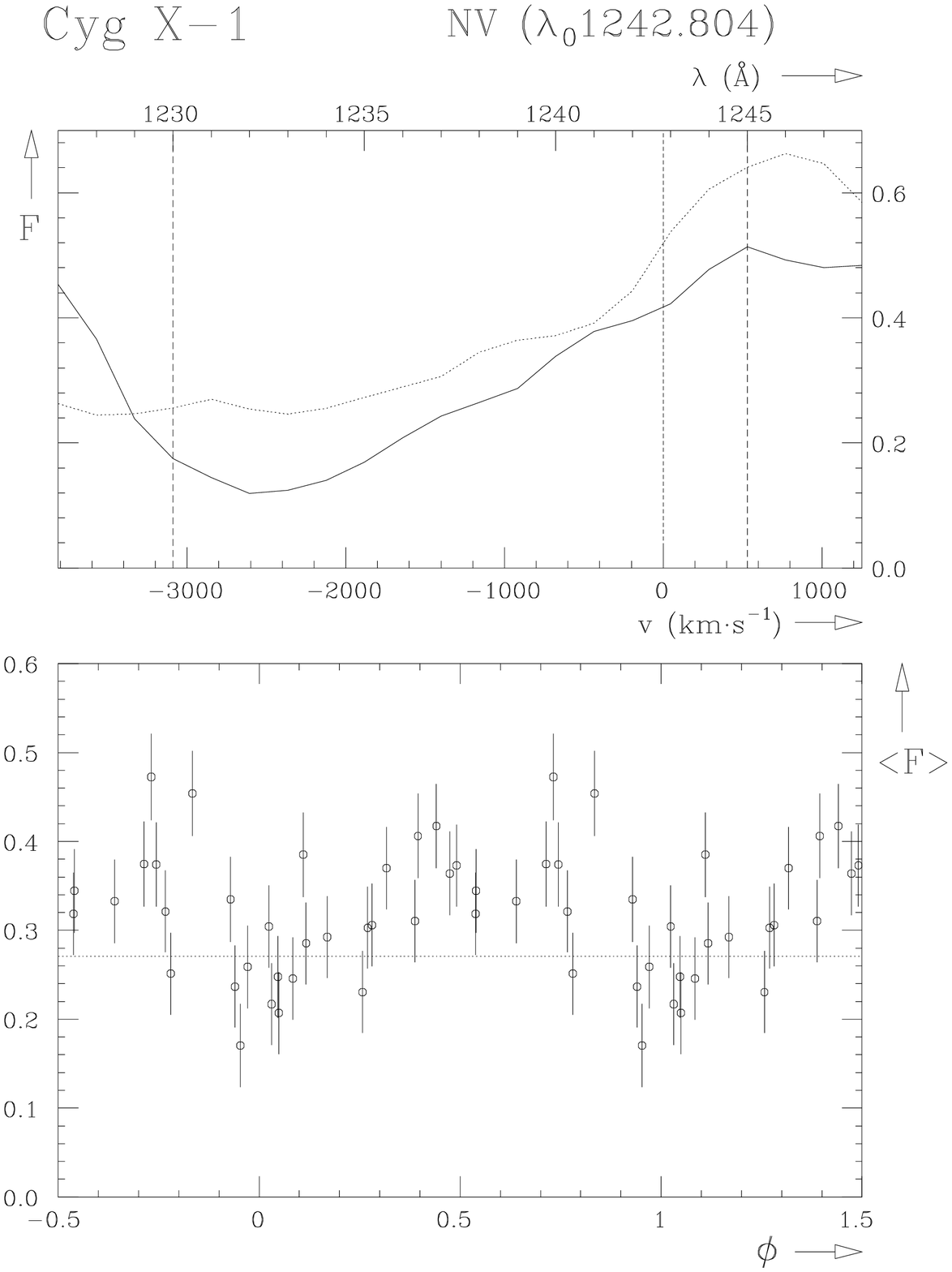,width=57mm}}
\hbox{\psfig{figure=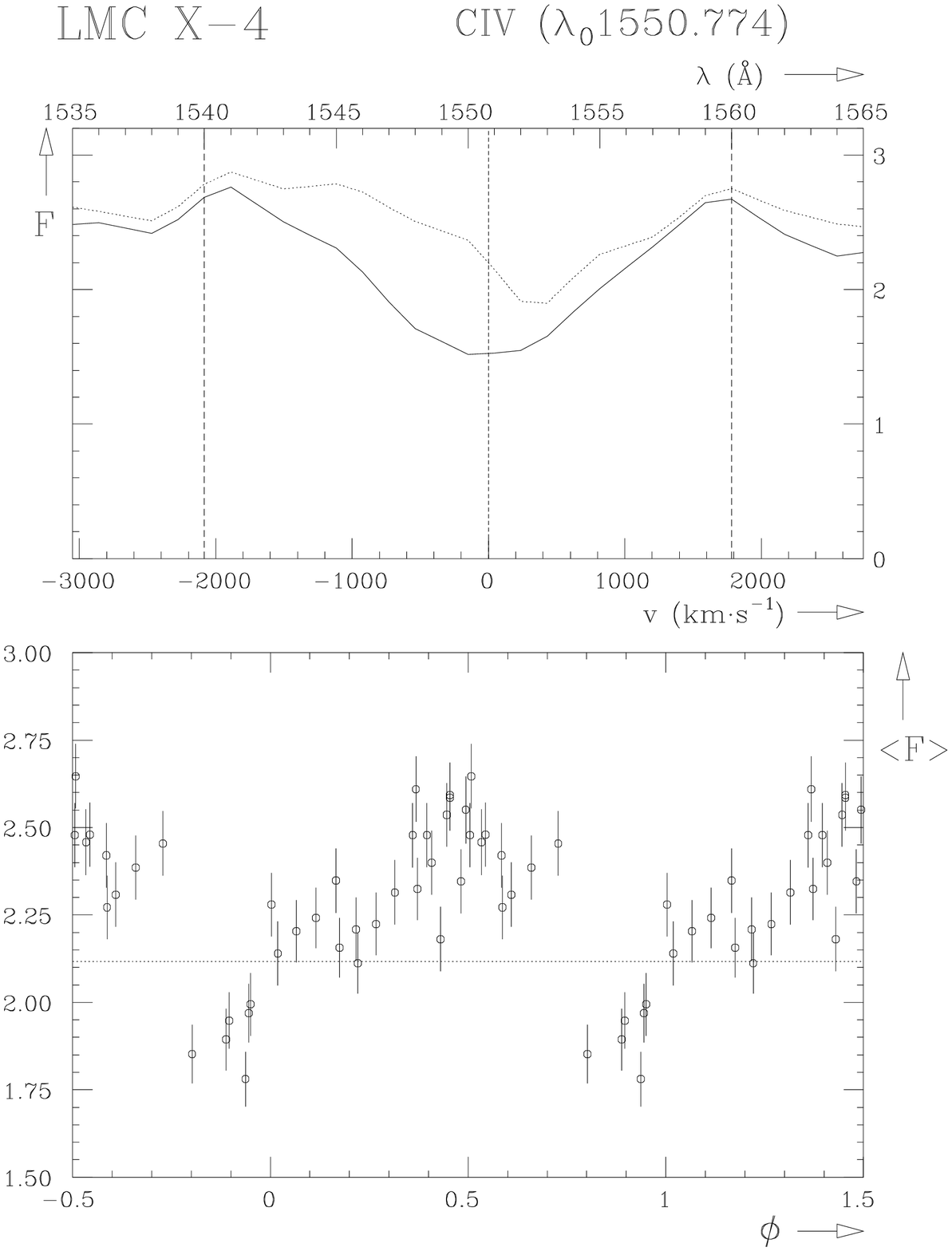,width=57mm}
      \psfig{figure=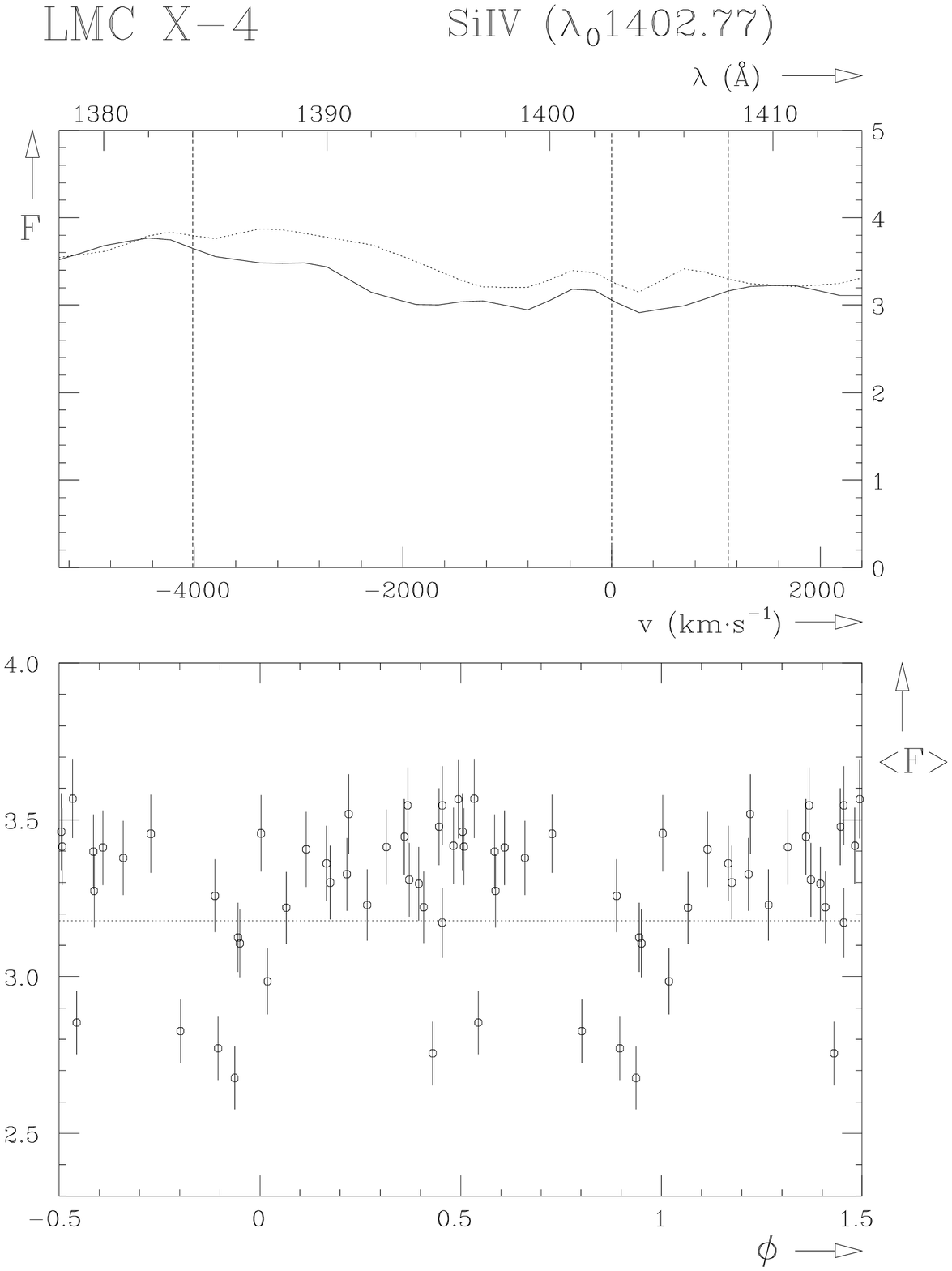,width=57mm}
      \psfig{figure=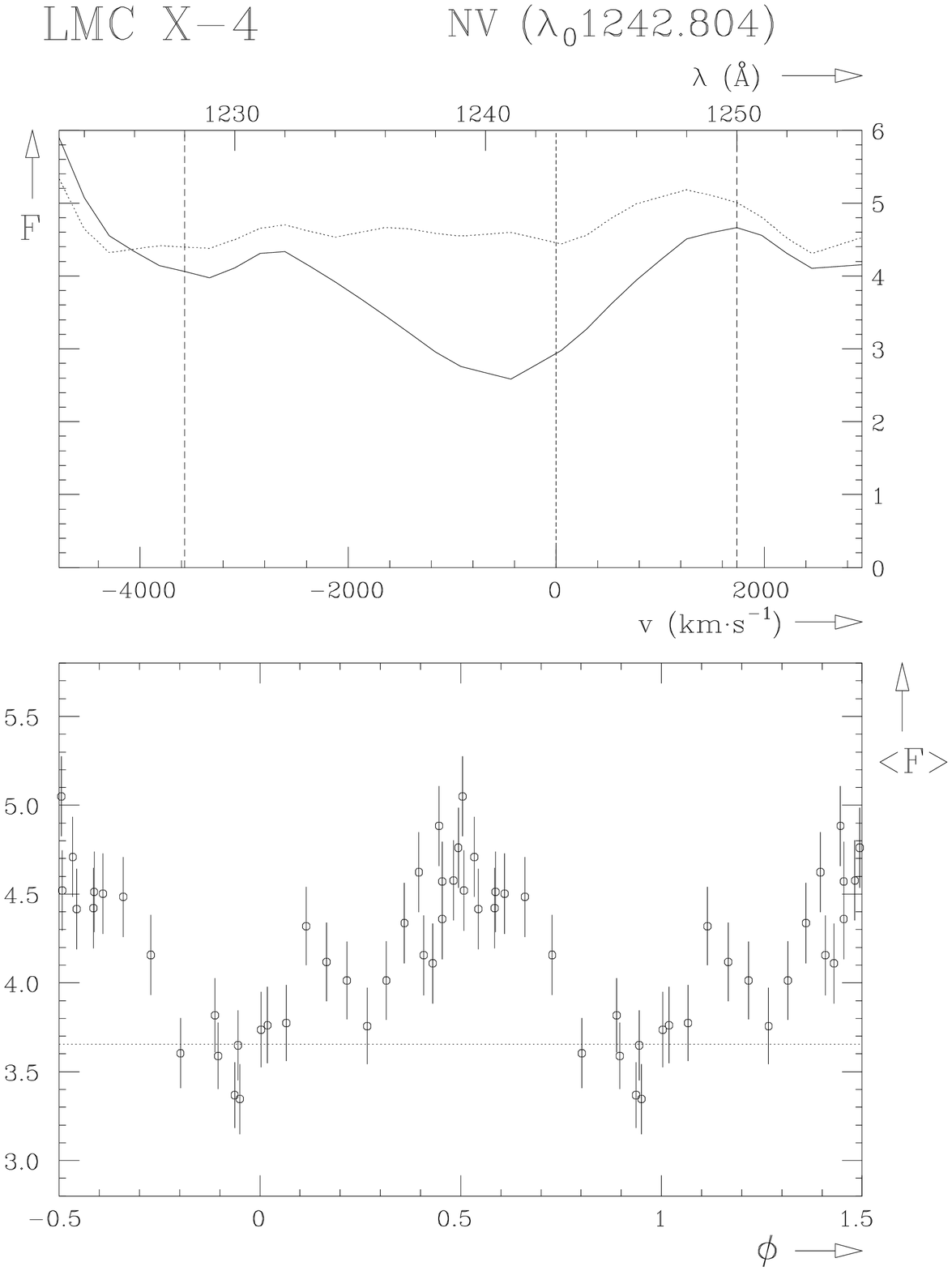,width=57mm}}
\hbox{\psfig{figure=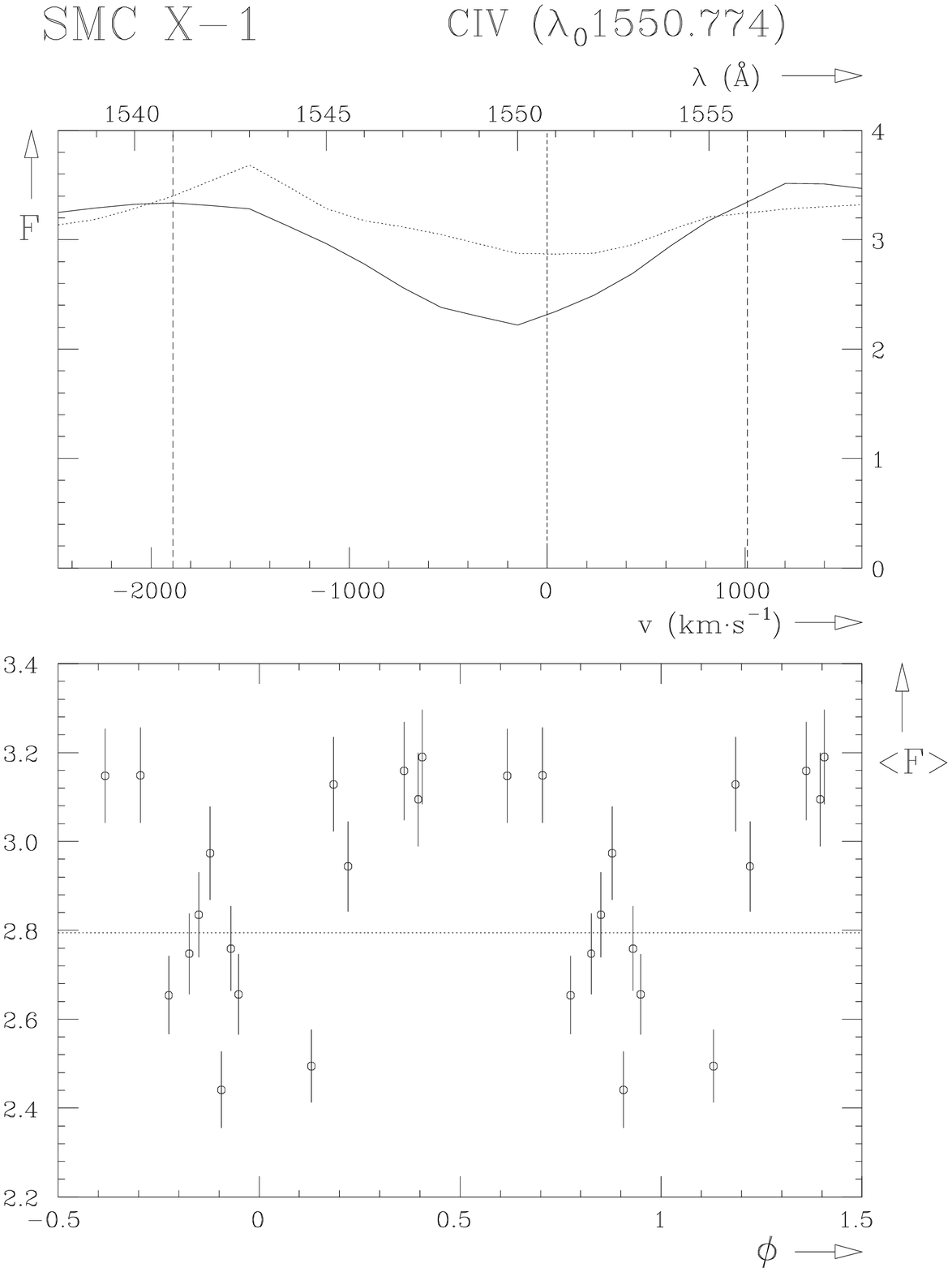,width=57mm}
      \psfig{figure=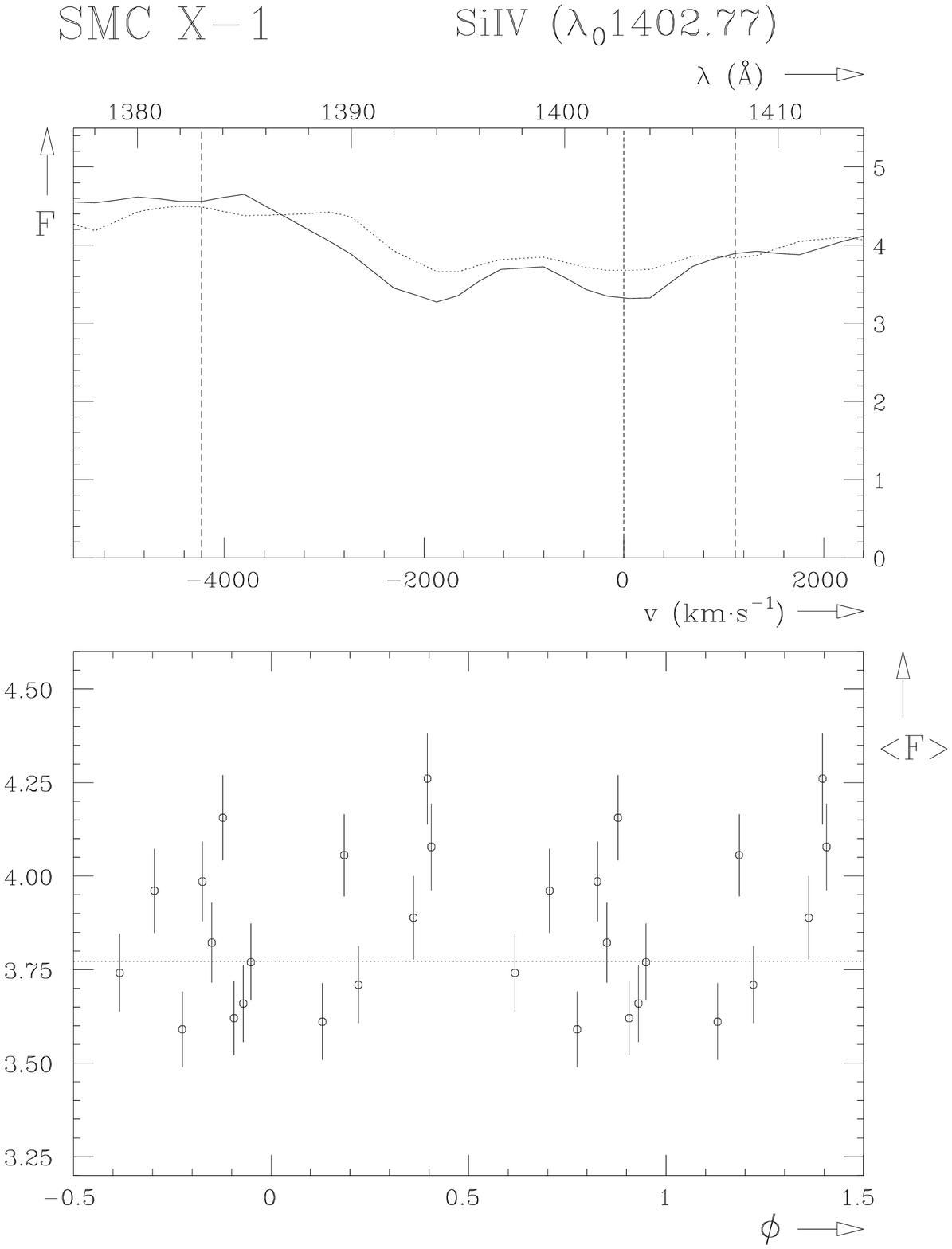,width=57mm}
      \psfig{figure=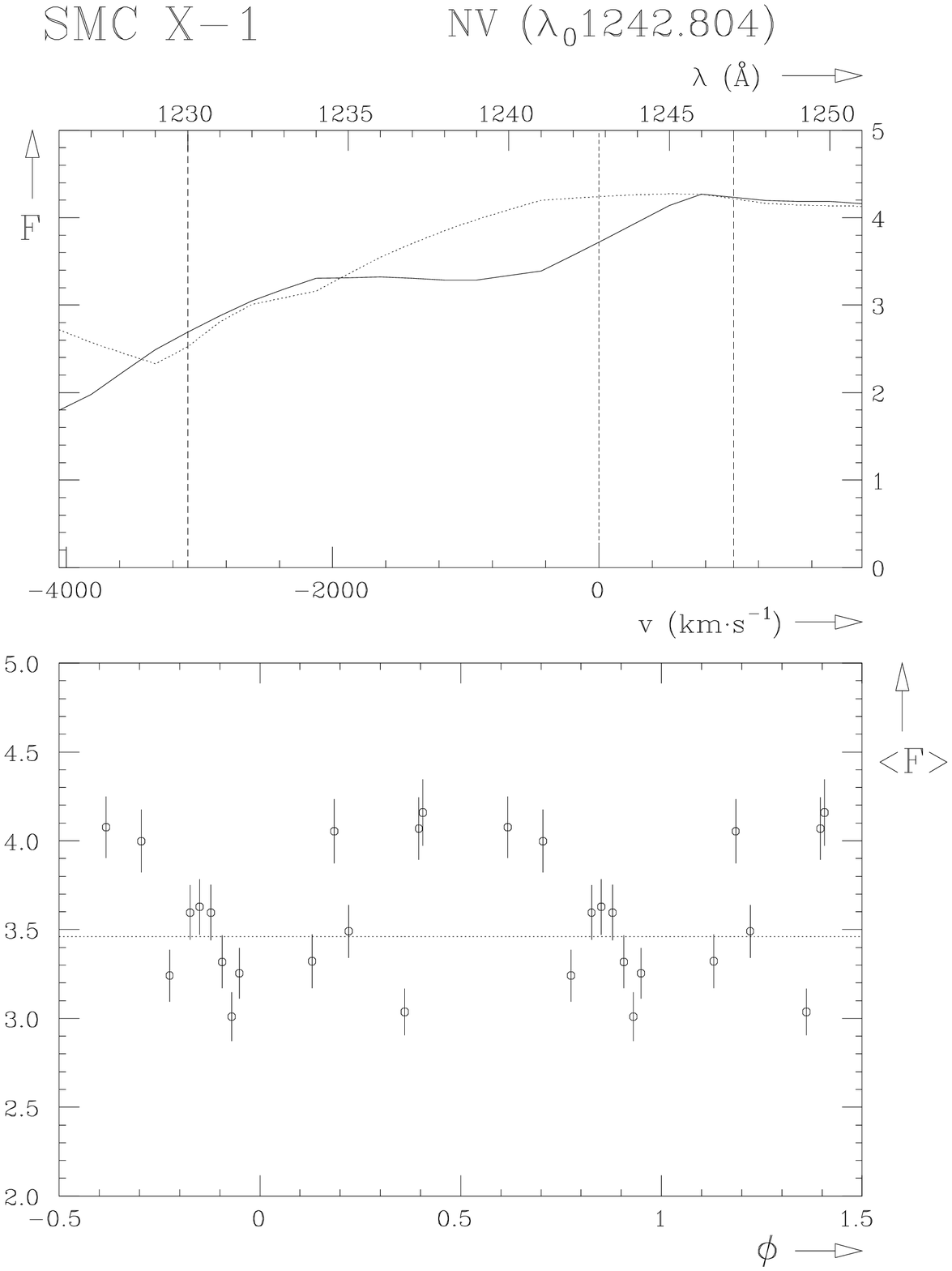,width=57mm}}
}}
\caption[]{Resonance lines of \Civ\ (left), \Siiv\ (middle) and \Nv\ (right)
for HDE226868/Cyg X-1 (top), Sk 160/SMC X-1 (middle) and Sk-Ph/LMC X-4
(bottom). Each upper panel displays the average spectrum near $\phi=0$ (drawn)
and $\phi=0.5$ (dotted). The short-dashed vertical line marks the
restwavelength, while the two long-dashed lines mark the boundaries of the
region in which the mean flux is determined and plotted in the lower panel
against orbital phase. The dotted horizontal line in each lower panel
represents the mean flux level in the $\phi=0$ average spectrum.}
\end{figure*}

%
%
\begin{figure*}[]
\centerline{\vbox{
\hbox{\psfig{figure=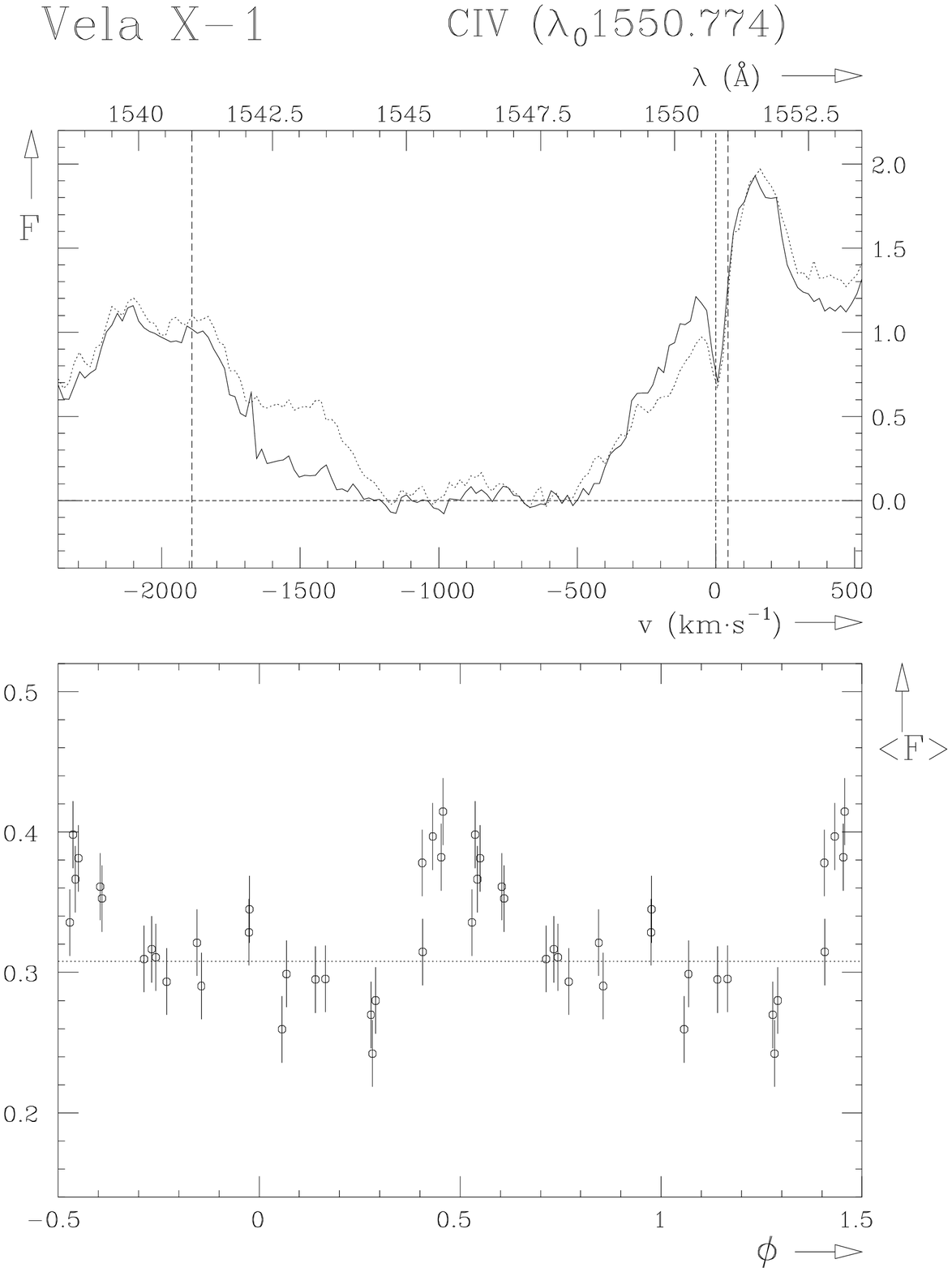,width=57mm}
      \psfig{figure=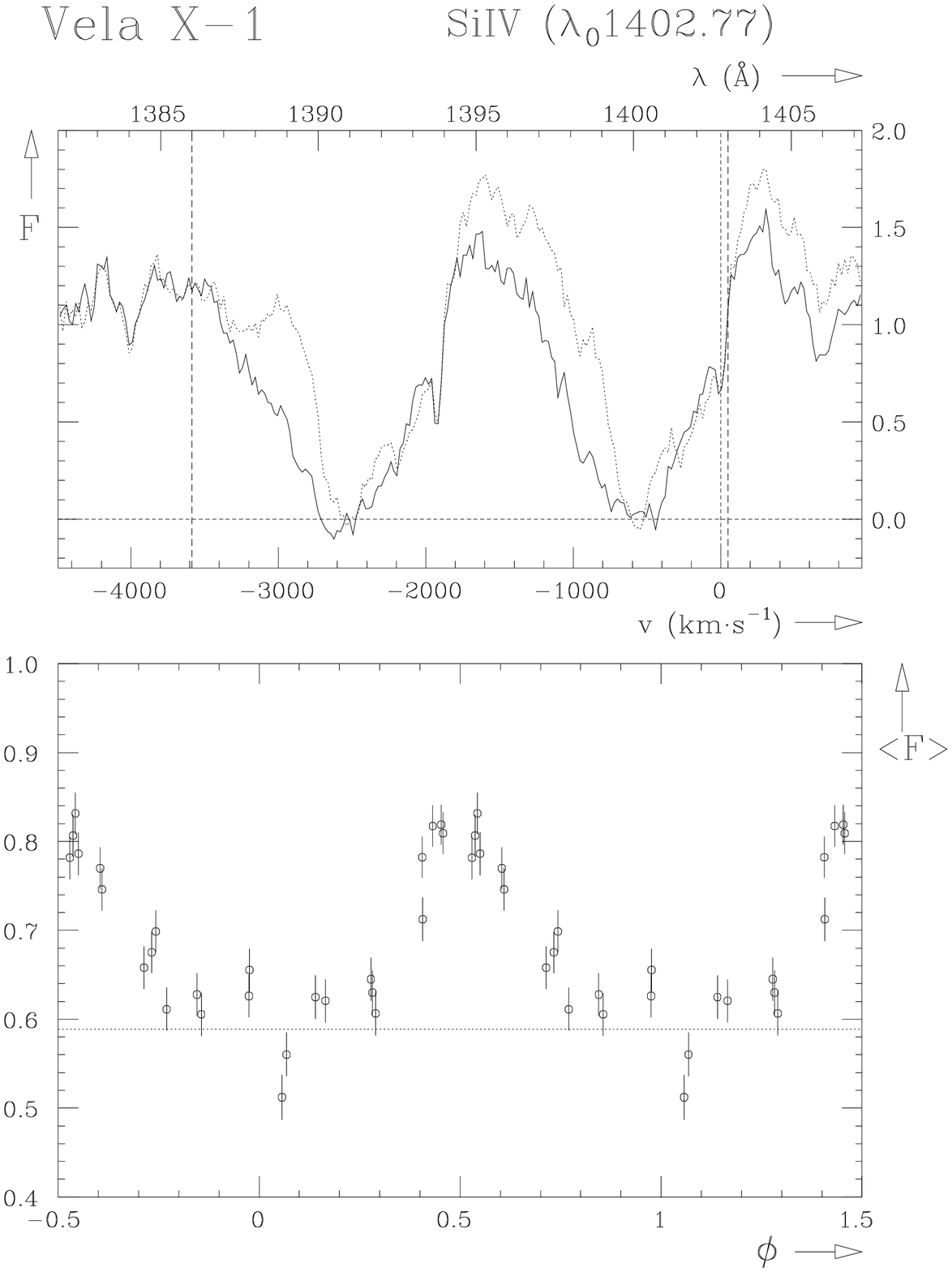,width=57mm}
      \psfig{figure=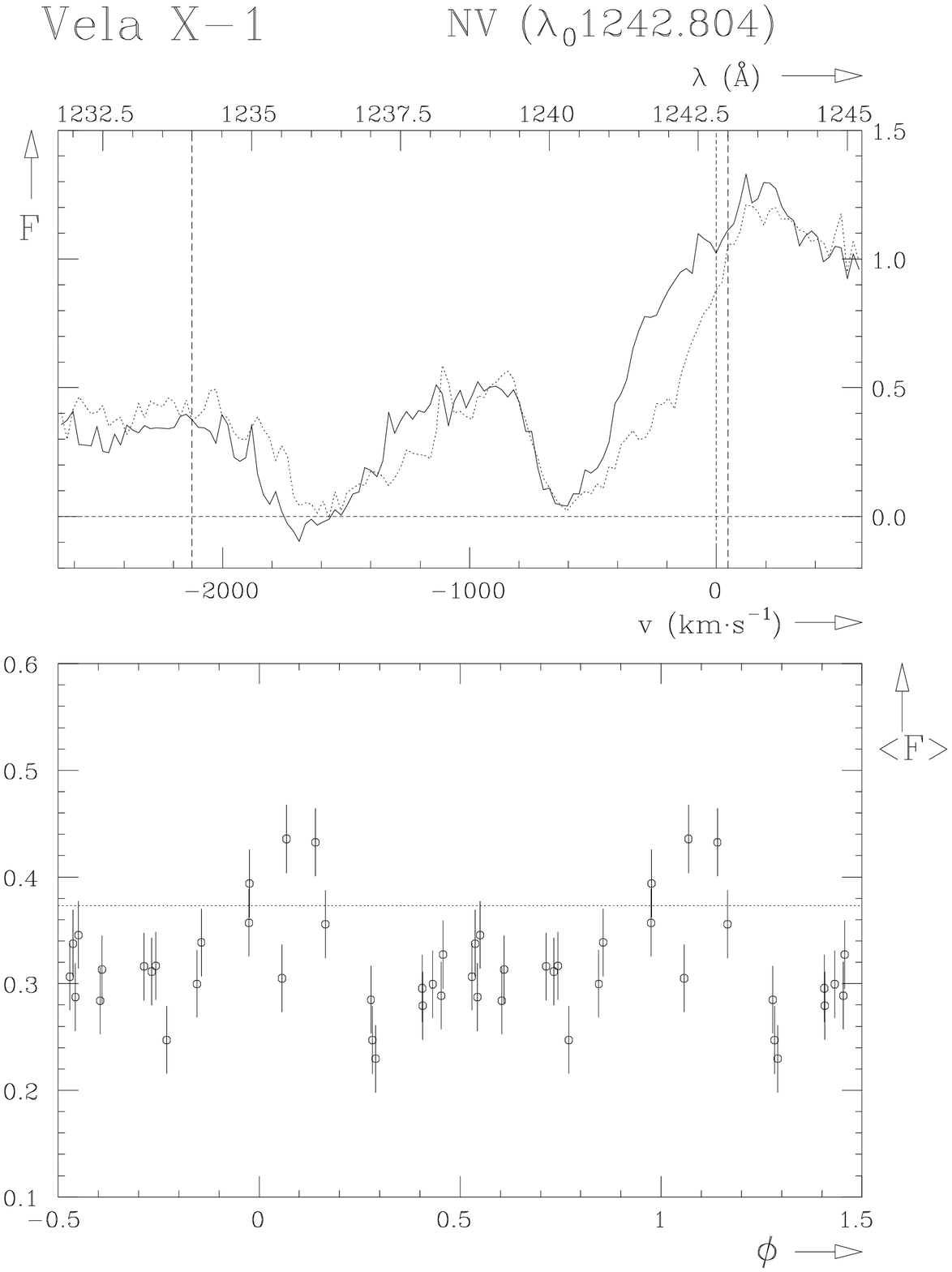,width=57mm}}
\hbox{\psfig{figure=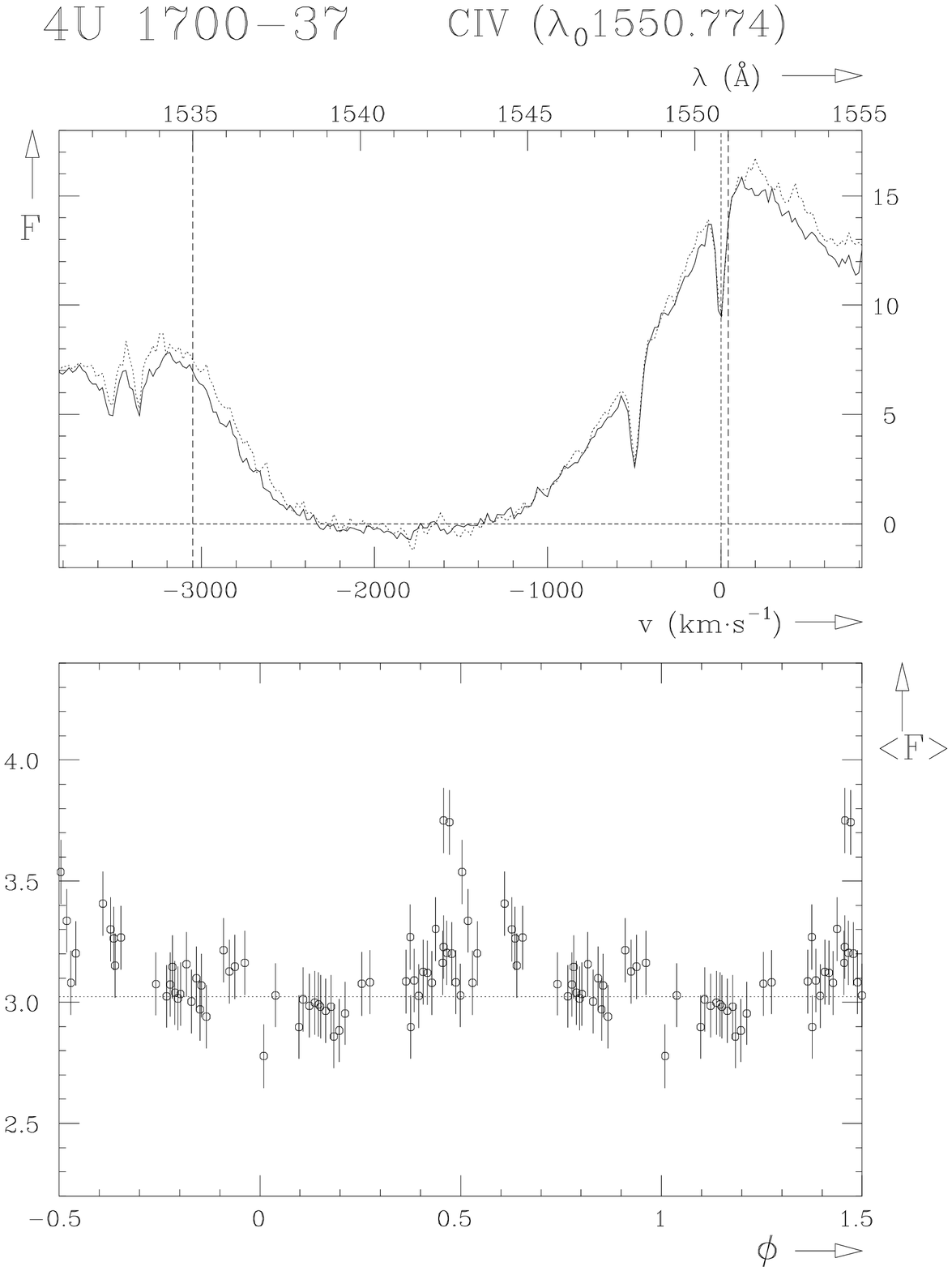,width=57mm}
      \psfig{figure=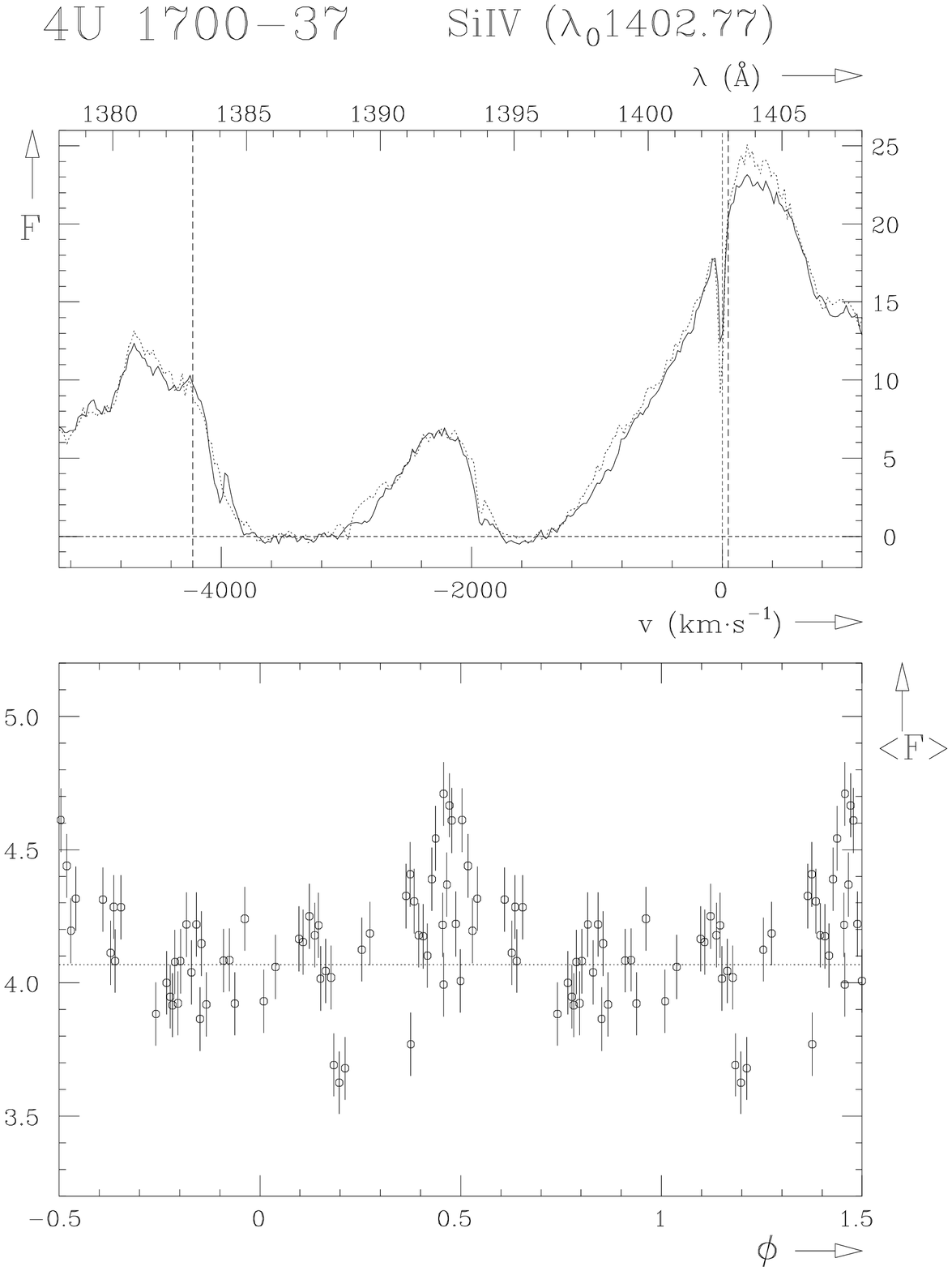,width=57mm}
      \psfig{figure=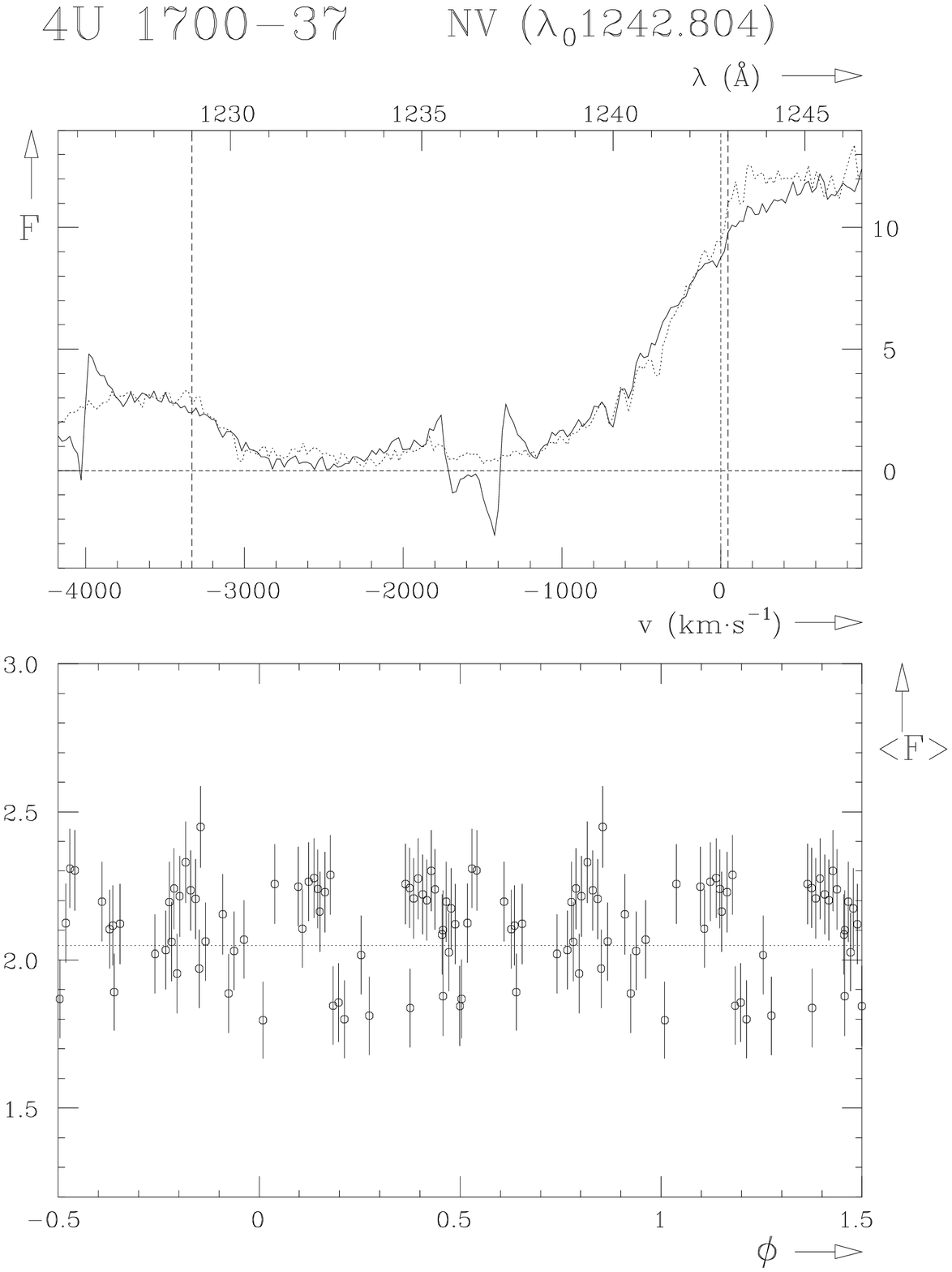,width=57mm}}
}}
\caption[]{Same as Fig.\ 1, but now for HD77581/Vela X-1 (top) and
HD153919/4U1700$-$37 (bottom).}
\end{figure*}

\subsection{Cyg X-1}

In HDE226868/Cyg X-1 (Fig.\ 1) all three lines show clear modulation of the
strength of the P-Cygni absorption with orbital phase due to the HM-effect.
For \Civ\ and \Siiv\ this was already observed by Treves et al.\ (1980). The
\Nv\ line shows the HM-effect too, although less pronounced. The modulation is
very strong and not confined to orbital phases near $\phi=0.5$: apparently, a
large fraction of the stellar wind is strongly ionized by the X-ray source,
suggesting the presence of a shadow wind in this system. The low resolution
does not allow a detailed study of the wind velocity structure, but it is
clear that the modulation of the absorption is visible up to the highest wind
velocities.

\subsection{LMC X-4}

In Sk-Ph/LMC X-4 (Fig.\ 1) the HM-effect is seen in all three lines, in
agreement with van der Klis et al.\ (1982). The \Nv\ profile is completely
absent at $\phi=0.5$, whereas part of the \Civ\ line remains visible.
High-resolution HST GHRS/STIS spectra (Boroson et al.\ 2001) show that the
remaining \Civ\ absorption at $\phi=0.5$ is mostly due to the intervening
interstellar medium. These HST data also show some \Nv\ absorption to persist
around $\phi=0.5$, which is probably photospheric. The strength of the P-Cygni
lines as a function of orbital phase indicate that the stellar wind is
confined to the X-ray shadow behind the OB star. The orbital modulation of the
\Civ\ and \Nv\ profiles demonstrates a complexity beyond that of a smooth
single wave, indicating a more complex mass flow. This has been confirmed with
recent HST GHRS/STIS spectra (Boroson et al.\ 1999; Kaper et al.\ in
preparation).

\subsection{SMC X-1}

In Sk 160/SMC X-1 (Fig.\ 1) the HM-effect is seen in all three lines (see also
van der Klis et al.\ 1982). The shape and amplitude of the orbital modulation
of the mean flux suggests that absorption is present in a tight orbital phase
range around $\phi=0$ only, i.e.\ a shadow wind. This is consistent with the
high X-ray luminosity of SMC X-1 (Hutchings 1974; van der Klis et al.\ 1982).

\subsection{Vela X-1}

In HD77581/Vela X-1 (Fig.\ 2) the HM-effect is observed in the \Civ\ and
\Siiv\ lines (Dupree et al.\ 1980) and the \Nv\ line (see Kaper et al.\ 1993).
The severely saturated \Civ\ line shows less orbital modulation than the just
saturated \Siiv\ line; remarkably, the \Nv\ line shows the inverse behaviour.
The variability in the UV lines of HD77581/Vela X-1 will be studied in more
detail in Sect.\ 4.1.

\subsection{4U1700$-$37}

In HD153919/4U1700$-$37 (Fig.\ 2) the P-Cygni lines are very strong and
saturated. The \Nv\ line does not show any significant variability with
orbital phase. The \Civ\ and \Siiv\ absorption lines are possibly slightly
weaker very near $\phi=0.5$. This would indicate a very small Str\"{o}mgren
zone, if the variability is due to the HM-effect. The variability in the \Civ\
line results from changes in the blue edge of the P-Cygni profile, while the
variability in the \Siiv\ line originates at wind velocities of about $-900$
km s$^{-1}$. However, the UV spectrum is affected by variable Raman scattered
emission lines which show a similar orbital-phase dependence (Kaper et al.\
1990, 1993).

\section{Modelling the UV resonance lines in HMXBs}

\subsection{Adapting the SEI code}

The high-mass X-ray binary is described as a spheroidal early-type star with a
spherically symmetric radially outflowing wind, and an orbiting X-ray point
source. The used coordinate system is based on the impact parameter $p$ and
the line-of-sight parameter $z$, in units of the stellar radius and with the
center of the early-type star at $(0,0)$:
\begin{equation}
\vec{x} = (p,z) \hspace{5mm} {\rm \&} \hspace{5mm} x = | \vec{x} | \, ,
\end{equation}
and with the spectrograph at $(0,-\infty)$. The velocity $v(x)$ of the
radiation driven stellar wind is normalised to $v(\infty)=1$ and parameterised
as (Lamers et al.\ 1987):
\begin{equation}
v(x) = v_0 + ( 1 - v_0 ) ( 1 - \frac{1}{x} )^{\gamma}
\end{equation}
where $v_0$ is the velocity at the base of the wind ($x=1$ and $\gamma > 0$):
\begin{equation}
v_0 = v (1) \simeq 0.01 v(\infty) \, ,
\end{equation}
which is of the order of the sound speed in the stellar wind.

%
%
\begin{figure}[]
\centerline{\psfig{figure=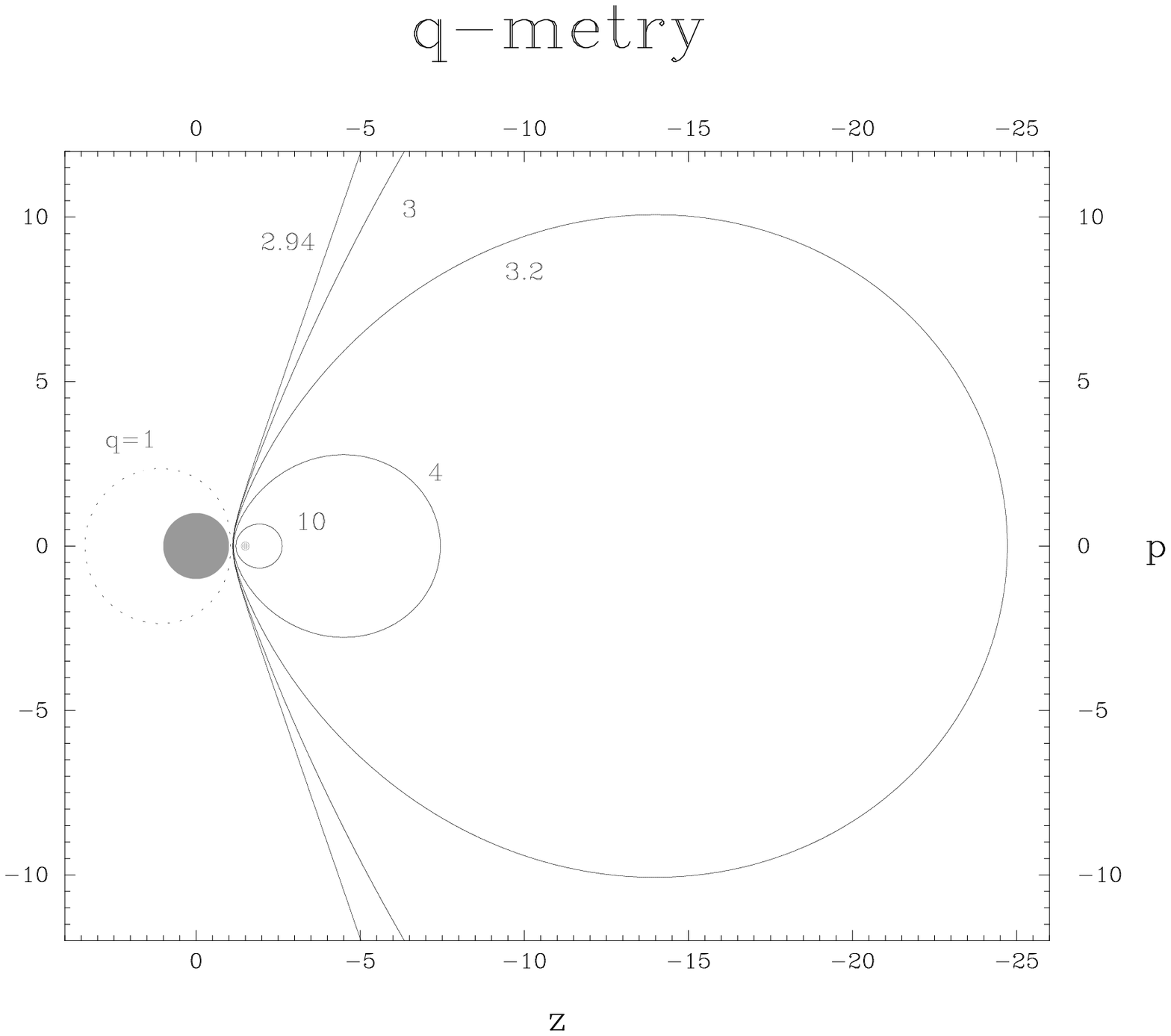,width=88mm}}
\caption[]{Sketch of the used coordinate system $(p,z)$ indicating the
location of the early-type star (large gray circle) and the X-ray point source
(tiny gray circle, not to scale) at a distance of 1.5 stellar radii, along
with surfaces of constant $q$ (using a velocity law with $\gamma=1$ and
$v_0=0.01$).}
\end{figure}

The X-ray point source at position ${\vec{x}}_{\rm X}$ affects the normal
abundance $q'(x)$ of a certain ion in the wind. The degree at which the wind
is disturbed is set by the parameter $q(\vec{x})$ (Hatchett \& McCray 1977):
\begin{equation}
q(\vec{x}) = \frac{ v(x) }{ v(x_{\rm X}) } 
             \frac{ x^2 }{ ( \vec{x} - {\vec{x}}_{\rm X} )^2 } \, .
\end{equation}
Approximately, at all points on a surface of constant $q$, the X-rays remove
the same fraction ${\eta}(q)$ of the ion under consideration (Tarter et al.\
1969). This leaves only a $1-{\eta}(q)$ contribution of this ion to the source
function $S(\vec{x})$ and optical depth ${\tau}(\vec{x})$. In our computations
we adopt a region enclosed by a surface set by the value of $q$, with
\begin{equation}
{\eta}( q(\vec{x}) ) = \left\{ \begin{array}{ll}
                                \eta & \mbox{ for $q(\vec{x}) > q$ } \\
                                0    & \mbox{ else }
                               \end{array}
                       \right.
\end{equation}
A negative value for $\eta$ enhances the abundance of the ion within the
Str\"{o}mgren zone. The X-ray point source is located inside a closed surface
of constant $q$ if
\begin{equation}
q < q_{\rm critical}
\end{equation}
with
\begin{equation}
q_{\rm critical} =\lim_{\begin{array}{l}
                         x \rightarrow \infty \\
                         p = 0
                        \end{array} } q(\vec{x}) = \frac{1}{v(x_{\rm X})} \, .
\end{equation}
Because of the singularity in ${\vec{x}}_{\rm X}$ when calculating
$q(\vec{x})$, we excluded a small spherical region around the X-ray point
source from the grid with a radius of one millionth of a stellar radius, which
is of the order of the radius of a neutron star. As an example surfaces of
constant q are shown (Fig.\ 3), for a velocity law with
\begin{equation}
\gamma = 1 \hspace{5mm} {\rm \&} \hspace{5mm} v_0 = 0.01
\end{equation}
and putting the X-ray point source at
\begin{equation}
x_{\rm X} = 1.5 \, ,
\end{equation}
which corresponds to
\begin{equation}
v(x_{\rm X}) = 0.34
\hspace{5mm} {\rm \&} \hspace{5mm}
q_{\rm critical} = 2.941 \, .
\end{equation}

The escape probability method (Castor 1970) makes a local approximation for
the source function $S_{\nu}(x)$ at comoving frequency $\nu$, namely the
Sobolev approximation
\begin{equation}
S_{\nu}(x) = \frac{ {\beta}_{c}(x) I_{\nu}^{\ast} + \epsilon B_{\nu}(x) }
                  { \beta + \epsilon }
\end{equation}
Here, $\beta$ is the escape probability of a line photon; ${\beta}_c$ the
penetration probability of a continuum photon; $\epsilon$ the ratio of
collisional over radiative de-excitations; $B_{\nu}$ the Planck function at
frequency $\nu$; and $I_{\nu}^{\ast}$ the stellar photospheric intensity at
frequency $\nu$. Due to the relatively low density of in the stellar wind, the
contribution of collisional (de)excitations to the source function of the UV
resonance lines can be neglected:
\begin{equation}
\epsilon = 0
\end{equation}

The SEI method (Lamers et al.\ 1987) incorporates an exact integration of the
radiative transfer equation, yielding the observed flux $F_v$ at the observed
velocity $v$:
\begin{equation}
\frac{ {\rm d}I_{\nu} }{ {\rm d}{\tau}_{\nu} } = I_{\nu} - S_{\nu}
     \longmapsto F_v
\end{equation}

The optical depth $\tau$ is parameterised as a function of velocity $v$:
\begin{equation}
{\tau}(v) = \left\{ \begin{array}{ll}
                     T \frac{ f(v) }{ \int_{v_0}^{v_1} f(v) {\rm d}v }
                       & \mbox{ for $v \leq v_1$ } \\
                     0 & \mbox{ else }
                    \end{array}
            \right.
\end{equation}
with
\begin{equation}
f(v) = \left( \frac{ v }{ v_1 }
       \right)^{{\alpha}_1}
       \left( 1 -
       \left( \frac{ v }{ v_1 }
       \right)^{\frac{ 1 }{ \gamma }}
       \right)^{{\alpha}_2} \, .
\end{equation}
The parameters $\alpha_{1}$ and $\alpha_{2}$ are defined as:
\begin{equation}
{\alpha}_1 = \frac{ 1 }{ \gamma } - t -2 \hspace{5mm} {\rm \&} \hspace{5mm}
{\alpha}_2 = s \, ,
\end{equation}
where $t$ and $s$ parameterize the ionization fraction $\kappa_{\rm i}$ as a
function of radial distance and velocity: 
\begin{equation}
\kappa_{\rm i}(x) \propto x^{-s} v(x)^{-t} \, .
\end{equation}
If $m$ is the dominant stage of ionization and $i$ is the ionization stage of
the ion under consideration, then $t=m-i$. If the abundance of the dominant
ion in the wind is constant, one gets
\begin{equation}
{\alpha}_1 = \frac{ 1 }{ \gamma } -2 \hspace{5mm} {\rm \&} \hspace{5mm}
{\alpha}_2 = 0 \, .
\end{equation}
If $v_1=1$, the ion is present up to the regions where the wind has reached
its terminal velocity.

The intrinsic line profile resembles a Gau{\ss}ian, with isotropic broadening
${\sigma}_v$ due to the thermal and turbulent motions --- hereafter referred
to as the ``turbulence'' (Lamers et al.\ 1987):
\begin{equation}
{\Phi}( {\Delta}v ) = \frac{ 1 }{ \sqrt{ \pi } {\sigma}_v } \exp{ -
                      \left( \frac{ {\Delta}v }{ {\sigma}_v } \right)^2 } \, .
\end{equation}
This introduces multiple resonance zones in the wind, which results in broader
and stronger absorption effectively moving the emission peak towards longer
wavelengths (in agreement with P-Cygni line obserservations).

In the case of doublets (most of the important UV resonance lines are
doublets), depending on the doublet separation the source function of the red
component includes a contribution from the blue component. Except for this
coupling, the principle of the SEI code is the same as for singlets (Lamers et
al.\ 1987). However, since we now include a Str\"{o}mgren zone we can no
longer use a one-dimensional radial grid to calculate the source function of
the red component (as is the case for the SEI method). The contribution of the
source function of the blue component to the source function of the red
component depends on whether the coupled blue component points are situated
inside the Str\"{o}mgren zone. Hence the radial profile of the source function
depends on the position of the Str\"{o}mgren zone. This is solved by replacing
the one-dimensional grid when calculating the source function of the red
component by a two-dimensional axi-symmetric grid.

The SEI method allows the inclusion of a photospheric component in the input
line spectrum, but we did not use this option in the model profiles shown
below.

%
%
\begin{figure*}[]
\centerline{\psfig{figure=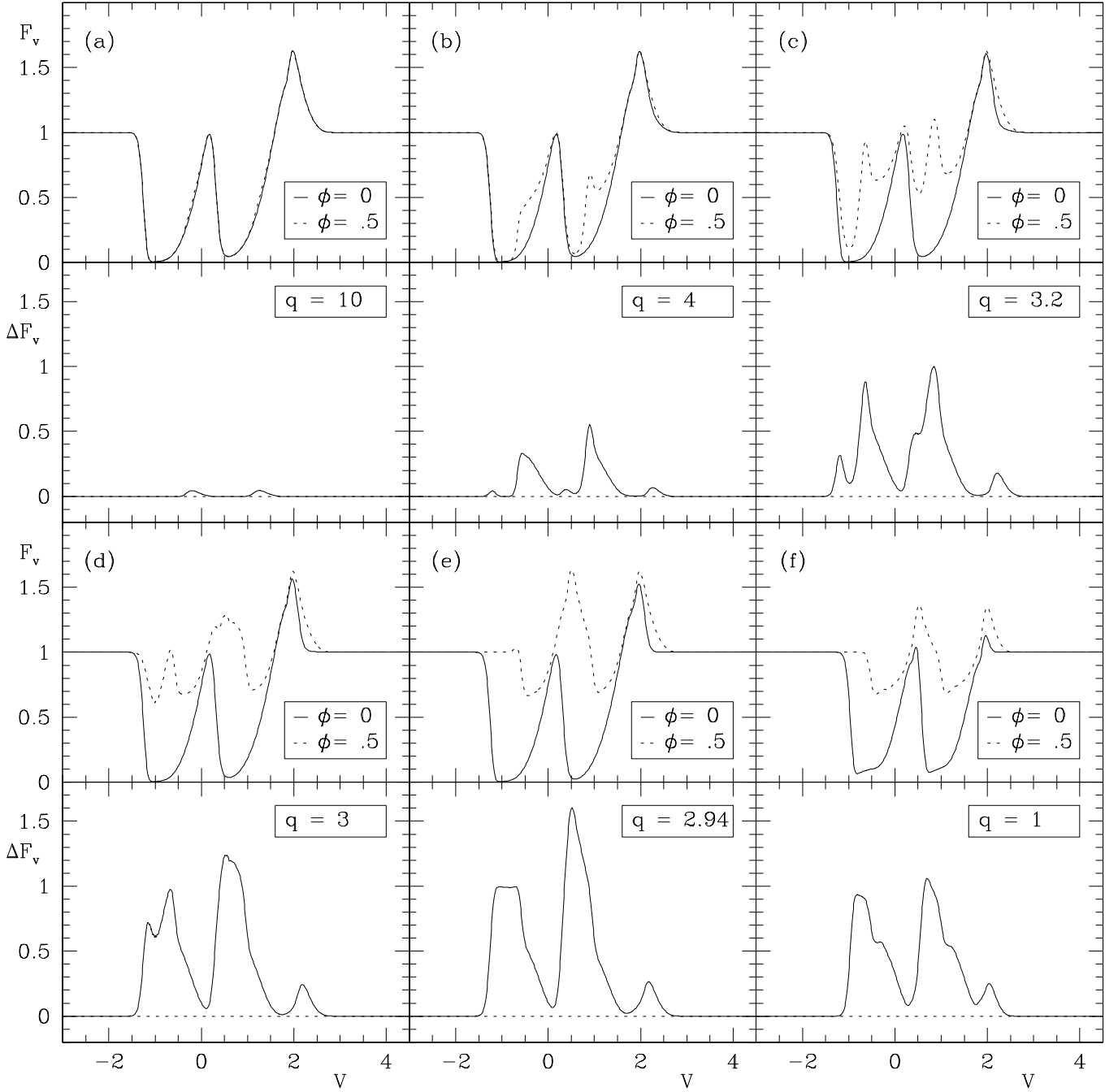,width=180mm}}
\caption[]{Changing the size of the X-ray ionization zone: SEI model profiles
for a HMXB with a wind velocity law with $\gamma=1$, $v_0=0.01$ and
${\sigma}_v=0.2$, $x_{\rm X}=1.5$ ($v_{\rm X}=0.34$), doublet separation of
$1.5$, integrated optical depth of the blue component of $T=100$ (twice that
of the red component), dominant ionization stage in the uniformly ionized
undisturbed wind, and no additional source of emission. The effect of
different sizes $q$ of the Str\"{o}mgren zone is illustrated (closed surfaces
for $q>2.941$). The top panel shows the profile at X-ray eclipse (drawn line,
$\phi=0$) and at $\phi=0.5$; the bottom panel presents the difference profile.
Note that (much) less absorption is observed at $\phi=0.5$ when the X-ray
source is in the line of sight.}
\end{figure*}

%
%
\begin{figure*}[]
\centerline{\psfig{figure=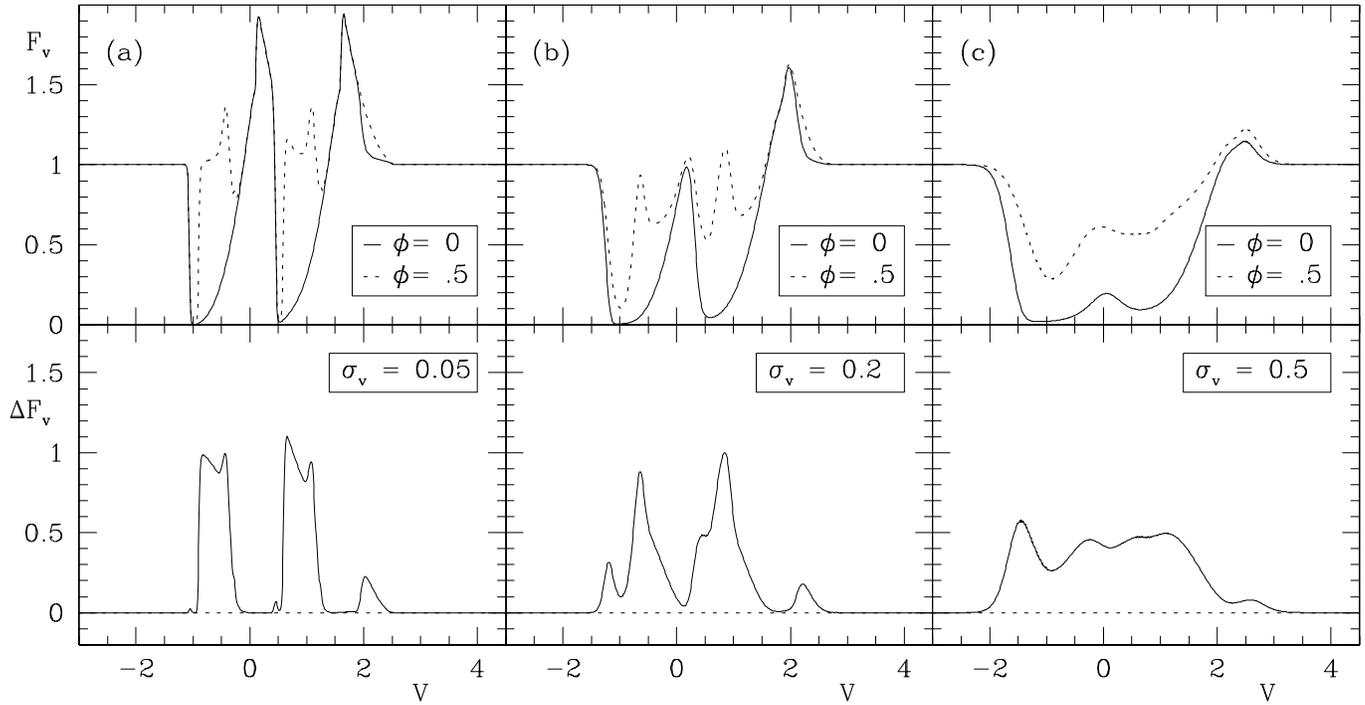,width=180mm}}
\caption[]{The effect of turbulence: same as Fig.\ 4, but for $q=3.2$ and
different levels of wind turbulence ${\sigma}_v$.}
\end{figure*}

%
%
\begin{figure*}[]
\centerline{\psfig{figure=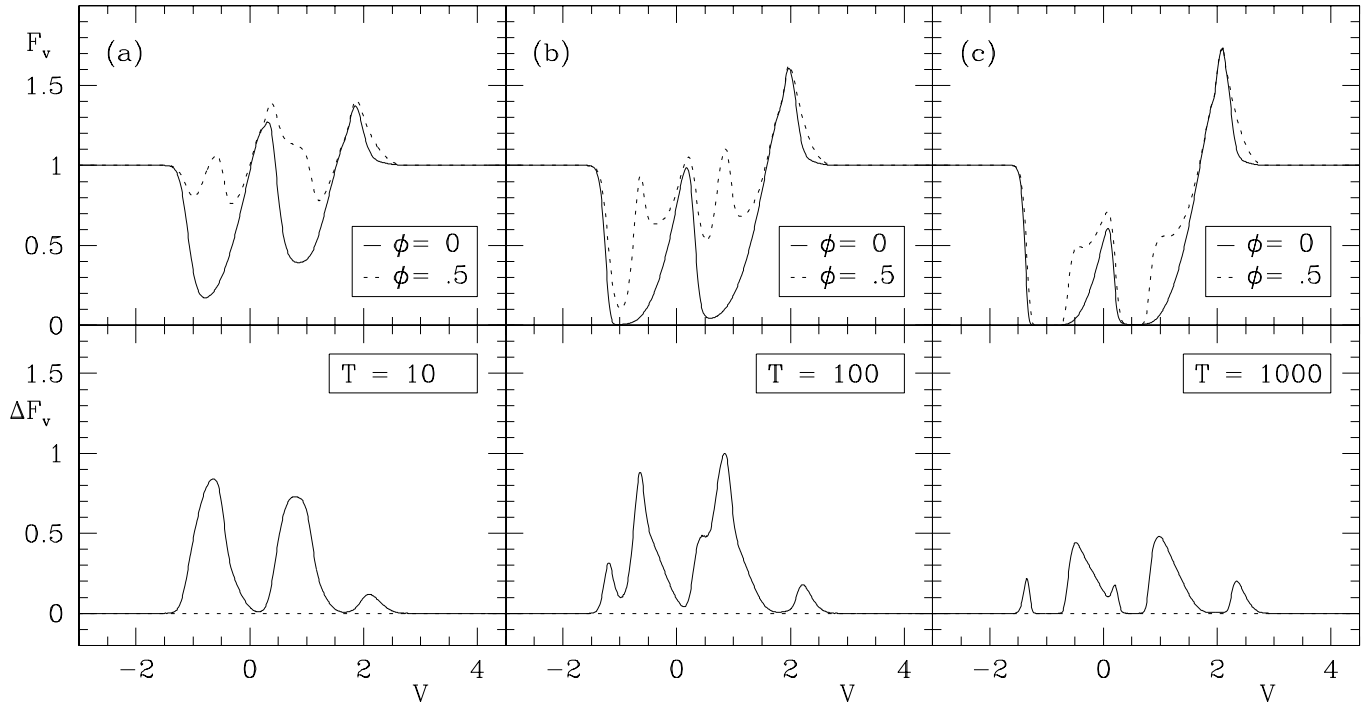,width=180mm}}
\caption[]{From weak to strong: same as Fig.\ 4, but for $q=3.2$ and
illustrating the effect of increased integrated optical depth $T$.}
\end{figure*}

%
%
\begin{figure*}[]
\centerline{\psfig{figure=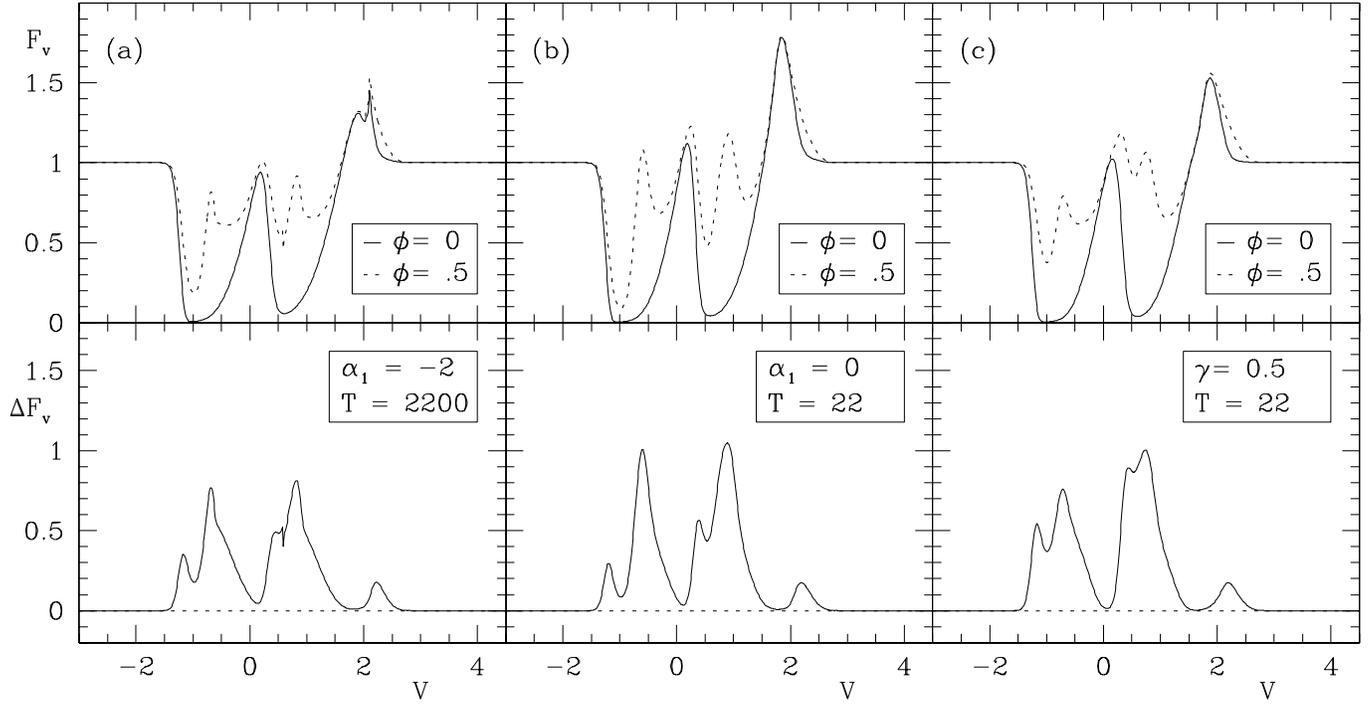,width=180mm}}
\caption[]{Same as Fig.\ 4, but for $q=3.2$ and illustrating the effect of
different parameterisations of the optical depth (via $\alpha_1$) and the
velocity law (via $\gamma$).}
\end{figure*}

%
%
\begin{figure*}[]
\centerline{\psfig{figure=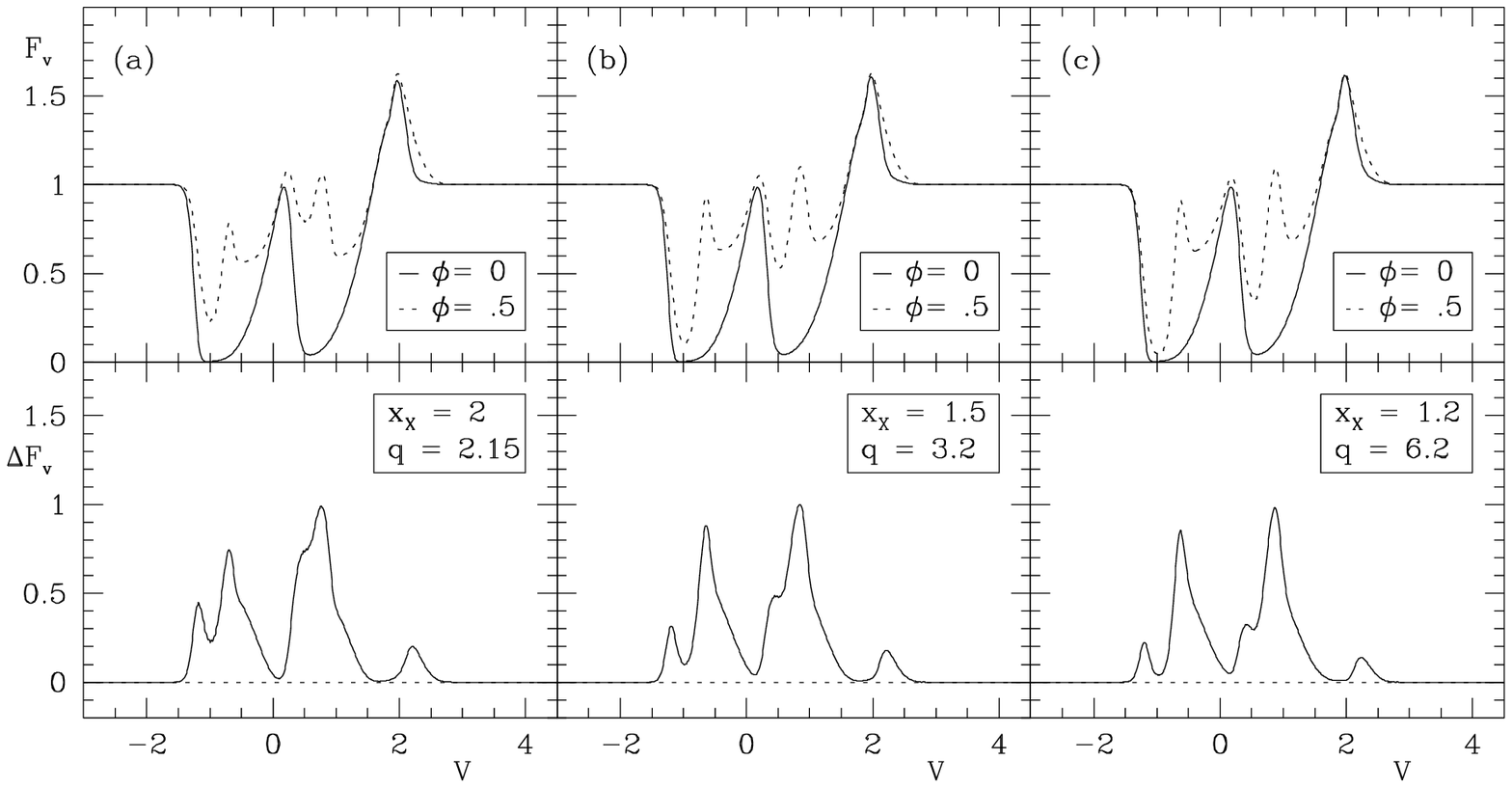,width=180mm}}
\caption[]{Same as Fig.\ 4, but for $q=3.2$ and illustrating the effect of
different distances $x_{\rm X}$ of the X-ray source.}
\end{figure*}

The accuracy of the calculations depends on the grid $I_{\rm g}$, which
samples the intrinsic line profile. The number of points used in the grid is
$I_{\rm g,max}$. In the original code a choice of $I_{\rm g,max}=11$ yields
satisfactory results, but if a Str\"{o}mgren zone is included this value
should be some two orders of magnitude larger to suppress the numerical noise.
Stronger turbulence demands larger $I_{\rm g,max}$.

\subsection{Application of the modified SEI model}

To illustrate how the important model parameters affect the shape of the
spectral line profile, model spectra are shown for a hypothetical resonance
doublet of a HMXB system. The canonical model consists of a HMXB with $x_{\rm
X}=1.5$, wind velocity law with $\gamma=1$, $v_0=0.01$ and ${\sigma}_v=0.2$,
doublet separation of $1.5$, integrated optical depth of the blue component of
$T=100$ (twice that of the red component), dominant ionization stage in the
uniformly ionized undisturbed wind ($t,s=0$), no additional source of emission
($\epsilon=0$), no photospheric absorption profile, and a Str\"{o}mgren zone
of size $q=3.2$.

Figs.\ 4 to 8 illustrate the effect of varying several of these parameters.
The top panels present the UV resonance profiles at two orbital phases: at
X-ray eclipse ($\phi=0$, drawn line) a large fraction of the Str\"{o}mgren
zone is behind the OB supergiant so that only a small part of the observable
stellar wind is affected. When the X-ray source is located in the absorbing
column in front of the OB supergiant ($\phi=0.5$), the blue-shifted absorption
trough is reduced in strength. Note that also the emission peak of the P-Cygni
profile is affected, though less pronounced because the wind volume
contributing to the P-Cygni emission is much larger in comparison to the
volume of the absorbing column. The bottom panels show the difference spectra
($\phi=0.5$ minus $\phi=0$).

The size $q$ of the Str\"{o}mgren zone strongly affects the spectral line
variability (Fig.\ 4). Small Str\"{o}mgren zones (large $q$) leave little
trace in the spectral line profile, apart from some diminished absorption at
$\phi=0.5$ at velocities between about $-v_{\rm X}$ and 0 (i.e.\ not centred
at $-v_{\rm X}$). A larger Str\"{o}mgren zone enhances the HM-effect
especially near velocities $<-v_{\rm X}$, also reducing the absorption in the
blue wing of the line profile, and reduces the emission at positive velocities
at $\phi=0$ (notably in the red component of the doublet). Saturation is not
maintained for Str\"{o}mgren zones that cover a significant fraction of the
hemisphere in front of the star. However, open Str\"{o}mgren zones leave only
a small area behind the primary unaffected (in the most extreme case this is a
shadow wind), and hence cause diminished absorption and emission at all
orbital phases, thereby reducing the contrast in line profile shape between
transit and eclipse of the X-ray source. The appearance of the HM-effect is
most prominent for $q{\sim}q_{\rm critical}$. The definition of $q$ allows a
Str\"{o}mgren zone extending into the X-ray shadow behind the supergiant;
obviously, this can only happen if scattering of X-rays by ions in the stellar
wind is important.

The presence of turbulence in the wind causes the line profile to broaden,
which results in less pronounced emission (Fig.\ 5). The HM-effect is also
smoothed in velocity space. The amplitude of variability at positive
velocities and between 0 and $-v_\infty$ decreases, whilst the amplitude of
variability in the blue wing ($v<-v_\infty$) increases. The blue wing
variability becomes dominant in the case of strong turbulence. In the absence
of turbulence it is difficult to hide the HM-effect unless the Str\"{o}mgren
zone is much smaller than the projected stellar disk.

Greater optical depth enhances the extent to which the absorption is
maintained when the Str\"{o}mgren zone moves through the line of sight (Fig.\
6). Note that for heavily saturated lines with turbulence, as for unsaturated
lines, it is not straightforward to determine the exact value of $v_\infty$,
and careful modelling is required.

The ionization stage of the ion, relative to the dominant ionization stage in
the (undisturbed) wind is of importance for the parameterisation and
normalisation of the optical depth. If the ion corresponds to an ionization
stage that is one level below the dominant ionization stage, then
$\alpha_1=-2$ instead of $-1$ in the canonical example (see Eq.\ 16). For the
same integrated optical depth, this would yield unsaturated and rather
triangular shaped absorption profiles with little emission. In the opposite
case, if the ion corresponds to one level above the dominant ionization stage,
then $\alpha_1=0$. This yields a heavily saturated line profile with stronger
emission and somewhat enhanced HM-effect in the absorption part between
$-v_{\rm X}$ and $-v_\infty$, compared to the canonical case. From the
normalisation of the optical depth (Eqs.\ 14 \& 15) it follows that $\tau/T$
at $v=v_\infty$ ($=v_1$) is 0.010, 0.22, or 1.0 for $\alpha_1=-2$, $-1$ and 0,
respectively. Scaling the total optical depth $T$ to 2200 and 22 for
$\alpha_1=-2$ and 0, respectively, the line profiles and appearance of the
HM-effect (Fig.\ 7) are similar to the canonical case. It may therefore be
difficult to distinguish between different values of $\alpha_1$, and prior
knowledge of the dominant ionization species is required.

The parameterisation of the optical depth involves the velocity law, and hence
the parameter $\gamma$. Changing $\gamma$ from 1 to 0.5 requires rescaling to
$q=1.87$, $\alpha_1=0$ and $T=22$ (Eqs.\ 4 \& 15). The resulting variability
is not exactly a scaled version of the canonical case (Fig.\ 7), because $q$
depends on both $v$ and $x$. Most notably, the blue wing HM-effect is stronger
and saturation more difficult to maintain.

A different distance $x_{\rm X}$ of the X-ray source implies that different
parts of the wind are probed. Changing to $x_{\rm X}=1.2$ and 2.0 requires
rescaling to $q=6.2$ and 2.15, respectively. Increasing the distance of the
X-ray source enhances the HM-effect in the blue wing at the expense of the
HM-effect in the absorption part between $-v_{\rm X}$ and $-v_\infty$, and
also makes it more difficult for a saturated line to remain saturated (Fig.\
8). If the X-ray source is located close to the primary, then the HM-effect in
the emission at positive velocities is less pronounced.

\section{Model fits to the ultraviolet resonance lines of HD77581/Vela X-1 \&
HD153919/4U1700$-$37}

%
%
\begin{table}[]
\caption[]{SEI model parameters for fits to the observed profiles and
variability of UV resonance lines in the spectra of HD77581/Vela X-1 and
HD153919/4U1700$-$37.}
\begin{tabular}{lrrrrrr}
\hline\hline
Line          &
$T$           &
$\alpha_1$    &
$\alpha_2$    &
$\sigma$      &
$v_\infty$    &
$q$           \\
\hline
\multicolumn{7}{l}{\it HD77581/Vela X-1:} \\
\Nv           &
            3 &
            1 &
         $-1$ &
         0.45 &
          600 &
          2.9 \\
\Siiv         &
          300 &
         $-1$ &
            0 &
         0.45 &
          600 &
          2.9 \\
\Civ          &
          200 &
            0 &
            0 &
         0.45 &
          600 &
          2.9 \\
\Aliii        &
           20 &
         $-1$ &
            0 &
         0.45 &
          600 &
          2.9 \\
\multicolumn{7}{l}{\it HD153919/4U1700$-$37:} \\
\Nv           &
           30 &
            0 &
            0 &
         0.15 &
         1700 &
         $>4$ \\
\Siiv         &
$2\times10^4$ &
         $-2$ &
            0 &
         0.15 &
         1700 &
         $>4$ \\
\Civ          &
         5000 &
         $-2$ &
            0 &
         0.15 &
         1700 &
         $>4$ \\
\hline
\end{tabular}
\end{table}

The modified SEI code is used to fit the line profiles at two orbital phases
($\phi=0$, X-ray eclipse, and $\phi=0.5$) of \Nv, \Siiv, \Civ\ and \Aliii\
resonance lines in HD77581/Vela X-1 and \Nv, \Siiv\ and \Civ\ in
HD153919/4U1700$-$37. The models are constrained by the following conditions:
(i) the values for $\gamma$, $v_\infty$, $\sigma$ and $q$ are the same for all
lines in a given HMXB; (ii) the value for $x_{\rm X}$ follows from the orbital
parameters (Table 1); (iii) the dominant ionization stage is estimated from
the spectral type (Lamers et al.\ 1999). In HD77581 N$^{2+}$, Si$^{3+}$ and
C$^{2+}$ are expected to dominate, while in HD153919 probably N$^{3+}$,
Si$^{4+}$ and C$^{4+}$ dominate. Al$^{2+}$ is assumed to dominate in HD77581,
as the observed \Aliii\ line is very strong. The ionization is uniform
throughout the undisturbed wind, i.e.\ $\alpha_2=0$ (except N$^{4+}$ in
HD77581). Note that the \Nv\ line suffers from strong absorption by the wing
of Ly-$\alpha$. A sudden jump in intensity near the maximum of the \Siiv\ and
\Civ\ emission in HD153919/4U1700$-$37 is due to numerical difficulties. These
lines were calculated by sampling the intrinsic line profile over $\pm5\sigma$
with $I_{\rm g,max}=3333$ (instead of $\pm3\sigma$ with $I_{\rm g,max}=1111$),
suppressing the spikey feature by half. The resulting line profiles are shown
in Figs.\ 9, 10 \& 14 and the model parameters are listed in Table 3.

%
%
\begin{figure*}[]
\centerline{\psfig{figure=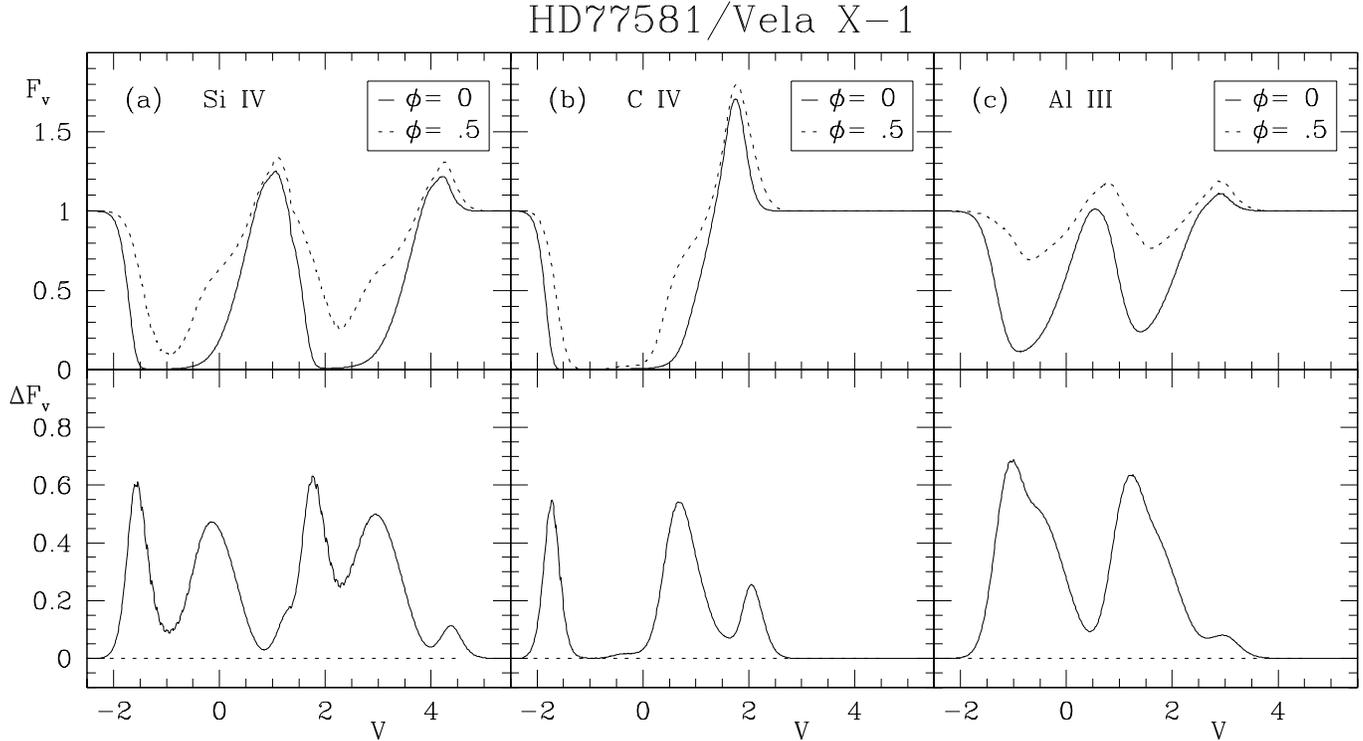,width=180mm}}
\caption[]{SEI models that approximately fit the observed line profiles and
variability of \Siiv, \Civ\ and \Aliii\ in the UV spectrum of HD77581/Vela
X-1. Parameters are summarised in Table 3, and $v=1$ corresponds to 600 km
s$^{-1}$.}
\end{figure*}

%
%
\begin{figure*}[]
\centerline{\psfig{figure=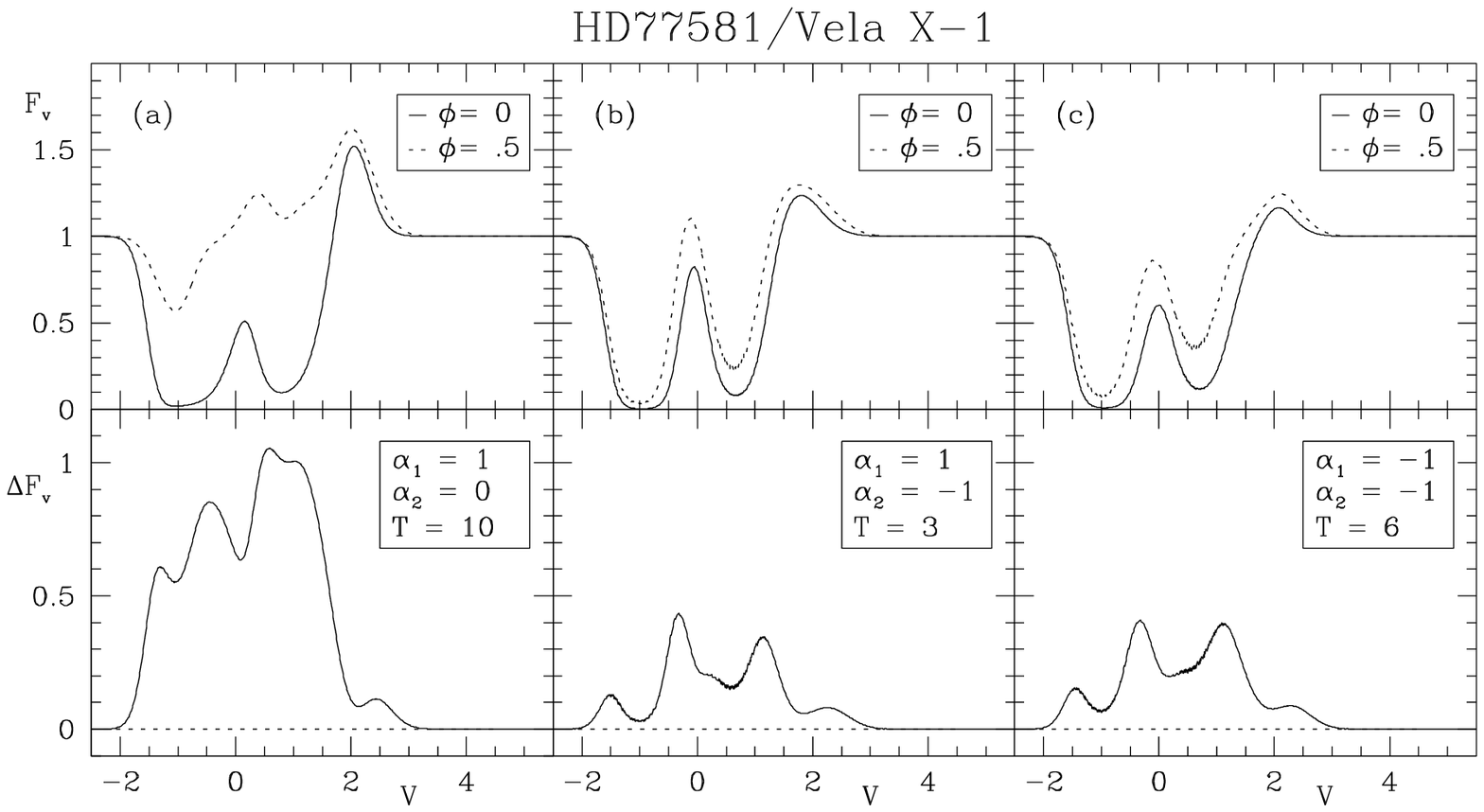,width=180mm}}
\caption[]{Same as Fig.\ 9, but for the \Nv\ line in the UV spectrum of
HD77581/Vela X-1 (see also Table 3).}
\end{figure*}

The orbital modulation of the strong wind lines in HD77581 (with the exception
of \Nv) is mainly due to the HM-effect. Some detailed changes are probably
related to the accretion flow in the system (Sadakane et al.\ 1985; Kaper et
al.\ 1994); since these structures are not included in the model, they cannot
be reproduced. The modified SEI model can naturally explain the (near) absence
of orbital modulation in the resonance lines of HD153919, which is one of the
main motivations for this study.

\subsection{HD77581/Vela X-1}

\subsubsection{Strong wind lines}

Adopting the ionization structure of the stellar wind as expected for single
stars of the same effective temperature as HD77581, and adopting
$v_\infty=600$ km s$^{-1}$, the shape and variability of the \Siiv, \Civ\ and
\Aliii\ (\siip\ resonance doublet: restwavelengths 1854.716 \& 1862.790,
separation 1302 km s$^{-1}$) line profiles are reproduced rather well (Fig.\
9, compare with Fig.\ 11).

%
%
\begin{figure*}[]
\centerline{\vbox{
\hbox{\psfig{figure=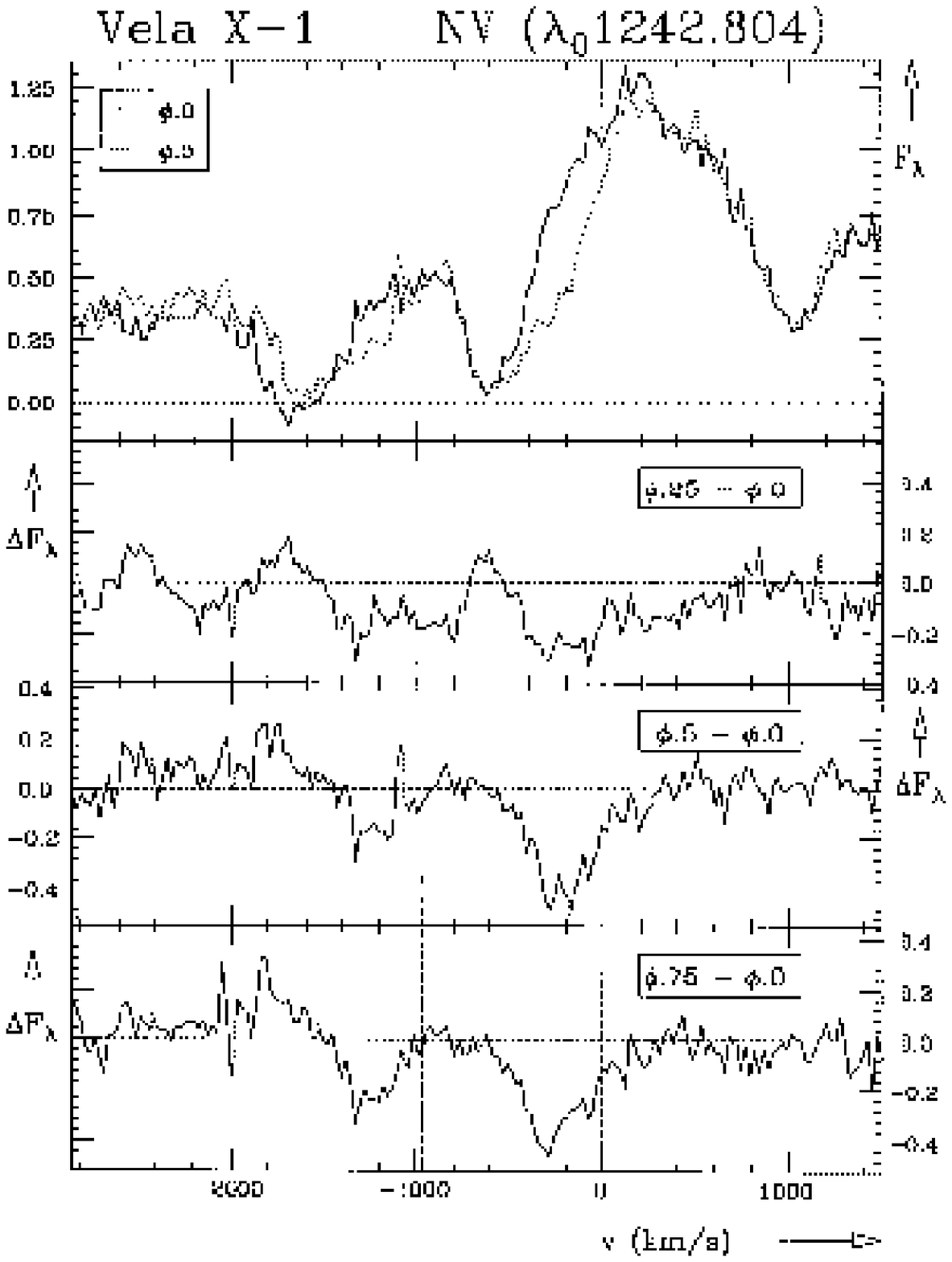,width=85mm}
      \hspace{9mm}
      \psfig{figure=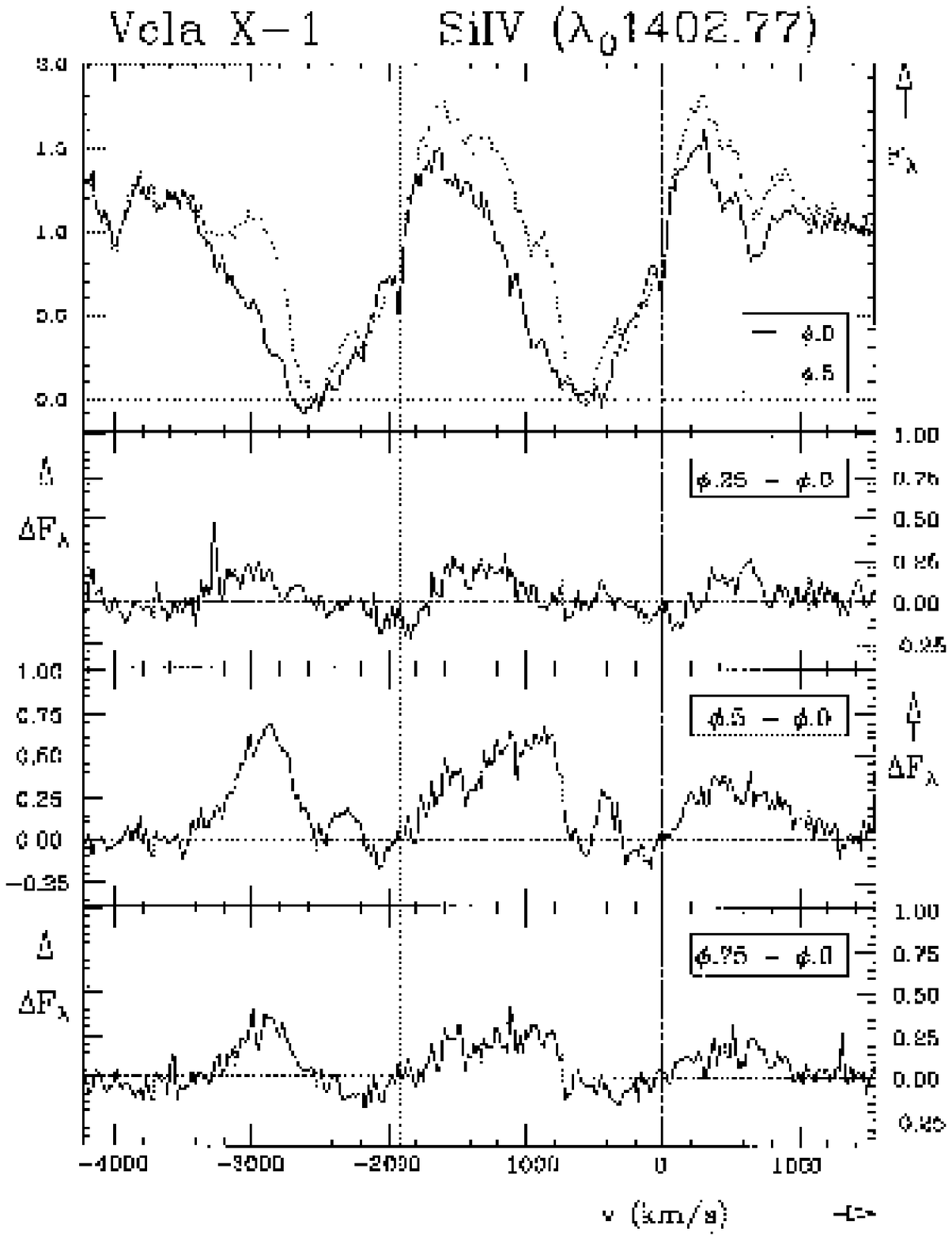,width=85mm}}
\vspace{0.8mm}
\hbox{\psfig{figure=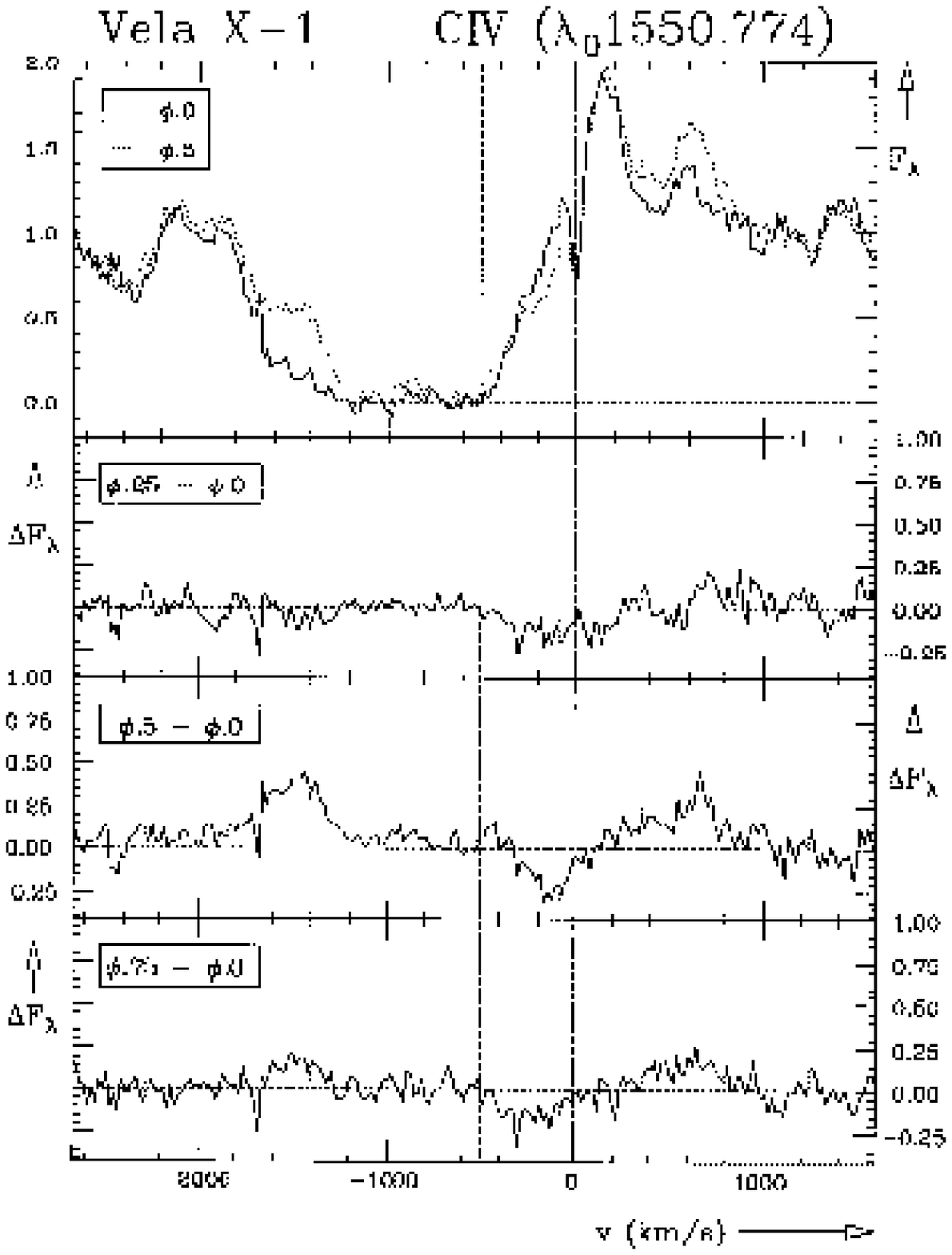,width=85mm}
      \hspace{9mm}
      \psfig{figure=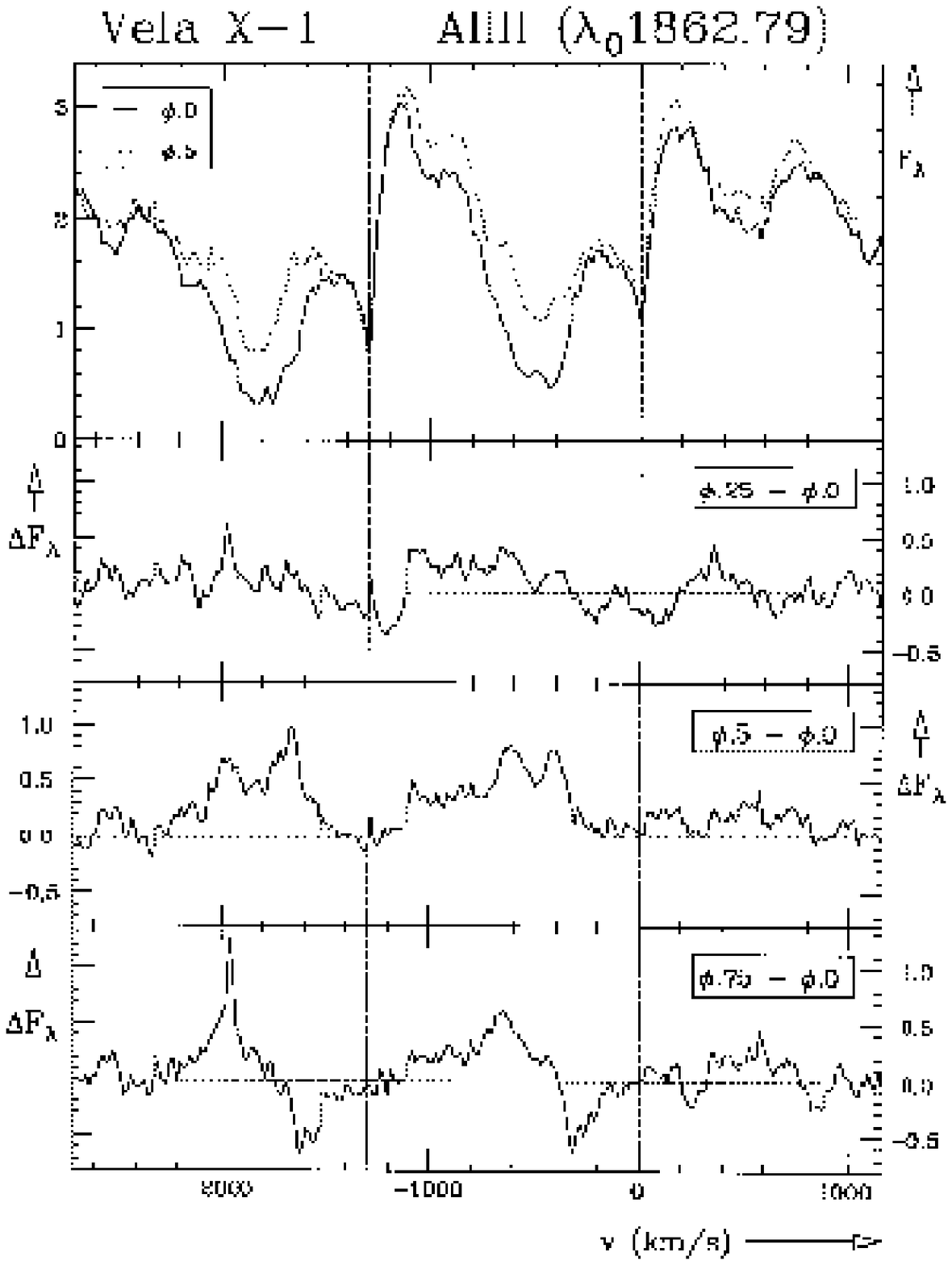,width=85mm}}
}}
\caption[]{Phase $\phi=0$ (solid) and $\phi=0.5$ (dotted) spectra for the \Nv,
\Siiv, \Civ\ and \Aliii\ resonance lines in HD77581/Vela X-1, with rest
wavelengths indicated by vertical dashed lines and the velocity axis defined
by the redmost component (labelled above the graph). The differences between
the $\phi=0.25$, 0.5 and 0.75 spectra relative to the $\phi=0$ spectrum are
plotted below.}
\end{figure*}

%
%
\begin{figure*}[]
\centerline{\vbox{ \hbox{\psfig{figure=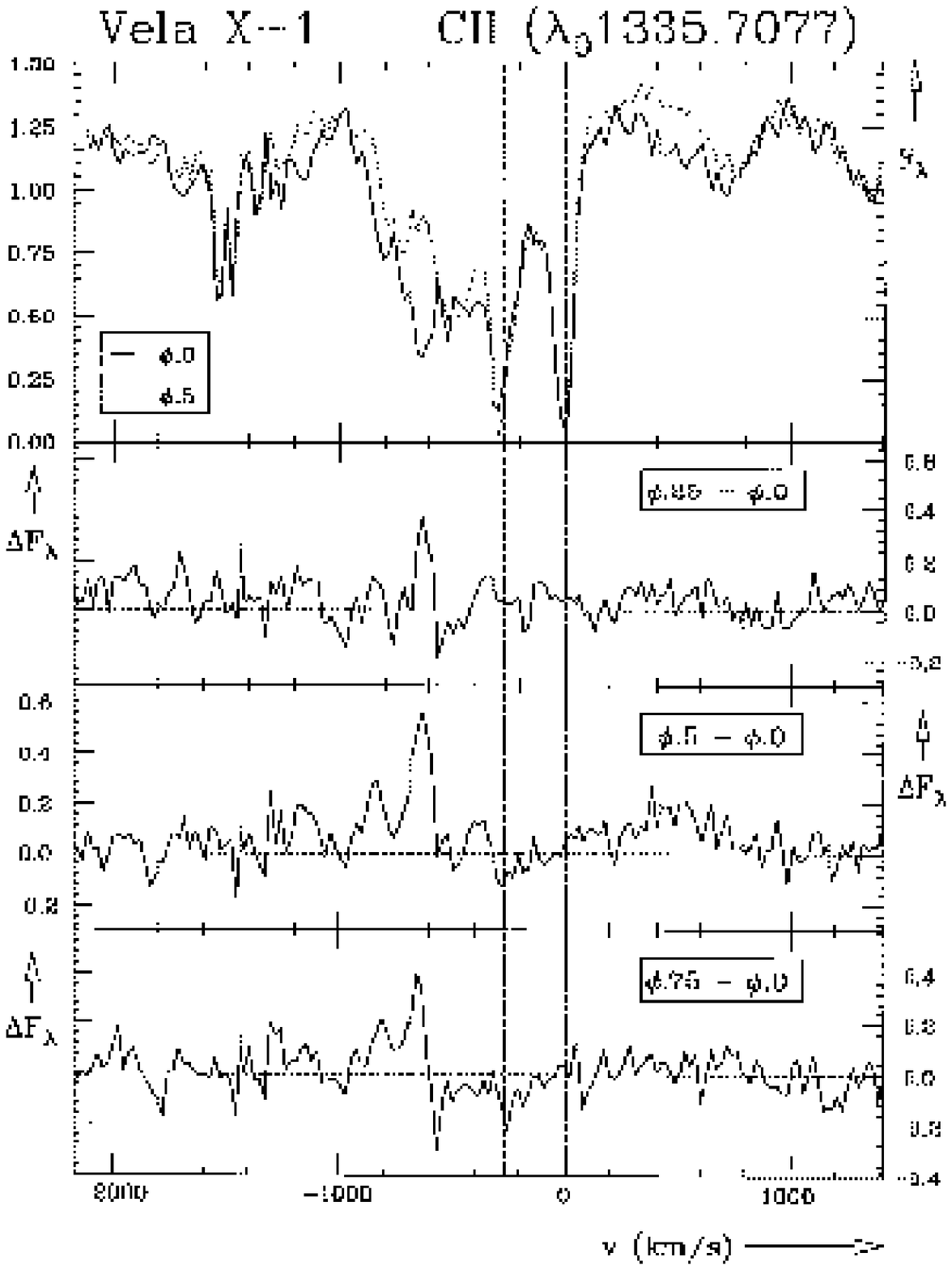,width=88mm}
\hspace{3mm} \psfig{figure=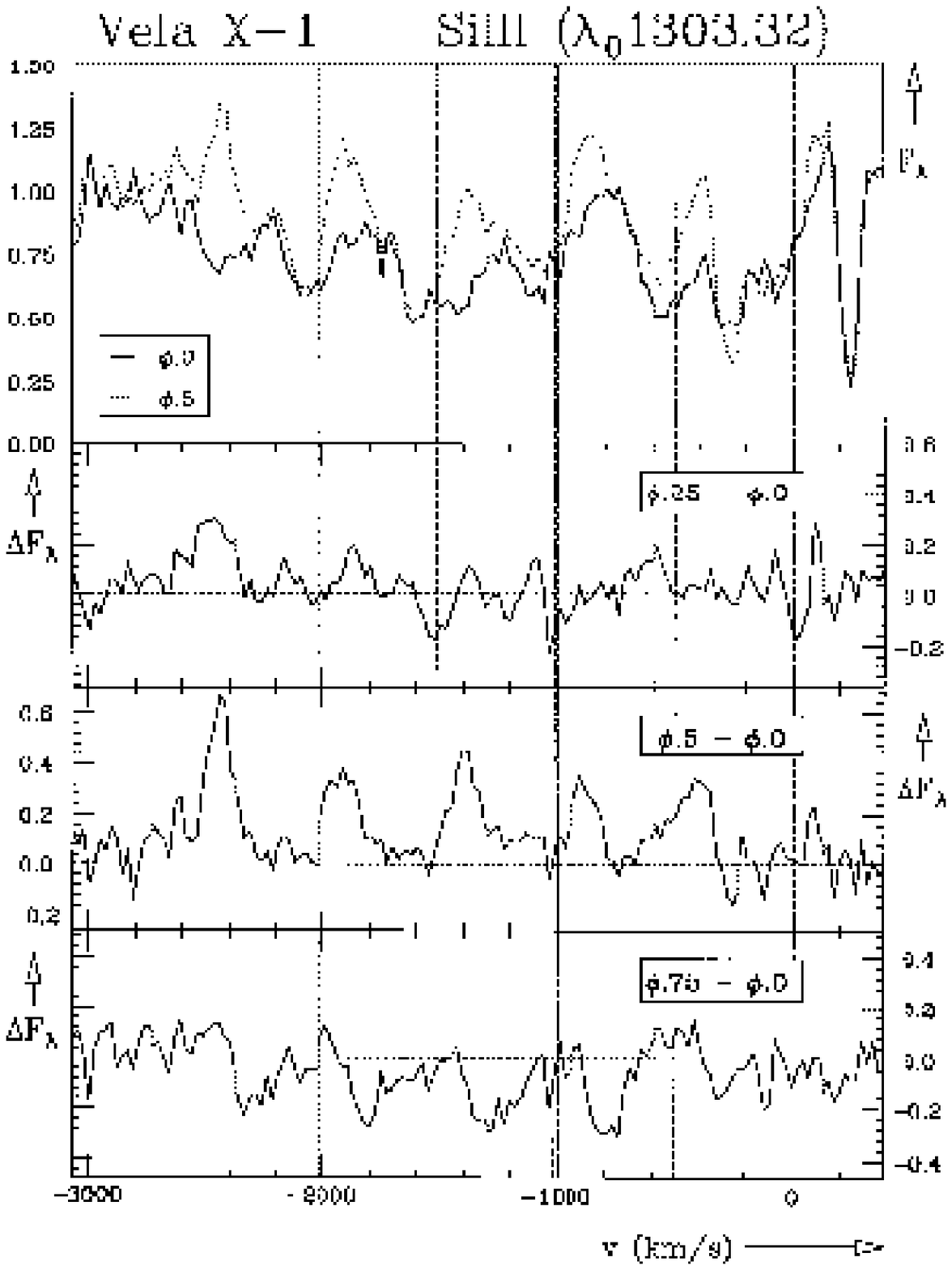,width=88mm}}
\vspace{0.8mm}
\hbox{\psfig{figure=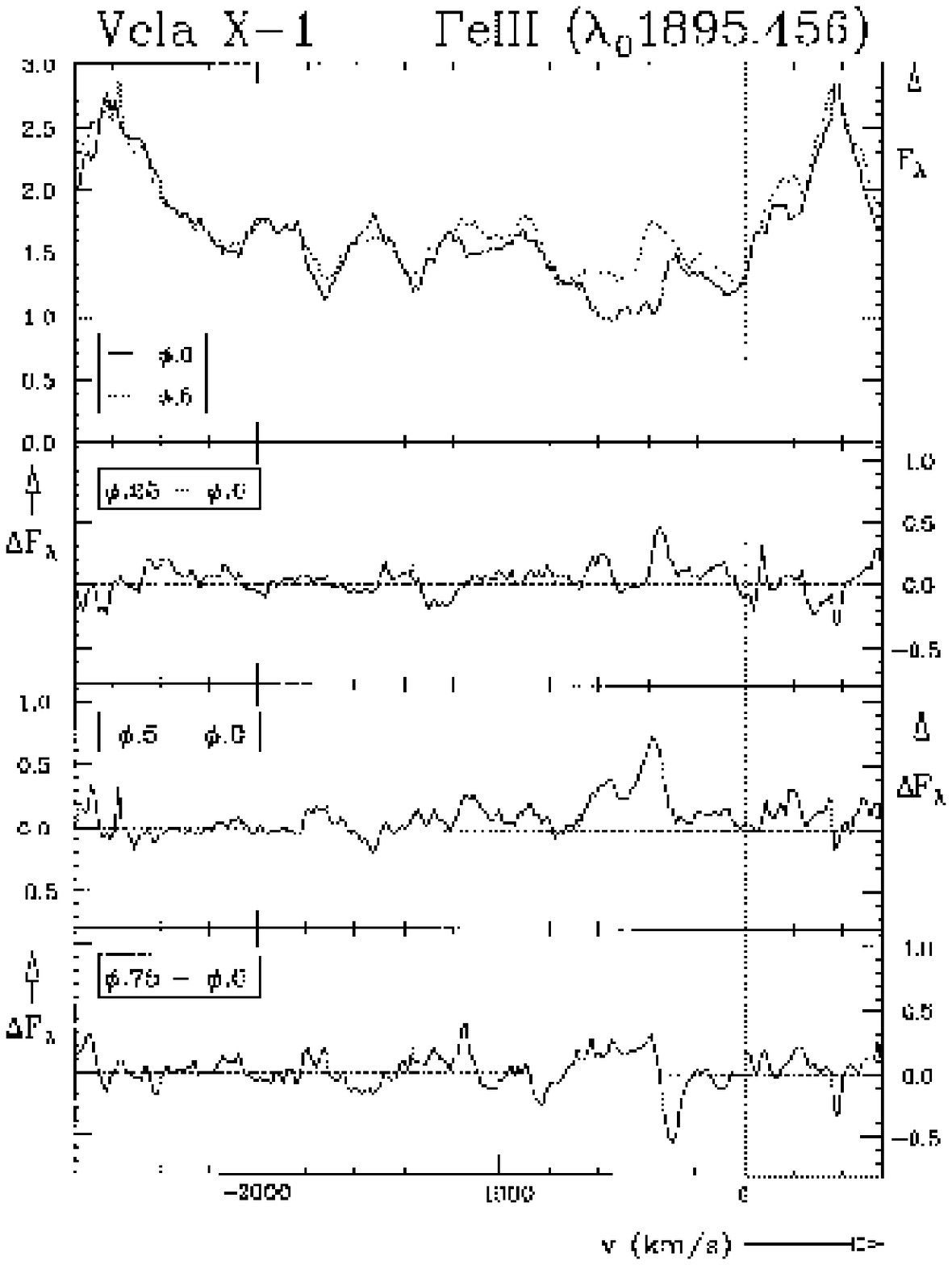,width=88mm}
      \hspace{3mm}
      \psfig{figure=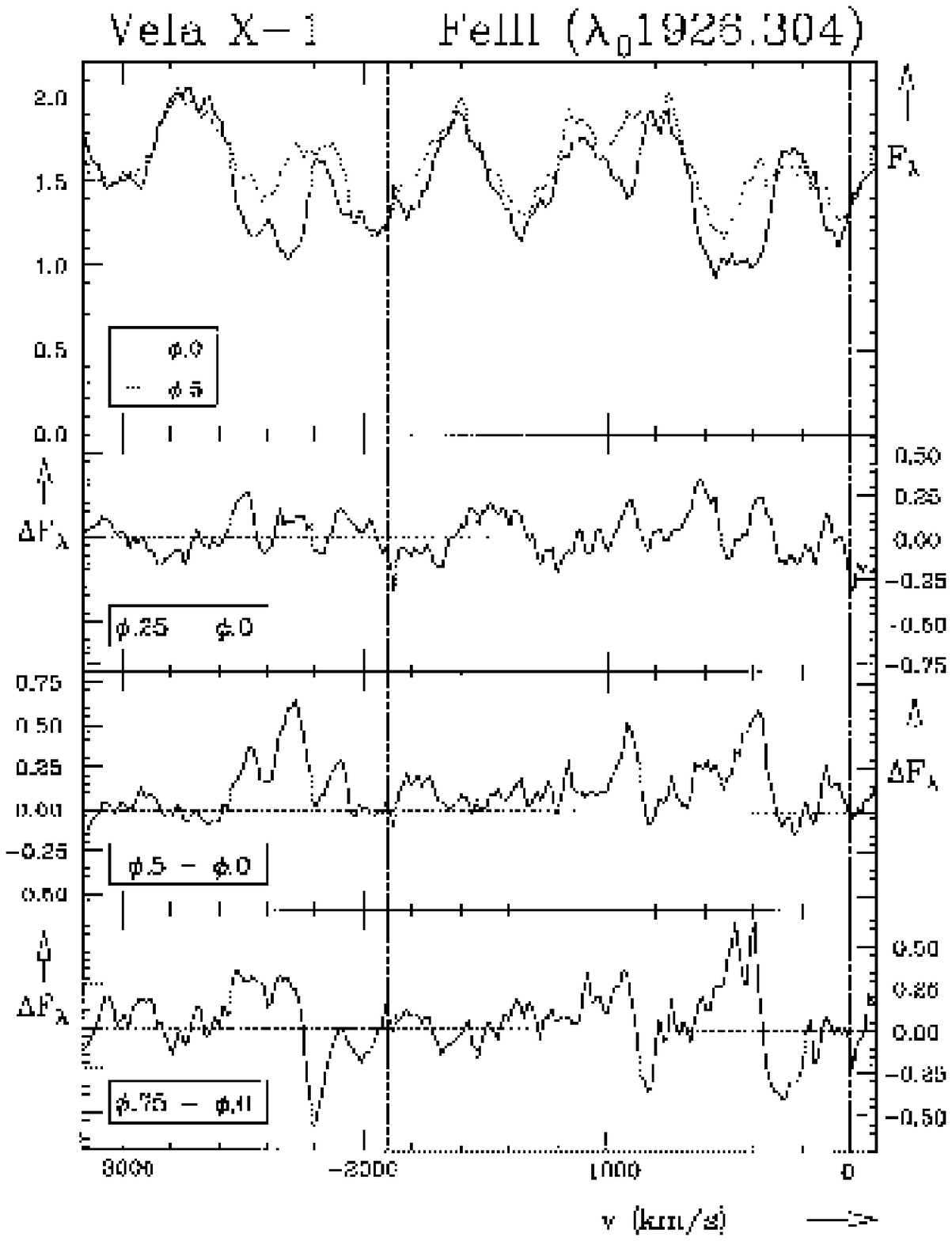,width=88mm}}
}}
\caption[]{Same as Fig.\ 11, but for \Cii, \Siiii\ and \Feiii.}
\end{figure*}

The \Nv\ line profile (Fig.\ 10, calculated with $I_{\rm g,max}=5555$),
however, cannot be modelled using a normal ionization balance that would yield
an extremely profound HM-effect, especially at the inner part of the
absorption (Fig.\ 10a), which is not observed (Fig.\ 11). Instead, the
observations seem to favour the degree of ionization to increase with distance
to the primary: $\alpha_2=-1$ gives an acceptable fit, suppressing the orbital
modulation of the absorption trough, while preserving some of the emission
(Fig.\ 10b). The line profile and its modulation are very similar for
$\alpha_1=0$ or $-1$ (Fig.\ 10c), only requiring a slightly larger integrated
optical depth ($T$). If $\alpha_1{\lsim}-2$, however, much of the absorption
arises from material at extreme velocities, causing the absorption profile to
extend up to large positive velocities beyond the severely diminished
emission. The absorption observed longward of the \Nv\ line can be attributed
to a \Ciii\ line.

The slope of the blue absorption wing (especially in \Siiv\ and \Civ, see
Fig.\ 2) indicates that the level of saturation (deeper and at velocities
slightly more negative than $-v_\infty$) is not reproduced well by the SEI
turbulence recipe (Eq.\ 19). The shocked wind structure produced by numerical
simulations (Owocki 1994), however, seem to predict more material at extreme
velocities (rather than gau{\ss}ian turbulence). Furthermore, the turbulence
may not be uniform throughout the wind, as it is assumed here.

\subsubsection{Variations in addition to the HM-effect: evidence for a
photo-ionization wake}

The main discrepancy between the model and observations of HD77581/Vela X-1 is
the $-400$ to 0 km s$^{-1}$ region, where the observed intensity is often
lower at $\phi=0.5$ than at $\phi=0$, contrary to what the HM-effect predicts.
This may be due to the presence of a photo-ionization wake that enhances
absorption at phases $\phi{\gsim}0.5$ (Kaper et al.\ 1994). The
photo-ionization wake, as observed in strong optical lines like the hydrogen
Balmer lines and \Hei\ lines in the spectrum of HD77581, is situated close to
the star at the trailing border of the ionization zone.

This effect is especially pronounced in \Nv\ (Fig.\ 11), possibly as a result
of a change in ionization degree within the photo-ionization wake. Enhanced
absorption is seen between $+200$ and $-600$ km s$^{-1}$ ($=-v_\infty$), with
its maximum evolving towards more negative velocities from $\phi=0.25$ to
$\phi=0.75$. Absorption at $\phi=0.25$ (or enhanced emission at $\phi=0$)
extends up to $+600$ km s$^{-1}$.

Similar enhanced absorption is seen in \Siiv, \Civ\ and \Aliii\ (Fig.\ 11) at
$-200$ and $-300$ to $-400$ km s$^{-1}$ around $\phi=0.5$ and 0.75,
respectively, and perhaps already around $\phi=0.25$. The enhanced absorption
due to the photo-ionization wake may have reduced the amplitude of the
HM-effect seen between $-600$ and $-400$ km s$^{-1}$.

Several weaker UV lines also show orbital modulation. The \piid\ resonance
doublet of \Cii\ at 1334.5323 and 1335.7077 \AA\ (Fig.\ 12) is sufficiently
strong to show the HM-effect between 0 and about $+700$ km s$^{-1}$ and
between about $-400$ and $-600$ km s$^{-1}$, but slightly enhanced absorption
at $-300$ km s$^{-1}$ around $\phi=0.75$ may be attributed to the
photo-ionization wake.

%
%
\begin{figure}[]
\centerline{\psfig{figure=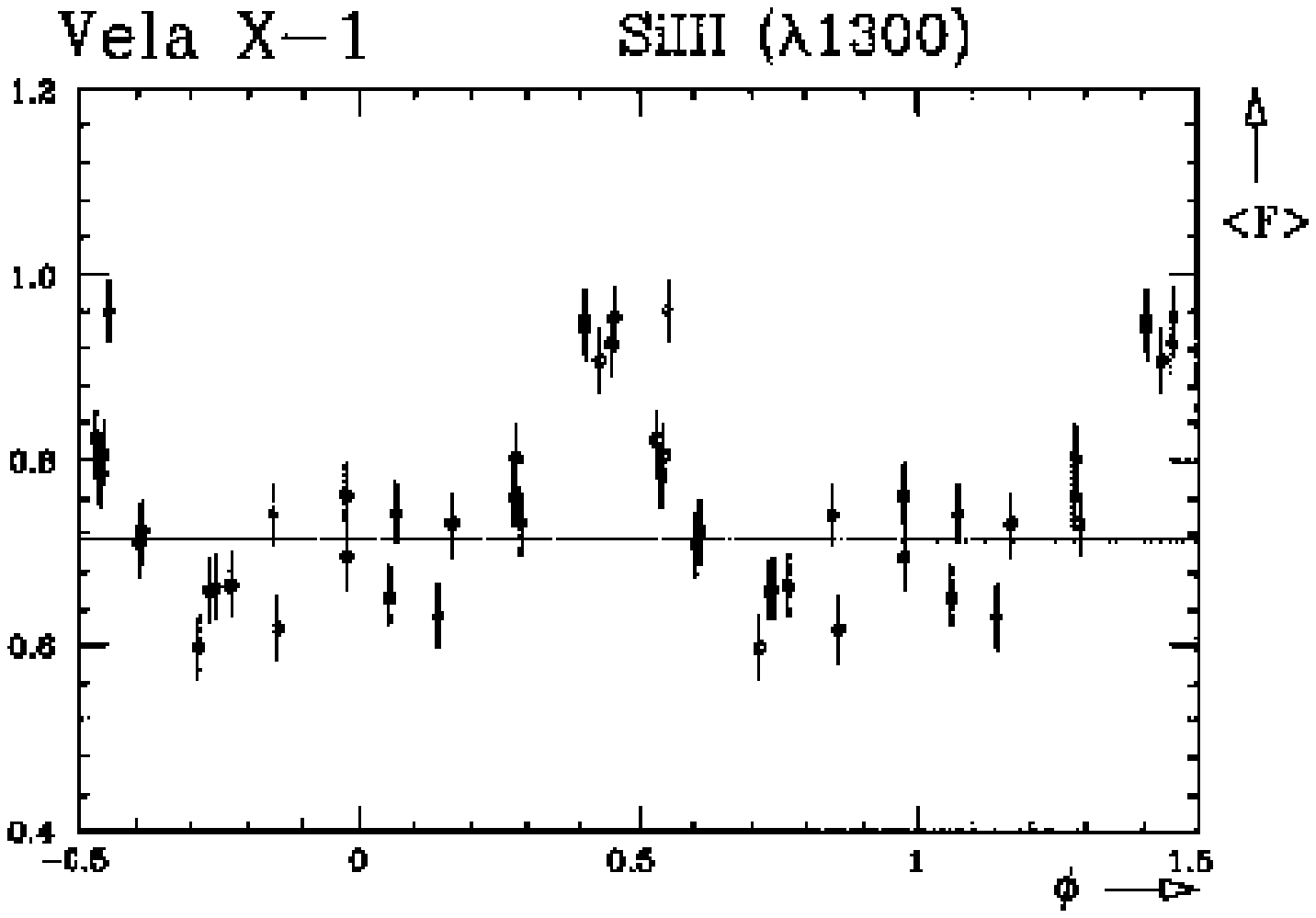,width=88mm}}
\caption[]{Lightcurve for the integrated flux in the \Siiii\ resonance lines
around 1300 \AA\ in HD77581/Vela X-1. The dotted line indicates the flux level
around $\phi=0$.}
\end{figure}

The strongly variable \piiip\ multiplet UV 4 of \Siiii\ (Fig.\ 12) consists of
6 components, of which 2 are blended: the pattern of variability is repeated
five-fold. The absorption is diminished at $-400$ km s$^{-1}$ at $\phi=0.5$
due to the HM-effect, but enhanced at $-300$ km s$^{-1}$ at $\phi=0.75$ due to
the photo-ionization wake. In Fig.\ 13 a lightcurve is plotted for the
spectral region between 1292 and 1303 \AA\ --- corresponding to $-2600$ and
$-70$ km s$^{-1}$, respectively. Absorption is minimal between $\phi=0.4$ and
0.5, and strongest between $\phi=0.7$ and 0.8. The blending of the \Siiii\
components makes it difficult to detect variability at $-600$ km s$^{-1}$,
although there is a hint for diminished absorption at $\phi=0.5$ at this
velocity in the bluemost component at 1294.543 \AA. The \sip\ (multiplet UV 2)
resonance singlet of \Siiii\ at 1206.51 \AA\ could not be studied, due to the
severe interstellar extinction and the strong Ly$\alpha$ geocoronal line at
1215.34 \AA\ in the neighbouring echelle order. The subordinate singlets of
\pis\ multiplets UV 9 and UV 10 of \Siiii\ at 1417.24 and 1312.59 \AA,
respectively, do not show variability, although the lines are clearly present
in the spectrum --- particularly the UV 9 singlet. The \piis\ resonance
doublet of \Siii\ at 1304.372 \AA\ ($+240$ km s$^{-1}$ in Fig.\ 12) does not
show any variability either, as it is of interstellar origin.

The spectral region around 1900 \AA\ is dominated by numerous lines of \Feiii,
of which the \sviip\ multiplet 34 is the strongest. It consists of 3 widely
separated components at 1895.456, 1914.056 and 1926.304 \AA\ (Fig.\ 12). They
all clearly show the diminished absorption at $-400$ and $-600$ km s$^{-1}$
around $\phi=0.5$ due to the HM-effect, and enhanced absorption at $-300$ km
s$^{-1}$ around $\phi=0.75$ due to the photo-ionization wake, exactly as
observed in the \Aliii\ resonance doublet. Some confusion arises from the
presence of 2 lines of the \gvh\ multiplet 51 of \Feiii\ at 1922.789 and
1915.083 \AA, corresponding to $-550$ and $-1750$ km s$^{-1}$ with respect to
1926.304 \AA, respectively. Multiplet 51 shows the same behaviour of the
absorption near $-300$ and $-400$ km s$^{-1}$ as multiplet 34, but not the
$-600$ km s$^{-1}$ absorption.

%
%
\begin{figure*}[]
\centerline{\psfig{figure=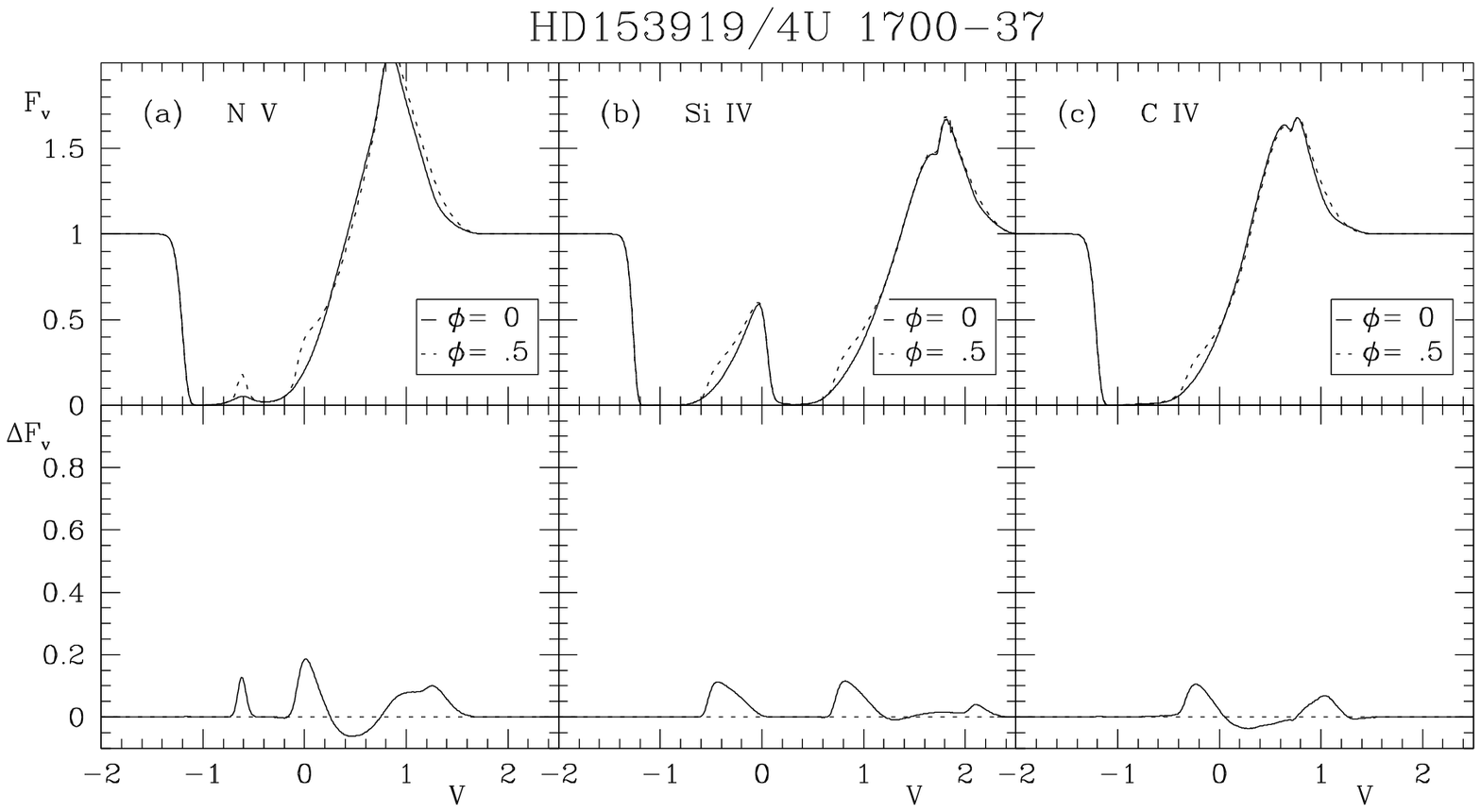,width=180mm}}
\caption[]{SEI models that approximately fit the observed line profiles and
variability of \Nv, \Siiv\ and \Civ\ in the UV spectrum of
HD153919/4U1700$-$37. Parameters are summarised in Table 3, and $v=1$
corresponds to 1700 km s$^{-1}$.}
\end{figure*}

In conclusion, the orbital modulation of both the strong wind lines and other
lines in the UV spectrum of HD77581 indicates the presence of two absorption
components that are not included in the SEI modelling. The most prominent of
these can be explained by a photo-ionization wake, causing additional
absorption in the line-of-sight at $v\sim-200$ km s$^{-1}$ around $\phi=0.5$
and at $v\sim-300$ km s$^{-1}$ around $\phi=0.75$. In addition, at $v\sim-600$
km s$^{-1}$, i.e.\ the terminal velocity of the wind of HD77581, more
absorption is present outside the Str\"{o}mgren zone than the SEI model
reproduces. This might reflect the stellar wind structure at larger distances
from the star: the velocity may be lower and the density higher after the wind
has passed through the Str\"{o}mgren zone compared to an undisturbed wind.

\subsection{HD153919/4U1700$-$37}

The shape and (lack of) orbital modulation of the line profiles in
HD153919/4U1700$-$37 are reproduced well (Fig.\ 14). The exact value for $q$
is hard to determine because of the (near) absence of the HM-effect due to a
combination of strong absorption and the presence of turbulence, yet it is
clear that $q<4$ would definitely yield a too strong HM-effect. Also, the
HM-effect is predicted to appear strongest at velocities between $-v_\infty$
and 0, implying that any variations in the blue absorption wing (at
$|v|>v_\infty$) without accompanying HM-effect at smaller velocities must be
due to some other mechanism, e.g.\ Raman-scattered far-UV emission lines
(Kaper et al.\ 1990).

\section{Discussion}

The adapted SEI model reproduces the HM-effect in HD77581/Vela X-1, and
naturally explains the lack of any (clear) HM-effect observed in the dense
wind system HD153919/4U1700$-$37. Absorption components that are seen both in
the strong wind lines and additional weaker lines in the UV spectrum of
HD77581/Vela X-1 and that could not be reproduced by the adapted SEI model can
be explained by a photo-ionization wake. The orbital modulation of the
resonance lines in the UV spectra of the five HMXBs studied here shows a clear
trend of the size of the Str\"{o}mgren zone to increase with higher X-ray
luminosity.

\subsection{Terminal velocities and turbulence}

The terminal velocity $v_\infty$ may be estimated from a comparison of an
appropriate model line profile with the observed line profile. It is best
derived from moderately saturated profiles, in which case the maximum
absorption in the $\phi=0.5$ line profile occurs at the terminal velocity.

The terminal velocity for HD77581/Vela X-1 is thus estimated to be
$v_\infty=600$ km s$^{-1}$, much slower than previously reported (1105 km
s$^{-1}$ Prinja et al.\ 1990). For HD153919/4U1700$-$37 we estimate
$v_\infty=1700$ km s$^{-1}$ (Prinja et al.\ 1990: 1820 km s$^{-1}$). These
values should be correct to within $100$ km s$^{-1}$. The low resolution IUE
spectra show that the wind in HDE226868/Cyg X-1 is faster (marginally
resolved: $1000<v_\infty<1500$ km s$^{-1}$) than in Sk-Ph/LMC X-4 and Sk
160/SMC X-1 (unresolved: ${\lsim}600$ km s$^{-1}$, see also
Hammerschlag-Hensberge et al.\ 1984 for Sk 160/SMC X-1). HST/STIS spectra of
the \Nv, \Siiv\ and \Civ\ lines in Sk-Ph/LMC X-4 around $\phi=0$ suggest a
terminal velocity of $\sim500$ km s$^{-1}$, even though wind velocities up to
$\sim1200$ km s$^{-1}$ occur (Kaper et al.\ in preparation).

Single (galactic) O-type and early-B-type stars are known to have a fairly
constant ratio of terminal over escape velocity $v_\infty/v_{\rm esc}\sim2.5$,
where
\begin{equation}
v_{\rm esc} = \sqrt{ \frac{2 G M (1-\Gamma)}{R} }
\end{equation}
with
\begin{equation}
\Gamma =
2.658 \times 10^{-5} \frac{L}{L_\odot} \left( \frac{M}{M_\odot} \right)^{-1}
\end{equation}
for a typical Population {\sc I} star (e.g.\ Groenewegen et al.\ 1989). This
is also true at temperatures between $\sim1.1$ and $2.1\times10^4$ K but then
$v_\infty/v_{\rm esc}\sim1.3$ (Lamers et al.\ 1995). The terminal velocities
for the primary stars of the five HMXBs studied here are, with the possible
exception of HDE226868, low compared to single stars of the same effective
temperature (Table 4 \& Fig.\ 15). Especially the two magellanic stars have
very low terminal velocities, which may at least partly be due to their
subsolar metallicity and hence lower radiation-force multiplier (Garmany \&
Conti 1985; Prinja 1987; Kudritzki et al.\ 1987) --- which is also found for
dust-driven winds of Asymptotic Giant Branch stars (e.g.\ van Loon 2000).

The Str\"{o}mgren zone could inhibit acceleration of the stellar wind material
flowing through it, not only resulting in an ionization wake but possibly also
leading to a slower terminal velocity of the wind. This scenario only works if
the Str\"{o}mgren zone is sufficiently large and the orbital period
sufficiently short that all wind material leaving the star passes through the
Str\"{o}mgren zone before having reached the terminal velocity. The magellanic
systems indeed must have very extended Str\"{o}mgren zones leaving only a
small region at the opposite side of the primary unaffected by the X-ray
source (i.e.\ a shadow wind), and their orbital periods are rather short.
However, the Str\"{o}mgren zones in HD77581/Vela X-1 and HD153919/4U1700$-$37
are closed surfaces, and certainly do not occupy more than half of the
circumstellar space. Within half their orbital periods of 4.5 and 1.7 days,
respectively, their stellar winds ought to already have accelerated to
velocities approaching the terminal velocity as if it were from a single star.

A viable alternative is that the $\Gamma$ factors for single stars are not
applicable to stars in interacting binaries. HMXBs primaries may have already
lost a significant fraction of their initial mantle mass, causing them to be
undermassive for their luminosity (Conti 1978; Kaper 2001). This seems to be
confirmed when comparing the primary star masses (Table 1) with Table 3 in
Howarth \& Prinja (1989). Hence we may have under-estimated $\Gamma$ and
over-estimated $v_{\rm esc}$.

%
%
\begin{table}[]
\caption[]{Ratios of terminal velocity $v_\infty$ and escape velocity $v_{\rm
esc}$, and turbulent velocities $\sigma_v$ of the primary stars in the HMXBs,
together with their effective temperatures $T_{\rm eff}$ and $\Gamma$ values
estimated from Lamers et al.\ (1995).}
\begin{tabular}{lllrr}
\hline\hline
Primary                &
$T_{\rm eff}$ (K)      &
$\Gamma$               &
$v_\infty/v_{\rm esc}$ &
$\sigma_v$             \\
\hline
HDE226868              &
30500                  &
0.34                   &
1.8---2.8              &
                       \\
Sk-Ph                  &
36000                  &
0.16                   &
$\lsim$0.8             &
                       \\
Sk 160                 &
26000                  &
0.18                   &
$\lsim$1.1             &
                       \\
HD77581                &
23400                  &
0.18                   &
1.3                    &
0.45                   \\
HD153919               &
37200                  &
0.34                   &
2.0                    &
0.15                   \\
\hline
\end{tabular}
\end{table}

%
%
\begin{figure}[]
\centerline{\psfig{figure=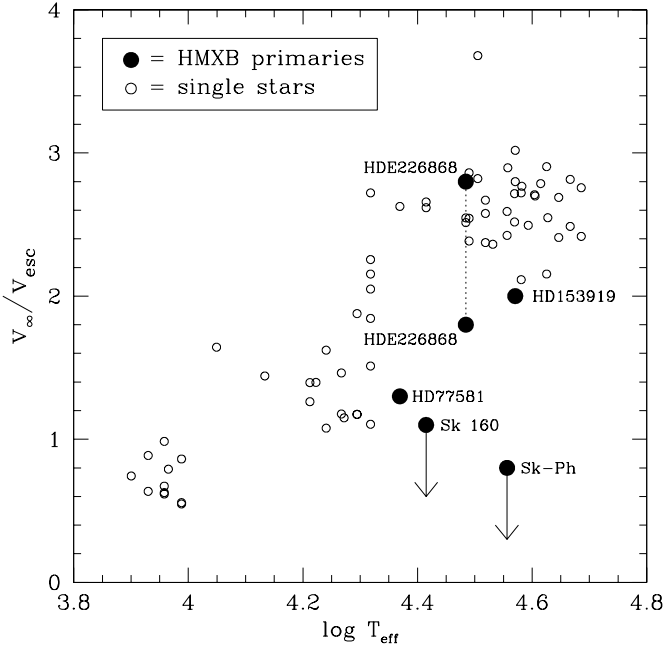,width=88mm}}
\caption[]{Ratio of terminal over escape velocity for the primary stars in the
HMXBs studied here (solid dots), compared to single stars (circles) from
Lamers et al.\ (1995). HMXB members have low terminal velocities for their
effective temperatures.}
\end{figure}

A turbulence description has been employed to account for material at
velocities deviating from the bulk flow that obeys a velocity law according to
standard radiation-driven wind theory. The $\sigma$ is the typical deviation
in units of $v_\infty$. To approximately describe both the undisturbed wind
and X-ray ionized components in the line profiles of HD77581/Vela X-1 a very
large value for $\sigma$ needs to be invoked ($\sigma_v\sim0.45$), whereas for
HD153919/4U1700$-$37 a much smaller value suffices ($\sigma_v\sim0.15$). In
absolute terms the turbulence in these two HMXBs is very similar, though:
$v_{\rm turb}=270$ and 255 km s$^{-1}$, respectively. These values are typical
for winds from O-type stars for which $v_{\rm turb}$ has been found to be
largely independent of $T_{\rm eff}$ (Groenewegen et al.\ 1989). As mentioned
before, we find indications for the deviations from the monotonic velocity law
to resemble a shocked wind structure rather than uniform turbulence.

\subsection{Ionization fractions in the stellar wind}

%
%
\begin{table*}[]
\caption[]{Ionization fractions $\kappa_{\rm i}$ for several ions in the
stellar winds of HD77581/Vela X-1 and HD153919/4U1700$-$37 as derived from SEI
model parameters, adopting mass-loss rates $\dot{M}$ from H$\alpha$ line
profile modelling (Schr\"{o}der et al.\ in preparation) and using solar
element abundances $A$. Ionization fractions from work on single stars (Lamers
et al.\ 1999) are given for comparison.}
\begin{tabular}{llcrcccr}
\hline\hline
Line                                            &
$\lambda_0$ (\AA)                               &
$f_{\rm line}$                                  &
$T$                                             &
$A$                                             &
$\dot{M}_{{\rm H}\alpha}$ (M$_\odot$ yr$^{-1}$) &
$\kappa_{\rm i}$                                &
$\kappa_{\rm i, Lamers}$                        \\
\hline
\multicolumn{8}{l}{\it HD77581/Vela X-1:} \\
\Nv                 &
1238.821            &
0.152               &
3                   &
$1.1\times10^{-4}$  &
$1\times10^{-6}$    &
$5\times10^{-4}$    &
$3\times10^{-5}$    \\
\Siiv               &
1393.755            &
0.528               &
300                 &
$3.5\times10^{-5}$  &
$1\times10^{-6}$    &
$4\times10^{-2}$    &
$1\times10^{-2}$    \\
\Civ                &
1548.20             &
0.194               &
200                 &
$3.6\times10^{-4}$  &
$1\times10^{-6}$    &
$6\times10^{-3}$    &
$1\times10^{-2}$    \\
\Aliii              &
1854.716            &
0.560               &
20                  &
$3.0\times10^{-6}$  &
$1\times10^{-6}$    &
0.3                 &
$<1$                \\
\multicolumn{8}{l}{\it HD153919/4U1700$-$37:} \\
\Nv                 &
1238.821            &
0.152               &
30                  &
$1.1\times10^{-4}$  &
$1\times10^{-5}$    &
$2\times10^{-3}$    &
$3\times10^{-3}$    \\
\Siiv               &
1393.755            &
0.528               &
$2\times10^4$       &
$3.5\times10^{-5}$  &
$1\times10^{-5}$    &
$\sim1$             &
$1\times10^{-3}$    \\
\Civ                &
1548.20             &
0.194               &
5000                &
$3.6\times10^{-4}$  &
$1\times10^{-5}$    &
$6\times10^{-2}$    &
$3\times10^{-3}$    \\
\hline
\end{tabular}
\end{table*}

In principle, the mass-loss rate may be estimated from the integrated optical
depth $T$ of a resonance line:
\begin{equation}
T = \frac{\pi e^2}{m_e c} f_{\rm line} \lambda_0 N_{\rm i} v_{\infty}^{-1}
\end{equation}
with $f_{\rm line}$ the oscillator strength of the transition in terms of the
harmonic oscillator $\pi e^2/m_e c$, $\lambda_0$ the rest wavelength of the
transition, and $N_{\rm i}$ the column density of the ion:
\begin{equation}
N_{\rm i} = \int_{R_\star}^{\infty} n_{\rm i}(r) {\rm d}r
\end{equation}
with $n_{\rm i}$ the number density of the ion. The continuity equation yields
\begin{equation}
\dot{M} = 4 \pi r^2 v(r) \left<m\right> \frac{n_{\rm i}(r)}{A_{\rm i}}
\end{equation}
with $\rho$ the mass density, $\left<m\right>$ the mean mass of an ion in the
wind, and $A_{\rm i}$ the ion abundance by number. The mass-loss rate
$\dot{M}$ can now be estimated from observed quantities in the following way:
\begin{equation}
\dot{M} = \frac{4 m_e c}{e^2 f_{\rm line} \lambda_0} \left<m\right> R_\star
v_{\infty}^2 \psi \frac{T}{A_{\rm i}}
\end{equation}
with
\begin{equation}
\psi = \left\{ \begin{array}{lll} (1-\gamma) & \mbox{\hspace{3mm} for
\hspace{1mm} $0{\leq}\gamma<1$} & \mbox{\hspace{1mm} \& \hspace{1mm} $v_0=0$}
\\ (v_0-1)/\ln{v_0} & \mbox{\hspace{3mm} for \hspace{5.5mm} $\gamma=1$} &
\mbox{\hspace{1mm} \& \hspace{1mm} $v_0>0$} \end{array} \right.
\end{equation}
For our choice of $\gamma=1$ \& $v_0=0.01$ the numerical factor
$\psi=0.215$.  Adopting a mean nucleus mass
$\left<m\right>=2.2624\times10^{-24}$ g (Lamers et al.\ 1999) the product of
mass-loss rate $\dot{M}$ and ion abundance $A_{\rm i}$ can be calculated for
the resonance lines in HD77581/Vela X-1 and HD153919/4U1700$-$37.

In practice, however, the limited knowledge of the ionization balance in the
winds of OB supergiants makes the derived mass-loss rates highly unreliable.
Instead, the ionization fractions of the ions may be derived when other, more
reliable estimates for the mass-loss rate are available. For resonance lines
the excitation fraction of the ion that produces the line is unity, and the
ion abundance $A_{\rm i}$ is the product of the ionization fraction
$\kappa_{\rm i}$ and elemental abundance $A$ (by number). Assuming solar
abundances (Anders \& Grevesse 1989), the ionization fraction may be derived
using Eq.\ 25.

Lamers et al.\ (1999) found $\kappa_{\rm i}$ to depend on both radiation
temperature and density, with the radiation temperature scaling approximately
linearly with $T_{\rm eff}$. The empirical dependencies of $\kappa_{\rm i}$ on
density were contrary to the expected ionization balance, however, and it was
argued that the empirical relations may suffer from selection effects.
Therefore, we only consider their empirical relations between $\kappa_{\rm i}$
and $T_{\rm eff}$.

Schr\"{o}der et al.\ (in preparation) model the H$\alpha$ profiles of OB
supergiants in HMXBs. They derive mass-loss rates for HD77581
($\dot{M}\sim1.0\times10^{-6}$ M$_\odot$ yr$^{-1}$) and HD153919
($\dot{M}\sim1.0\times10^{-5}$ M$_\odot$ yr$^{-1}$), which are comparable to
those derived for single stars of similar spectral type (Howarth \& Prinja
1989). Combining their results with the results from the SEI models,
ionization fractions $\kappa_{\rm i}$ are estimated for the N$^{4+}$,
Si$^{3+}$ and C$^{3+}$ (and Al$^{2+}$) ions in the (undisturbed) stellar winds
of HD77581 and HD153919 (Table 5). These are then compared with the ionization
fractions estimated from Fig.\ 3 in Lamers et al.\ (1999), taking into account
their lower and upper limits.

The ion fractions derived from the SEI modelling for HD77581/Vela X-1 are in
agreement with the predictions for single stars by Lamers et al.\ (1999),
except for the N$^{4+}$ abundance which is observed to be an order of
magnitude higher than predicted. This may be due to either super-ionization or
nitrogen over-abundance (or both). Auger ionization (Cassinelli \& Olson 1979)
is sometimes invoked to explain strong \Nv\ resonance lines. Unless the
density becomes very high, Auger ionization increases with the velocity in the
wind, and may originate in the high-velocity extrema of a shocked wind. This
might mimic a moderate increase of ionization fraction with distance, as
required to reproduce the observed line profile and variability of the \Nv\
line in HD77581/Vela X-1. On the other hand, nitrogen over-abundance at the
surface of HD77581 may have resulted from (i) the transfer in the past of
nitrogen-enriched material from the progenitor of Vela X-1 onto HD77581, or
(ii) strong mass loss exposing deeper layers mixed with the products of
nuclear burning. Kaper et al.\ (1993) note that HD77581 might be a BN star.

For HD153919/4U1700$-$37 the opposite is found: the observed ionization
fraction of N$^{4+}$ agrees very well with the predictions, whereas the
ionization fractions of Si$^{3+}$ and C$^{3+}$ are observed to be (much)
higher than predicted. Perhaps these are the dominant ionization states for
silicon and nitrogen in the stellar wind of HD153919, rather than Si$^{4+}$
and C$^{4+}$. Still, an ionization fraction $\kappa_{\rm i}\sim1$ for
Si$^{3+}$ is unrealistic for a multi-level atom, and hence the SEI model must
have over-estimated the integrated optical depth of the \Siiv\ line in
HD153919/4U1700$-$37. If indeed the degree of ionization in the wind of
HD153919 is lower than that predicted for single stars, this would mean that
--- like in the wind of HD77581 --- N$^{4+}$ is in fact overabundant in the
wind of HD153919.

\subsection{Size of the Str\"{o}mgren zone}

The size of the Str\"{o}mgren zone, indicated by a particular value of the
parameter $q$, is related to the ionization parameter $\xi$ as
\begin{equation}
\xi = \frac{q L_{\rm X}}{n_{\rm X} a^2}
\end{equation}
where $a$ is the distance between the centres of the primary and the X-ray
source. The ionization parameter can be interpreted as a measure for the
number of X-ray photons per particle. The ionization balance in the stellar
wind is mostly affected by soft X-ray photons. In common with previous models
of the X-ray ionization of stellar winds in HMXBs, we assume the dominant
source of opacity for soft X-rays is oxygen (Hatchett \& McCray 1977). Masai
(1984) has objected that this assumption may not be valid if \Heii is present,
but incorporating the effects of a {\Heii}/{\Heiii} ionization front would be
beyond the scope of this paper. As a consequence of our assumption, the sharp
boundaries between the X-ray ionized and the undisturbed stellar wind coincide
for ions like C$^{3+}$, Si$^{3+}$ and N$^{4+}$ (cf.\ Kallman \& McCray 1982;
McCray et al.\ 1984). Hence the edge of the Str\"{o}mgren zone is given by a
particular value for $\xi$ that corresponds to the boundary at which oxygen is
being completely ionized by the X-ray source, and depends on the shape of the
X-ray spectrum and the oxygen abundance (Hatchett et al.\ 1976).

Accretion of matter onto a star moving through a medium was first described by
Bondi \& Hoyle (1944). Their concept was applied to HMXBs by Davidson \&
Ostriker (1973). In an HMXB the compact object has a velocity $v_{\rm rel}$
relative to the stellar wind flow:
\begin{equation}
v_{\rm rel}^2 = v_{\rm wind}^2 + ( |v_{\rm X}-v_\star| - \frac{R_\star}{a}
v_{\rm rot} )^2
\end{equation}
with stellar wind velocity $v_{\rm wind}$ near the X-ray source (which may be
lower than implied by the undisturbed wind velocity law; see Section 5.1),
orbital velocity $v_{\rm X}$ of the X-ray source, and orbital velocity
$v_\star$, stellar radius $R_\star$ and rotation velocity $v_{\rm rot}$ of the
primary. Kaper (1998) finds that generally $v_{\rm rel}\sim\frac{1}{2}v_{\rm
wind}$. Stellar wind matter is accreted onto the compact object if it
approaches within an accretion radius approximately given by
\begin{equation}
r_{\rm acc} = \frac{2 G M_{\rm X}}{v_{\rm rel}^2}
\end{equation}
with $M_{\rm X}$ the mass of the compact object. The X-ray luminosity due to
the release of gravitation energy of the matter being accreted onto the
surface of the compact object with radius $R_{\rm X}$ is
\begin{equation}
L_{\rm X} = \pi \zeta r_{\rm acc}^2 v_{\rm rel} \rho_{\rm X} \frac{G M_{\rm
X}}{R_{\rm X}}
\end{equation}
where $\rho_{\rm X}$ is the mass density of the stellar wind near the compact
object, and $\zeta$ is an efficiency parameter ($\zeta \sim 0.1$ for accretion
onto a neutron star). Hence we obtain
\begin{equation}
q = \frac{\xi R_{\rm X} a^2}{4 \pi \zeta \left<m\right>} \left(\frac{v_{\rm
rel}}{G M_{\rm X}}\right)^3
\end{equation}
The size of the Str\"{o}mgren zone depends on: (1) the dimensions of the
compact object via $R_{\rm X}$, $M_{\rm X}$ and $\xi$ (by the shape of the
X-ray spectrum); (2) the orbit via $a$ and $v_{\rm rel}$ (by the orbital
velocity); (3) the wind flow via $v_{\rm rel}$; (4) and the abundances in the
wind via $\left<m\right>$ and $\xi$ (by the oxygen abundance). The size of the
Str\"{o}mgren zone does not, however, explicitly depend on the mass-loss rate
nor on the X-ray luminosity.

%
%
\begin{table}[]
\caption[]{Critical, expected and observed sizes of the Str\"{o}mgren zones in
the five HMXBs (the smaller $q$, the larger the Str\"{o}mgren zone).}
\begin{tabular}{llrr}
\hline\hline
HMXB                 &
$q_{\rm critical}$   &
$q_{\rm expected}$   &
$q_{\rm observed}$   \\
\hline
HDE226868/Cyg X-1    &
1.7                  &
1.7---5.0            &
${\lsim}1.7$         \\
Sk-Ph/LMC X-4        &
2.7                  &
$<$1.6---4.5         &
$<2.7$               \\
Sk 160/SMC X-1       &
2.9                  &
$<$6.3---17          &
${\ll}2.9$           \\
HD77581/Vela X-1     &
2.7                  &
10---24              &
$2.9$                \\
HD153919/4U1700$-$37 &
2.0                  &
177---215            &
$>4.0$ \\
\hline
\end{tabular}
\end{table}

From the time sequence of the integrated line variability (Figs.\ 1 \& 2) and
the modelling of the line profile variability it is already clear that the
Str\"{o}mgren zone is largest for Sk-Ph/LMC X-4 and especially Sk 160/SMC X-1:
their zones must extend far beyond the primary as seen from the X-ray source,
and the stellar wind is undisturbed only in a small shadow region behind the
primary. The Str\"{o}mgren zone is smaller but still very extended for
HDE226868/Cyg X-1 where the variability due to the HM-effect is continuous
over the orbit. HD77581/Vela X-1 has an again smaller and now closed
Str\"{o}mgren zone, but it still occupies a considerable fraction of the wind
volume. The (by far) smallest Str\"{o}mgren zone is found for
HD153919/4U1700$-$37.

The critical and expected sizes of the Str\"{o}mgren zones for the five HMXBs
are listed in Table 6. The $q_{\rm critical}$ is calculated assuming a
velocity law according to Eq.\ (2) with $v_0=0.01$ and $\gamma=1$. The $q_{\rm
expected}$ is calculated assuming $\xi=10^3$ (Hatchett \& McCray 1977) for a
typical X-ray spectrum, $\zeta=0.1$, and $R_{\rm X}=10$ km for $M_{\rm X}=1.4$
M$_\odot$ and $R_{\rm X}{\propto}M_{\rm X}$. The range in values results from
either assuming co-rotation of the primary with the orbit or no rotation of
the primary (and the uncertainty about $v_\infty$ for Cyg X-1, LMC X-4 and SMC
X-1). The observed sizes of the Str\"{o}mgren zone are in reasonable agreement
with the expected sizes that, in general, seem to have been somewhat
under-estimated. This could easily be solved if for instance the accretion
efficiency $\zeta$ were twice as large. In particular, $\zeta$ increases with
larger mass and smaller radius of the compact object (Shakura \& Sunyaev
1973), which would yield an accretion efficiency for the black-hole candidate
Cyg X-1 significantly larger than 10\%.

\section{Summary}

An analysis is presented of a large set of IUE spectra for five HMXBs:
HDE226868/Cyg X-1, Sk-Ph/LMC X-4, Sk 160/SMC X-1, HD77581/Vela X-1 and
HD153919/4U1700$-$37. We compared their spectra and variability, and adapted
the SEI radiation transfer code for modelling the variability of resonance
lines in HMXBs. From model fits for HD77581/Vela X-1 and HD153919/4U1700$-$37
we derive terminal velocities and ionization fractions of their stellar winds,
as well as the sizes of the Str\"{o}mgren zone created in the wind by the
X-ray source. For three other HMXBs (HDE226868/Cyg X-1, Sk-Ph/LMC X-4 and Sk
160/SMC X-1) only rough estimates are derived for the terminal velocity and
size of the Str\"{o}mgren zone.

The adapted SEI model reproduces the \Siiv, \Civ\ and \Aliii\ resonance line
profiles and their orbital modulation in HD77581/Vela X-1, with additional
absorption components attributed to a photo-ionization wake. The \Nv\
resonance line in HD77581/Vela X-1 is more difficult to model, and it seems
to be severely affected by the photo-ionization wake. The \Nv, \Siiv\ and
\Civ\ line profiles in HD153919/4U1700$-$37 are well reproduced by the adapted
SEI model; their lack of orbital modulation is due to the dense stellar wind
and low X-ray luminosity.

OB supergiants in HMXBs have lower terminal velocities than single stars of
similar spectral type, possibly due to a lower effective gravity of the
Roche-Lobe filling primary. Ionization fractions in the stellar wind of
HD77581/Vela X-1 agree with predictions for single stars, except for an
overabundance of N$^{4+}$ ions in the wind of HD77581. This may be due to
super-ionization possibly resulting from X-ray emission from shocks at the
base of the wind, or nitrogen overabundance due to mass-transfer from the
progenitor of Vela X-1 or extensive mass loss from HD77581 itself. The
N$^{4+}$ abundance in the stellar wind of HD153919/4U1700$-$37 agrees very
well with predictions for single. However, the ionization fractions of the
other ions (Si$^{3+}$ and C$^{3+}$) are much lower than predicted, suggesting
a generally lower ionization of the stellar wind of HD153919 than what is
predicted for single stars. This would then infer an N$^{4+}$ {\it
over}-abundance in the stellar wind of HD153919. The sizes of the
Str\"{o}mgren zones are in fair agreement with the expectations from standard
Bondi \& Hoyle accretion, and they are larger for more luminous X-ray sources.

\begin{acknowledgements}
We thank Henny Lamers for interesting discussions and for supplying the
original SEI code, and an anonymous referee for her/his remarks. LK is
supported by a fellowship of the Royal Academy of Sciences in The Netherlands.
Much of the work presented here was carried out while JvL was in Amsterdam for
his MSc. Mas este trabalho nunca podia estar feito sem o encargo pelo anjo
Joana.
\end{acknowledgements}

\appendix

\section{X-ray eclipse spectra}

Averages of spectra near $\phi=0$ (eclipse of the X-ray source) are a fair
representation of the ``undisturbed'' stellar wind. The low-resolution spectra
of HDE226868/Cyg X-1, Sk 160/SMC X-1, and Sk-Ph/LMC X-4 all contain the strong
geocoronal Lyman-$\alpha$ emission line at 1215.34 \AA, and the resonance
lines of \Civ\ near 1550~\AA\ and \Siiv\ near 1400 \AA\ (Fig.\ 1). These are
much stronger in HDE226868/Cyg X-1 than in the two Magellanic Cloud members
that both have a lower metalicity than their counterparts in the Milky Way.

Also immediately clear is the difference between the continua of HDE226868/Cyg
X-1 and the two Magellanic Cloud members. Due to the high extinction towards
HDE226868/Cyg X-1, its continuum is tilted both at the 1200 \AA\ side (dust
and hydrogen) and at the 2100 \AA\ side (grains or large molecules). The
Magellanic Cloud members are much less affected by interstellar extinction.

%
%
\begin{figure}[]
\centerline{\vbox{
\psfig{figure=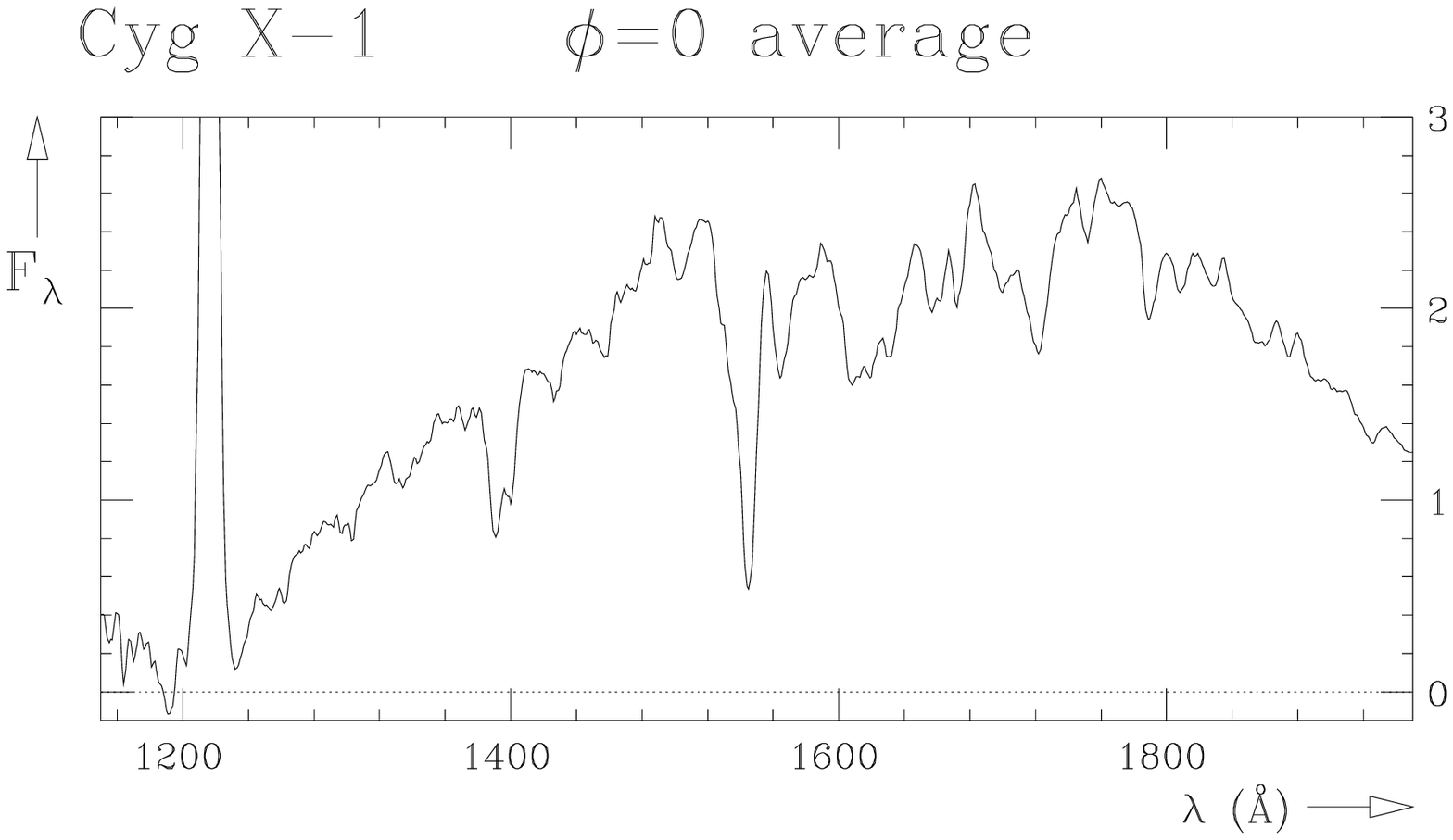,width=88mm}
\psfig{figure=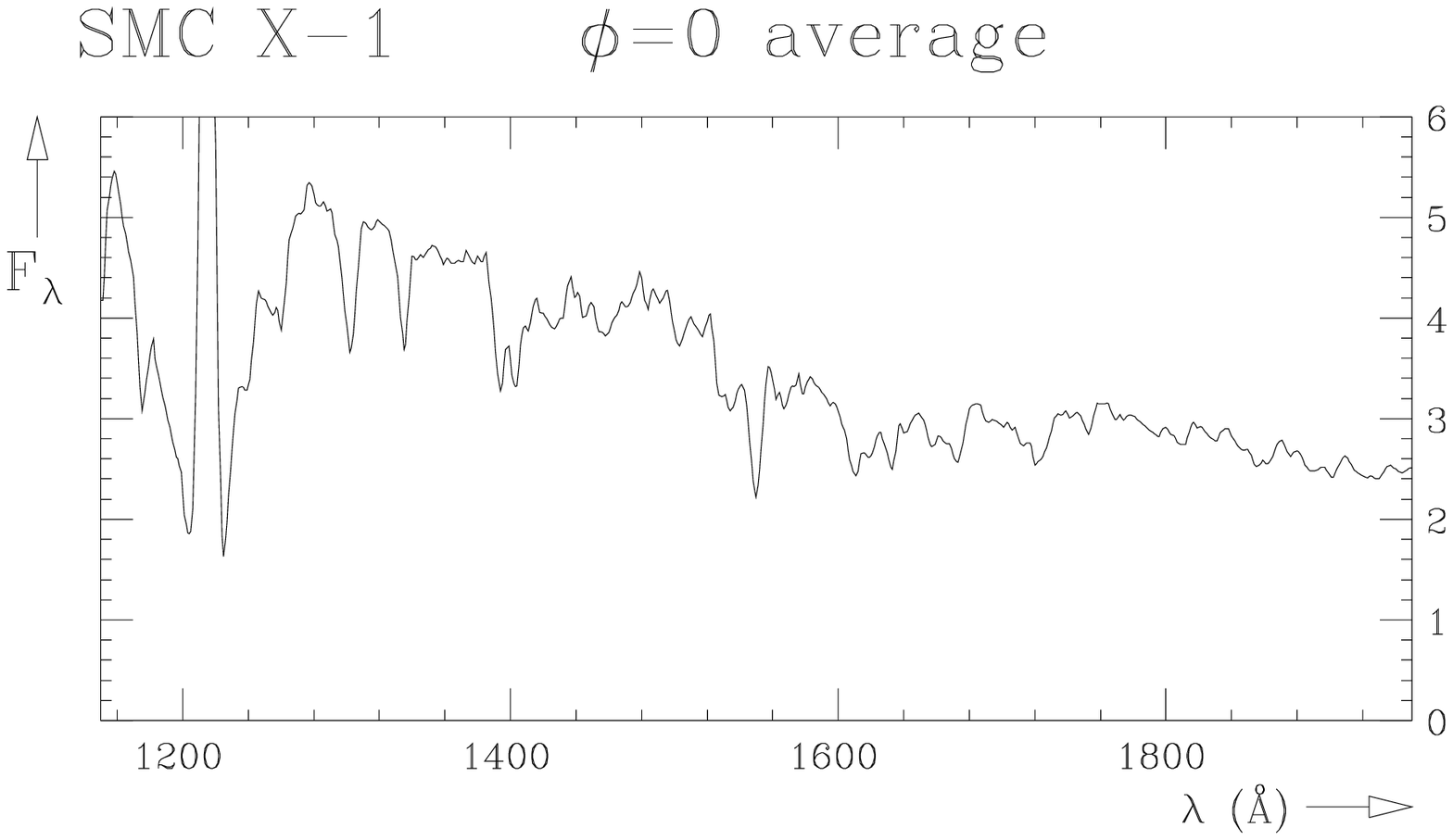,width=88mm}
\psfig{figure=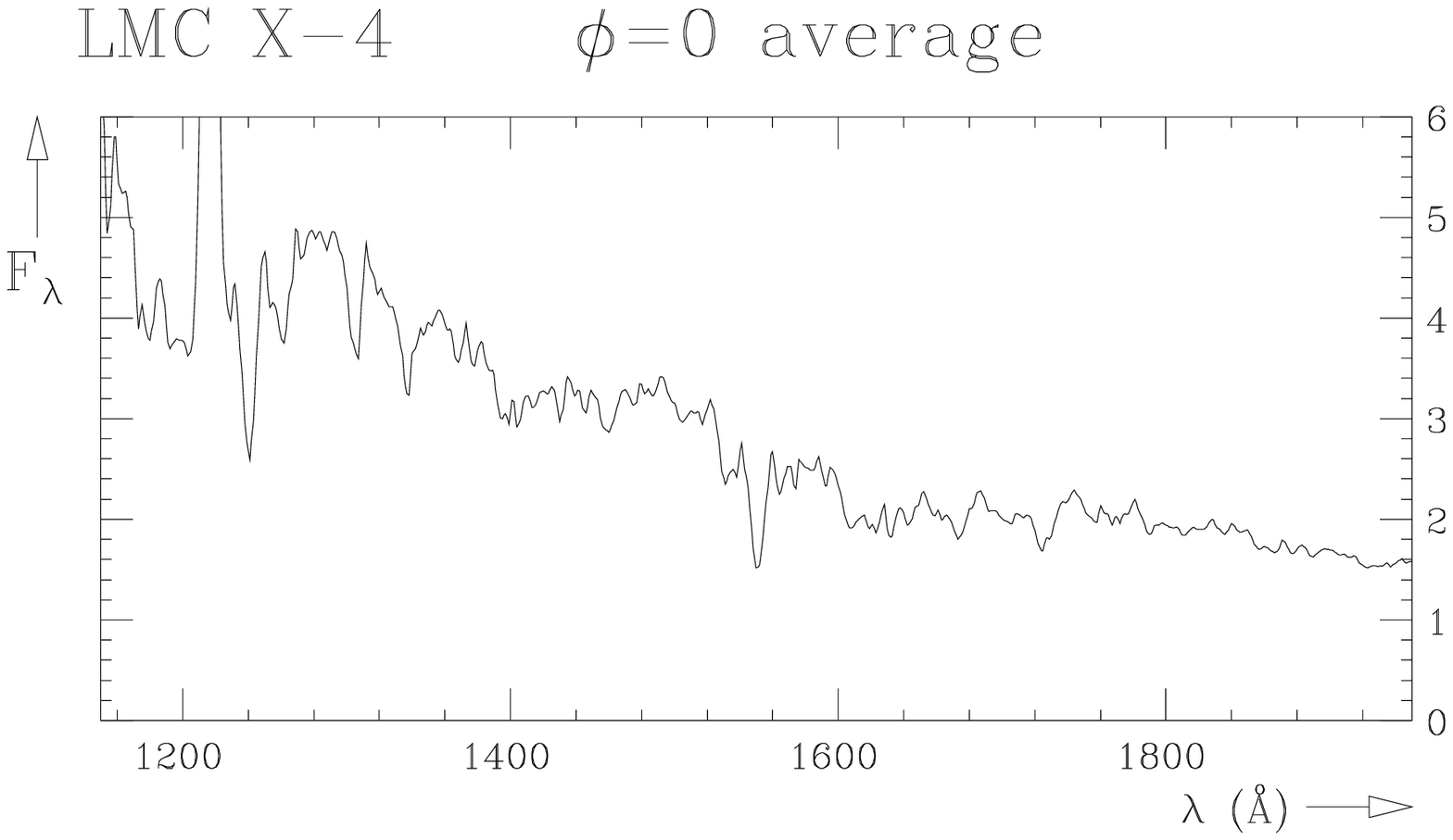,width=88mm}
}}
\caption[]{X-ray eclipse spectra of HDE226868/Cyg X-1, Sk 160/SMC X-1, and
Sk-Ph/LMC X-4.}
\end{figure}

The high resolution spectra have not been flux calibrated in an absolute
sense, therefore the shape of the continuum in Figs.\ A2 and A3 does not
represent the shape of the continuum as it entered the telescope.

The geocoronal Ly-$\alpha$ line and the resonance lines of \Nv, \Siiv, \Civ\
and \Aliii\ are the most prominent features in the spectrum of HD77581/Vela
X-1 (Fig.\ A2, top). \Nv\ and \Aliii\ are very strong for its spectral type.
Notice the rich \Feiii\ spectrum at wavelengths longer than $\sim1880$ \AA,
the \Feiv\ spectrum between $\sim1480$ \AA\ and 1880 \AA, and the lack of
\Fev\ and \Fevi\ at wavelengths $\lsim1480$ \AA.

%
%
\begin{figure*}[]
\centerline{\vbox{
\psfig{figure=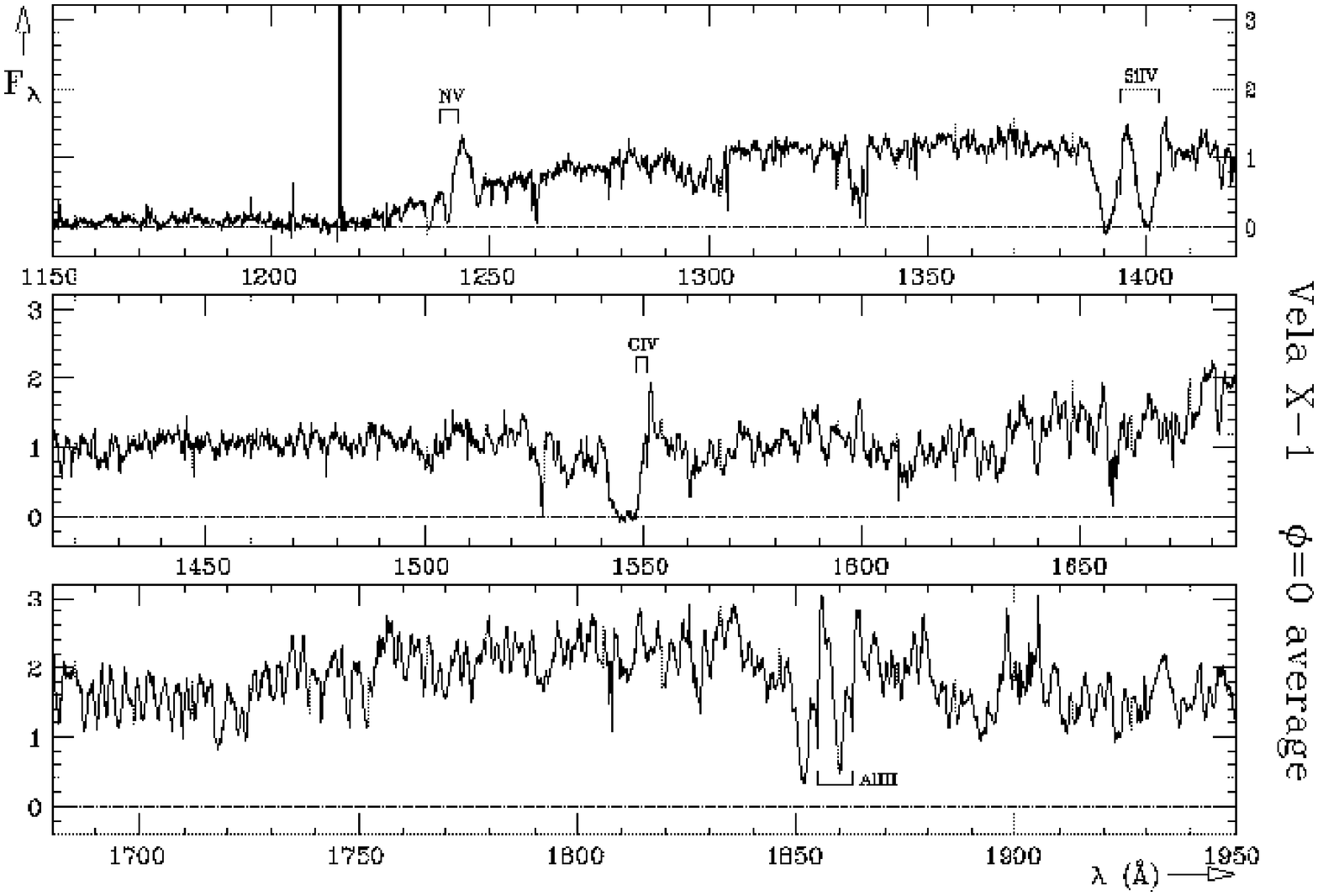,angle=90,width=172mm}
\psfig{figure=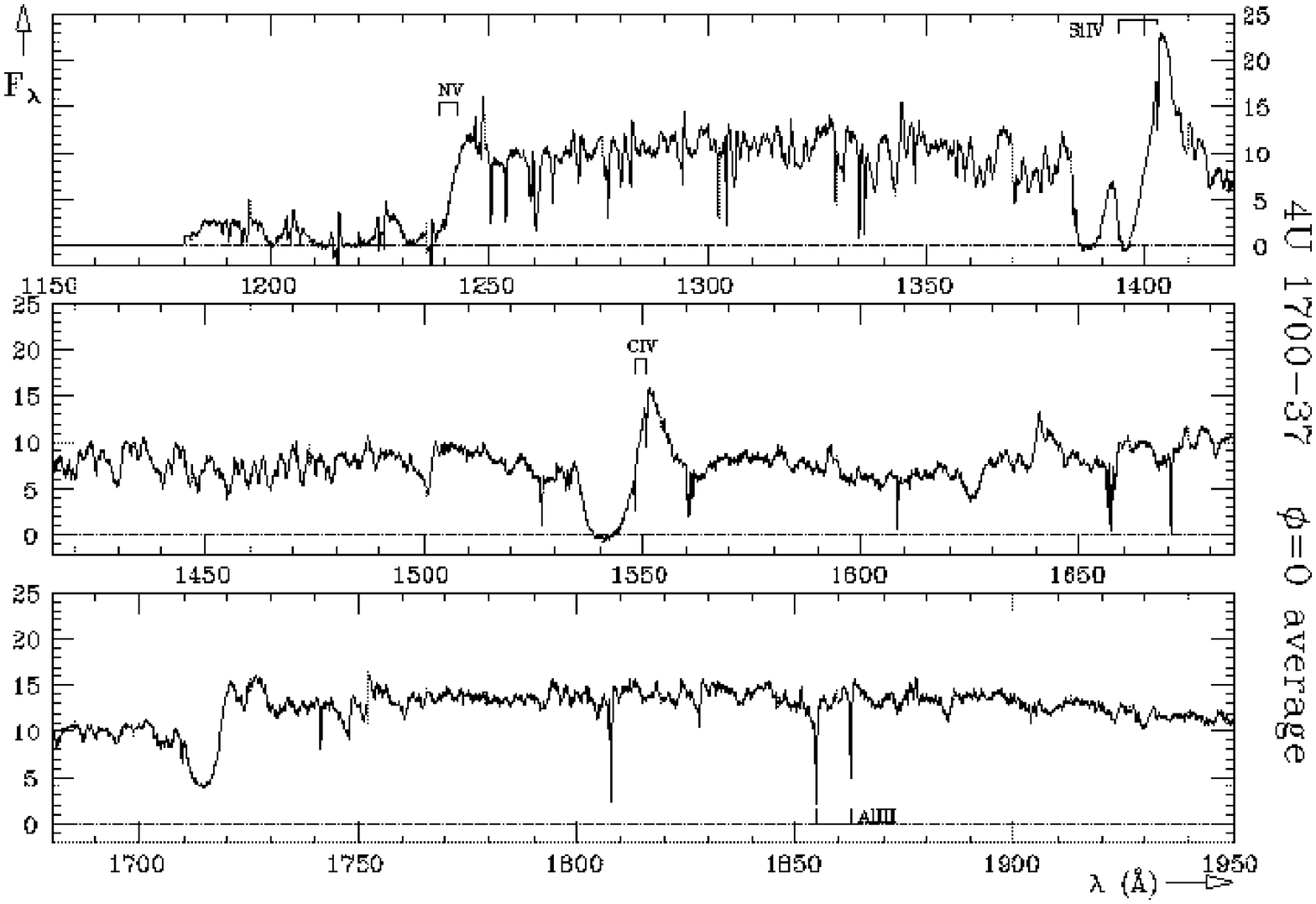,angle=90,width=172mm}
}}
\caption[]{X-ray eclipse spectra of HD77581/Vela X-1 (top) and
HD153919/4U1700$-$37 (bottom).}
\end{figure*}

In the spectrum of HD153919/4U1700$-$37 (Fig.\ A2, bottom) the resonance lines
of \Nv, \Siiv\ and \Civ\ and the subordinate line of \Niv\ at 1718.551 \AA\
are very strong. The resonance line of \Aliii\ is mainly interstellar. Notice
that the \Feiii\ and \Feiv\ spectra are absent, whilst at wavelengths
$\lsim1480$ \AA\ the spectrum is crowded with lines of \Fev\ and \Fevi. This
contrast with HD77581/Vela X-1 is due to the higher photospheric temperature
of the O6.5 Iaf$^{+}$ star HD153919 compared to the B0.5 Iab star HD77581. The
stronger P-Cygni lines of \Siiv\ and \Civ\ in HD153919/4U1700$-$37 reflect the
denser and faster stellar wind of HD153919 compared to HD77581. Note that the
\Nv\ emission is stronger in HD77581/Vela X-1.

\section{Normalisation and continuum variability}

Comparison of column densities towards the primary as derived from line
absorption requires the continuum to be normalised to eliminate continuum
variability of the primary --- which is mainly due to tidal deformation of the
primary, with a possible contribution of X-ray heating of the primary by the
X-ray source. It must be realised that normalising the continuum is in
principle incorrect when studying line emission: at a certain orbital phase,
the emission is due to scattered light originating from parts of the primary
that are not necessarily the part from which the observed continuum originates.

For each object, three wavelength intervals were chosen that show no intrinsic
variability other than a possible continuum variability. These are listed in
Table B1. To normalise the flux scales of the spectra relative to eachother,
each spectrum $k$ was divided by a normalisation factor $f_{k}$. This factor
was constructed from the spectral points $\{ x_{ki} \}_{k=1}^{ N}$ in the
three intervals $\{(A_{j},B_{j})\}_{j=1}^{ 3}$, with interval $j$ having
$n_{j}$ spectral points:
\begin{equation}
f_{k} = \sum_{j=1}^{3} \left[ \left( \frac{ n_{j} }{ \sum_{j=1}^{3} n_{j} }
\right) \times \frac{ \sum_{i=A_{j}}^{B_{j}} x_{ki} }{ \sum_{k=1}^{N}
\sum_{i=A_{j}}^{B_{j}} x_{ki} } \right]
\end{equation}
In this way the spectra are corrected for continuum variability as long as the
continuum slope remains constant.

The normalisation constants may in principle be used to construct UV continuum
lightcurves, after correcting for the change in sensitivity of the detector
according to the correction table of Bohlin \& Grillmair (1988). The
correction procedure does not discriminate between the two resolution modes
(Cassatella et al.\ 1994). Not all spectra can be used to construct a
lightcurve. For the HMXBs in low resolution all spectra available are used
except for SWP3989 (HDE226868/Cyg X-1) and SWP1458 and 1459 (Sk-Ph/LMC X-4)
that had a large (factor of 3 to 5) offset in flux level. For the HMXBs in
high resolution all spectra of HD77581/Vela X-1 are used, but SWP1476, 1714
and 1972 through 5180 (HD153919/4U1700$-$37) are omitted because they were
taken through the small aperture causing the loss of an unknown fraction of
the total light.

%
%
\begin{table}[]
\caption[]{Intervals $(A,B)$, in \AA, used for constructing the continuum
lightcurves.}
\begin{tabular}{llll}
\hline\hline
HMXB        & Band 1      & Band 2      & Band 3      \\
\hline
Cyg X-1     & (1265,1295) & (1759,1765) & (1820,1895) \\
LMC X-4     & (1650,1670) & (1690,1710) & (1820,1895) \\
SMC X-1     & (1605,1635) & (1746,1762) & (1820,1895) \\
Vela X-1    & (1272,1278) & (1492,1498) & (1696,1702) \\
4U1700$-$37 & (1465,1470) & (1515,1518) & (1772,1777) \\
\hline
\end{tabular}
\end{table}

%
%
\begin{figure*}[]
\centerline{\psfig{figure=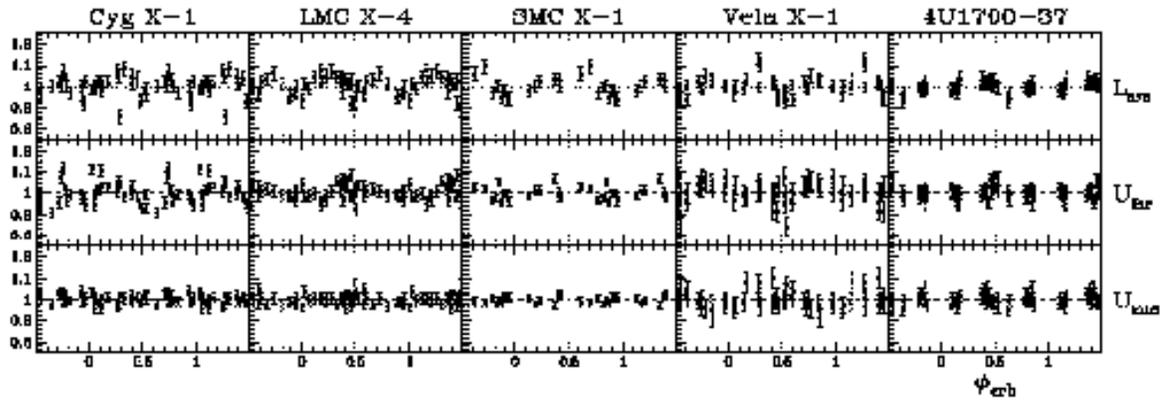,width=180mm}}
\caption[]{Lightcurves of the UV continuum flux and the UV colours as defined
in the text.}
\end{figure*}

To study variability in the spectrum tilt, we define UV colours in the
following way:
\begin{equation}
U_{\rm far} = L_1 - L_2 \hspace{10mm} \& \hspace{10mm} U_{\rm mid} = L_2 - L_3
\end{equation}
where $L_1$, $L_2$ and $L_3$ are the fluxes (normalised to unity) for the
wavelength bands 1, 2 and 3 in order of increasing wavelength. Because of the
difficulty to find suitable wavelength bands of the best quality for each
HMXB, the bands --- and therefore the colours --- are defined for each HMXB
individually.

The UV continuum lightcurves are displayed in Fig.\ B1. The $L_{\rm ave}$
lightcurve is an average of all points within the 3 wavelength bands. It is
not straightforward to determine the errors on the individual points in the
lightcurve. The photometric accuracy of IUE is reported to be $\sim$6\% at a
95\% confidence level (Bohlin et al.\ 1980). This is probably a conservative
estimate, judging from the small degree of scatter. For the error on the
colour measurement we adopt the standard deviation of the distribution
function of the measured $U_{\rm mid}$ points, which is a very conservative
error estimate, and the same value is assigned to the error of the $U_{\rm
far}$ colour.

\subsection{HDE226868/Cyg X-1}

The lightcurve of HDE226868/Cyg X-1 (Fig.\ B1) has a minimum to maximum
amplitude of $\sim$17\% (the deviating point at $\phi=0.281$ probably does not
contain all the flux of the source). This is consistent with the scatter of
similar amplitude in the UV photometry around 1500 and 1800 \AA\ obtained by
Wu et al.\ (1982) with the {\it Astronomical Netherlands Satellite} (ANS), but
considerably larger than the 8\% found by Treves et al.\ (1980) who analysed
only 7 spectra by integrating the entire spectrum between 1250 and 1900 \AA.
The absence of X-ray eclipses limits the inclination to $i\sim60\degr$ (Bolton
1975). Hence the deprojected amplitude is even larger, implying a severe
deformation of HDE226868.

The U$_{\rm far}$ colour shows clear orbital modulation similar to the average
flux: when the HMXB becomes fainter in the UV, the spectrum becomes redder.
This may reflect a lower photospheric temperature of HDE226868 at the side
that faces to and away from Cyg X-1, probably a result of the tidal
deformation of the primary (Hutchings 1974). HDE226868 is possibly filling its
tidal lobe during peri-astron of the slightly eccentric orbit ($e\sim0.05$:
Bolton 1975). No heating of the photosphere of the primary by the X-ray
companion is observed, in agreement with Str\"{o}mgren photometry by Hilditch
\& Hill (1974).

\subsection{Sk-Ph/LMC X-4}

The lightcurve of Sk-Ph/LMC X-4 (Fig.\ B1) has a minimum to maximum amplitude
of $\sim$19\%, and is an improvement over the lightcurve derived from only 15
spectra by van der Klis et al.\ (1982) (see also Vrtilek et al.\ 1997). The
variable depth of the $\phi=0.5$ minimum was explained by a precessing
accretion disk (van der Klis et al.\ 1982). The precession causes variations
in the amount of surface of Sk-Ph that is exposed to X-ray heating, as well as
variations in the fraction of Sk-Ph that is eclipsed by the disk. Heemskerk \&
van Paradijs (1989) confirmed the presence of an accretion disk that precesses
with a period of 30.36$\pm0.02$ days, whilst X-ray observations had already
revealed a 30.42$\pm0.03$ days period (Pakull et al.\ 1985); we adopt
$P=30.38\pm0.03$ days. Phase ${\phi}_{\rm prec}=0$ corresponds to the phase
when the accretion disk is seen edge-on.

%
%
\begin{figure}[]
\centerline{\psfig{figure=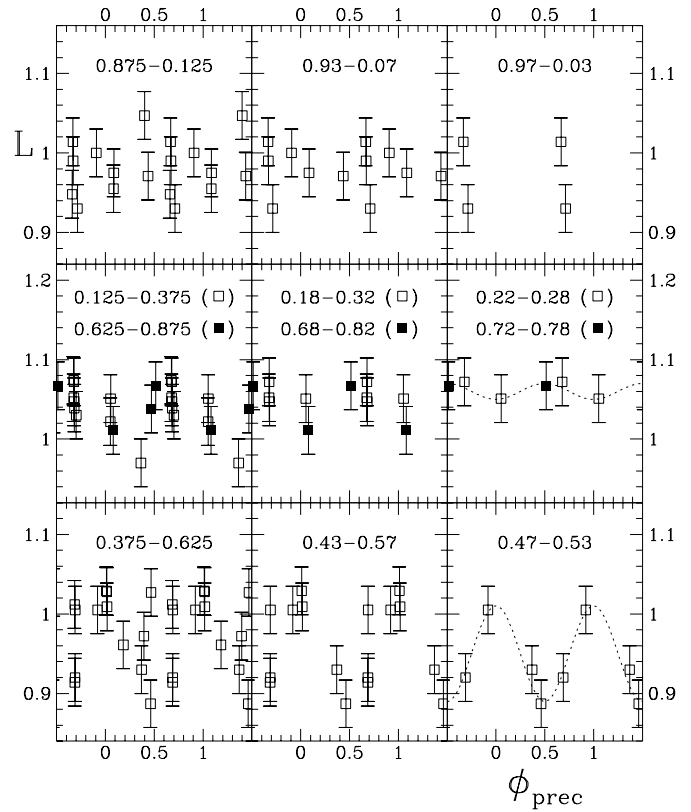,width=88mm}}
\caption[]{Continuum lightcurves of Sk-Ph/LMC X-4 for orbital phase bins as
indicated in each of the nine frames. The cycle parameters are those of the
precessing disk: $\phi_0 \equiv$ JD 2\,443\,392.52(2); $P=30{\fd}38$(3). The
dotted cosine waves are fits-by-eye to the data (see text).}
\end{figure}

Lightcurves are created according to the cycle parameters of the precession of
the accretion disk, for bins around orbital phases ${\phi}_{\rm orb}=0$, 0.25
\& 0.75, and 0.5 (Fig.\ B2). For orbital phases ${\phi}_{\rm orb}\sim0$ no
clear variability with the disk precession period is seen. For ${\phi}_{\rm
orb}\sim0.5$ the luminosity is lowest when the accretion disk eclipses the
largest part of Sk-Ph (${\phi}_{\rm prec}=0.5$), and highest when the
accretion disk is seen edge-on (${\phi}_{\rm prec}=0$). This is especially
clear when narrower orbital phase bins are taken. At ${\phi}_{\rm
orb}\sim0.25$ \& 0.75 the luminosity varies in anti-phase with the variation
at ${\phi}_{\rm orb}=0.5$. Our results agree well with those of Heemskerk \&
van Paradijs (1989) from optical data. The amplitude of variability of the UV
luminosity at ${\phi}_{\rm orb}=0.5$ is 10\% --- the same as in the optical.

The $U_{\rm far}$ colour (Fig.\ B1) is modulated with the orbital period. The
colour behaviour around orbital phase ${\phi}_{\rm orb}\sim0.5$ may be
dominated by X-ray heating, causing the UV spectrum to become hotter close to
${\phi}_{\rm orb}=0.5$. The scatter in the U$_{\rm far}$ colour very close to
${\phi}_{\rm orb}=0.5$ may be due to variable obscuration by the precessing
disk.

\subsection{Sk 160/SMC X-1}

The lightcurve of Sk 160/SMC X-1 (Fig.\ B1) has a minimum to maximum amplitude
of $\sim$17\%. In optical lightcurves the $\phi=0.5$ minimum is observed to be
weaker than the $\phi=0$ minimum, explained by X-ray heating by the bright
X-ray source (Hutchings 1974). Van Paradijs \& Zuiderwijk (1977) and Howarth
(1982), however, found that X-ray heating is not sufficient and that an
emitting accretion disk is required, in agreement with the suggestion that the
mass transfer is dominated by Roche-lobe overflow (Hutchings et al.\ 1977).
Despite using more spectra than van der Klis et al.\ (1982), the orbital phase
coverage of our UV lightcurve around $\phi=0.5$ is too poor to answer the
question of X-ray heating and accretion disk. An $\sim60$ day periodicity in
the X-ray characteristics of SMC X-1 has been attributed to a precessing
accretion disk (e.g.\ Wojdowski et al.\ 1998).

\subsection{HD77581/Vela X-1}

The lightcurve of HD77581/Vela X-1 (Fig.\ B1) has a minimum to maximum
amplitude of $\sim$20\% and clear orbital modulation, although Dupree et al.\
(1980) could not see continuum variability in the 5 low resolution spectra
they used. The $\phi=0.5$ minimum is significantly deeper than the $\phi=0$
minimum, much alike the visual lightcurve (Zuiderwijk et al.\ 1977) that has a
somewhat smaller amplitude of $\sim$15\%. Zuiderwijk et al.\ argue that
HD77581 is nearly filling its Roche lobe.

The $U_{\rm far}$ colour is modulated with the orbital period, becoming redder
around $\phi=0.5$. This may be explained by temperature gradients over the
photosphere of HD77581. The scatter of the $U_{\rm far}$ colour near
$\phi=0.5$ may result from variable X-ray heating of the part of the
photosphere of HD77581 that faces Vela X-1, due to the strongly variable X-ray
flux.

\subsection{HD153919/4U1700$-$37}

The lightcurve of HD153919/4U1700$-$37 (Fig.\ B1) is flat within $\sim$4 to
5\%, possibly with a marginable maximum at $\phi=0.5$. This is less than the
UV amplitude around 1500 and 1800 \AA\ of 8\% found by Hammerschlag-Hensberge
\& Wu (1977) using the ANS. Optical amplitudes are $\sim$4 to 8\%
(Hammerschlag-Hensberge \& Zuiderwijk 1977; van Paradijs et al.\ 1978). The
small amplitude may be due to a large mass ratio of the system and therefore
small distortion of HD153919. The optical lightcurve cannot be described by
tidal distortion of HD153919 alone (van Paradijs et al.\ 1978). In particular
the deepest minimum does not occur at $\phi=0.50$ but at $\phi=0.56$. Possible
explanations include extra absorption originating from the mass flow in the
system. The UV lightcurves presented here also suggest a minimum near
$\phi=0.6$. Hints for the signature of a photo-ionization wake in
Str\"{o}mgren photometry are presented in Appendix C. The $U_{\rm far}$ colour
is bluest at orbital phase $\phi=0.5$, which suggests X-ray heating of the
part of the photosphere of HD153919 facing 4U1700$-$37.

\section{Str\"{o}mgren $uvby\beta$ photometry of HD153919/4U1700$-$37}

%
%
\begin{figure}[]
\centerline{\psfig{figure=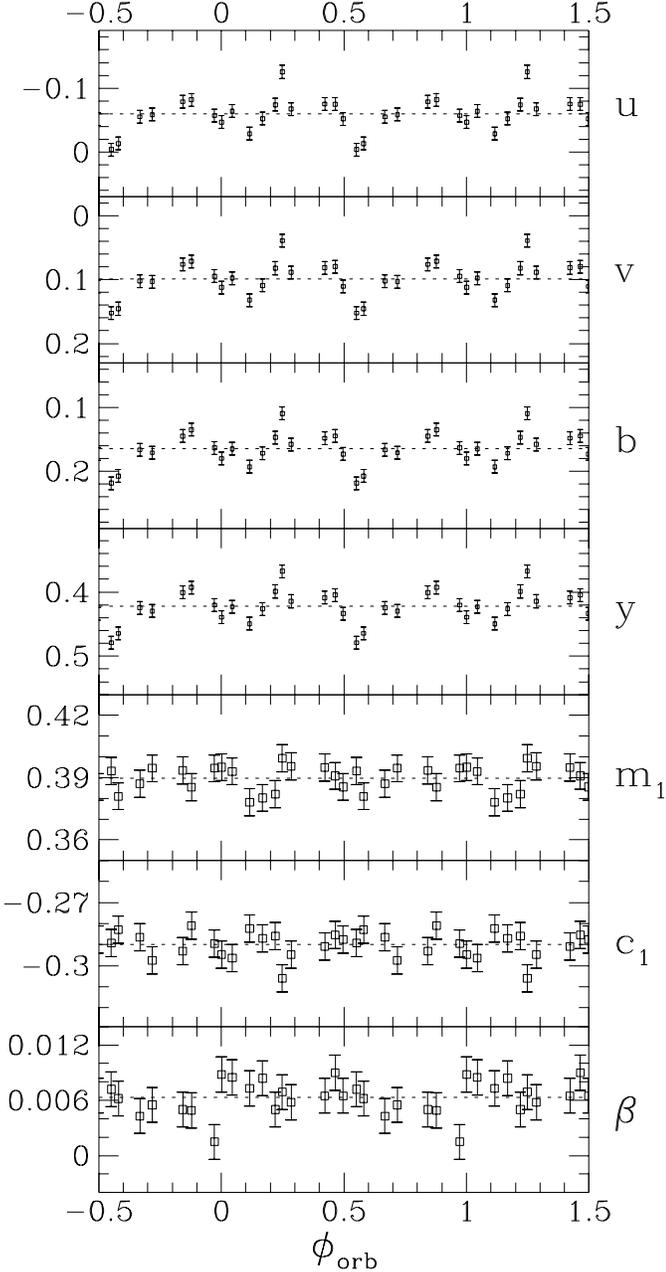,width=88mm}}
\caption[]{Str\"{o}mgren $uvby$ photometry, the derived colours $m_1$ and
$c_1$, and the $\beta$ parameter (see text) for HD153919/4U1700$-$37. The
$uvby\beta$ are presented differentially with respect to the comparison star
HD154368. Mean levels are represented by dotted lines.}
\end{figure}

In August 1993 we obtained Str\"{o}mgren $uvby$ plus broad and narrow band
H$\beta$ photometry using the 50 cm SAT at ESO/La Silla. The O9.5 Iab star
HD154368 was measured for comparison. The photometry is presented in Tables C1
\& C2. The magnitude differences between HD153919/4U1700$-$37 and HD154368 are
plotted in Fig.\ C1, together with the colours $m_1$ and $c_1$ for
HD153919/4U1700$-$37 as described in Sterken et al.\ (1995). The comparison
star did not vary with time. The standard deviation for each individual
measurement is typically 0.005 mag in $u$ and $y$, and 0.004 mag in $v$ and
$b$. From the scatter on small orbital phase scales in the resulting
lightcurves a formal, conservative estimate of 1-$\sigma \sim0.01$ mag is
derived. Because of the advantages of the multi-channel photometer, the errors
in $m_1$ and $c_1$ are smaller than this. The standard deviation of the
distribution of measured values gives an upper limit to the error on each
individual measurement of 1-$\sigma=0.0065$ mag for both colours. The $\beta$
parameter is derived from the H$\beta$ narrow- and wide-band magnitudes. It
increases with increasing absorption strength of the H$\beta$ line. The data
give an upper limit to the error in $\beta$ of 1-$\sigma=0.0019$ mag.

%
%
\begin{table}[]
\caption[]{Heliocentric Reduced Julian Date RJD$_\odot=$JD$-2\,449\,200$, $u$,
$v$, $b$ and $y$ magnitudes (with their standard deviation of the last
digit(s) within parentheses) and $\beta$ parameter of HD153919/4U1700$-$37.
Orbital phases may be calculated from: $\phi_0 \equiv$ JD
2\,446\,161.3400(30); $P=3{\fd}411652(26)$ (Haberl et al.\ 1989). The last row
gives the mean values and their 1-$\sigma$ errors.}
\begin{tabular}{llllll}
\hline\hline
RJD$_\odot$ & $u$ (${\sigma}_u$) & $v$ (${\sigma}_v$) &
$b$ (${\sigma}_b$) & $y$ (${\sigma}_y$) & $\beta$ \\
\hline
09.6414          & 7.172(4\rlap{)}  & 7.071(3\rlap{)} & 6.772(3\rlap{)} &
6.584(4\rlap{)}  & 1.230            \\
13.6344          & 7.153(4\rlap{)}  & 7.048(2\rlap{)} & 6.751(2\rlap{)} &
6.564(3\rlap{)}  & 1.249            \\
14.6736          & 7.130(4\rlap{)}  & 7.023(3\rlap{)} & 6.730(3\rlap{)} &
6.544(4\rlap{)}  & 1.516            \\
15.6159          & 7.086(2\rlap{)}  & 6.987(4\rlap{)} & 6.695(3\rlap{)} &
6.506(4\rlap{)}  & 1.195            \\
16.6497          & 7.182(5\rlap{)}  & 7.078(3\rlap{)} & 6.780(3\rlap{)} &
6.595(5\rlap{)}  & 1.380            \\
17.6424          & 7.101(5\rlap{)}  & 6.995(4\rlap{)} & 6.703(4\rlap{)} &
6.515(4\rlap{)}  & 1.352            \\
18.5755          & 7.177(6\rlap{)}  & 7.067(4\rlap{)} & 6.771(5\rlap{)} &
6.594(6\rlap{)}  & 1.096            \\
19.6214          & 7.114(5\rlap{)}  & 7.011(3\rlap{)} & 6.717(3\rlap{)} &
6.534(3\rlap{)}  & 1.268            \\
20.6311          & 7.137(7\rlap{)}  & 7.036(4\rlap{)} & 6.743(4\rlap{)} &
6.556(4\rlap{)}  & 1.335            \\
21.5959          & 7.155(3\rlap{)}  & 7.051(3\rlap{)} & 6.755(3\rlap{)} &
6.566(4\rlap{)}  & 1.183            \\
22.5641          & 7.129(3\rlap{)}  & 7.025(2\rlap{)} & 6.731(3\rlap{)} &
6.541(3\rlap{)}  & 1.096            \\
23.5721          & 7.187(7\rlap{)}  & 7.084(5\rlap{)} & 6.784(4\rlap{)} &
6.595(4\rlap{)}  & 1.124            \\
24.5867          & 7.132(4\rlap{)}  & 7.025(4\rlap{)} & 6.728(2\rlap{)} &
6.539(3\rlap{)}  & 1.180            \\
25.5770          & 7.183(7\rlap{)}  & 7.089(8\rlap{)} & 6.787(8\rlap{)} &
6.597(1\rlap{2)} & 1.158            \\
26.5891          & 7.161(8\rlap{)}  & 7.052(3\rlap{)} & 6.754(2\rlap{)} &
6.566(5\rlap{)}  & 1.212            \\
28.5688          & 7.141(3\rlap{)}  & 7.044(6\rlap{)} & 6.748(7\rlap{)} &
6.560(7\rlap{)}  & 1.159            \\
32.5778          & 7.143(6\rlap{)}  & 7.042(3\rlap{)} & 6.745(3\rlap{)} &
6.554(3\rlap{)}  & 1.236            \\
\hline
Aug'93           & 7.146(7\rlap{)}  & 7.043(7\rlap{)} & 6.747(7\rlap{)} &
6.559(7\rlap{)}  & 1.233(2\rlap{7)} \\
\hline
\end{tabular}
\end{table}

%
%
\begin{table}[]
\caption[]{Same as Table C1, but now for the comparison star HD154368.}
\begin{tabular}{llllll}
\hline\hline
RJD$_\odot$ & $u$ (${\sigma}_u$\rlap{)} & $v$ (${\sigma}_v$\rlap{)} &
$b$ (${\sigma}_b$\rlap{)} & $y$ (${\sigma}_y$\rlap{)} & $\beta$ \\
\hline
09.6355          & 7.224(4\rlap{)}  & 6.962(2\rlap{)}  & 6.600(2\rlap{)}  &
6.151(2\rlap{)}  & 1.197            \\
09.6475          & 7.224(4\rlap{)}  & 6.959(3\rlap{)}  & 6.598(2\rlap{)}  &
6.150(4\rlap{)}  & 1.250            \\
13.6283          & 7.210(3\rlap{)}  & 6.946(2\rlap{)}  & 6.585(3\rlap{)}  &
6.139(2\rlap{)}  & 1.214            \\
13.6405          & 7.207(5\rlap{)}  & 6.946(3\rlap{)}  & 6.584(2\rlap{)}  &
6.140(3\rlap{)}  & 1.271            \\
14.6676          & 7.188(4\rlap{)}  & 6.928(3\rlap{)}  & 6.566(3\rlap{)}  &
6.124(3\rlap{)}  & 1.463            \\
14.6795          & 7.186(7\rlap{)}  & 6.929(3\rlap{)}  & 6.567(3\rlap{)}  &
6.124(3\rlap{)}  & 1.566            \\
15.6101          & 7.215(9\rlap{)}  & 6.949(5\rlap{)}  & 6.587(5\rlap{)}  &
6.139(3\rlap{)}  & 1.165            \\
15.6237          & 7.207(4\rlap{)}  & 6.947(3\rlap{)}  & 6.583(2\rlap{)}  &
6.138(3\rlap{)}  & 1.219            \\
16.6434          & 7.189(1\rlap{1)} & 6.927(5\rlap{)}  & 6.562(5\rlap{)}  &
6.116(5\rlap{)}  & 1.334            \\
16.6560          & 7.182(4\rlap{)}  & 6.924(4\rlap{)}  & 6.560(5\rlap{)}  &
6.116(5\rlap{)}  & 1.418            \\
17.6364          & 7.183(5\rlap{)}  & 6.918(4\rlap{)}  & 6.558(4\rlap{)}  &
6.112(4\rlap{)}  & 1.310            \\
17.6484          & 7.177(6\rlap{)}  & 6.920(4\rlap{)}  & 6.558(4\rlap{)}  &
6.117(4\rlap{)}  & 1.384            \\
18.5696          & 7.203(5\rlap{)}  & 6.931(4\rlap{)}  & 6.575(4\rlap{)}  &
6.142(5\rlap{)}  & 1.075            \\
18.5832          & 7.209(6\rlap{)}  & 6.940(4\rlap{)}  & 6.582(4\rlap{)}  &
6.148(4\rlap{)}  & 1.107            \\
19.6155          & 7.193(3\rlap{3)} & 6.933(8\rlap{)}  & 6.570(8\rlap{)}  &
6.127(6\rlap{)}  & 1.232            \\
19.6283          & 7.185(6\rlap{)}  & 6.926(6\rlap{)}  & 6.567(5\rlap{)}  &
6.123(6\rlap{)}  & 1.296            \\
20.6244          & 7.199(6\rlap{)}  & 6.933(4\rlap{)}  & 6.572(4\rlap{)}  &
6.126(4\rlap{)}  & 1.291            \\
20.6374          & 7.192(6\rlap{)}  & 6.933(5\rlap{)}  & 6.572(4\rlap{)}  &
6.127(5\rlap{)}  & 1.368            \\
21.5897          & 7.198(3\rlap{)}  & 6.939(2\rlap{)}  & 6.575(3\rlap{)}  &
6.127(3\rlap{)}  & 1.153            \\
21.6021          & 7.206(6\rlap{)}  & 6.939(6\rlap{)}  & 6.575(5\rlap{)}  &
6.127(5\rlap{)}  & 1.199            \\
22.5579          & 7.198(4\rlap{)}  & 6.935(3\rlap{)}  & 6.572(2\rlap{)}  &
6.124(2\rlap{)}  & 1.074            \\
22.5707          & 7.195(3\rlap{)}  & 6.938(3\rlap{)}  & 6.574(3\rlap{)}  &
6.129(3\rlap{)}  & 1.104            \\
23.5659          & 7.204(6\rlap{)}  & 6.938(5\rlap{)}  & 6.577(4\rlap{)}  &
6.129(4\rlap{)}  & 1.099            \\
23.5784          & 7.197(4\rlap{)}  & 6.939(3\rlap{)}  & 6.576(3\rlap{)}  &
6.132(4\rlap{)}  & 1.134            \\
24.5792          & 7.215(3\rlap{)}  & 6.956(2\rlap{)}  & 6.594(2\rlap{)}  &
6.148(3\rlap{)}  & 1.146            \\
24.5931          & 7.213(4\rlap{)}  & 6.952(3\rlap{)}  & 6.592(2\rlap{)}  &
6.145(5\rlap{)}  & 1.197            \\
25.5690          & 7.234(9\rlap{)}  & 6.972(7\rlap{)}  & 6.607(7\rlap{)}  &
6.158(7\rlap{)}  & 1.124            \\
25.5837          & 7.236(1\rlap{2)} & 6.987(1\rlap{5)} & 6.622(1\rlap{5)} &
6.182(1\rlap{8)} & 1.172            \\
26.5828          & 7.241(8\rlap{)}  & 6.975(5\rlap{)}  & 6.611(4\rlap{)}  &
6.161(4\rlap{)}  & 1.179            \\
26.5956          & 7.230(6\rlap{)}  & 6.970(3\rlap{)}  & 6.607(3\rlap{)}  &
6.162(5\rlap{)}  & 1.231            \\
28.5619          & 7.206(1\rlap{1)} & 6.948(4\rlap{)}  & 6.584(4\rlap{)}  &
6.137(4\rlap{)}  & 1.128            \\
28.5756          & 7.204(4\rlap{)}  & 6.945(2\rlap{)}  & 6.582(2\rlap{)}  &
6.137(3\rlap{)}  & 1.173            \\
32.5704          & 7.224(6\rlap{)}  & 6.969(4\rlap{)}  & 6.607(4\rlap{)}  &
6.164(6\rlap{)}  & 1.196            \\
32.5859          & 7.210(8\rlap{)}  & 6.951(6\rlap{)}  & 6.589(7\rlap{)}  &
6.146(6\rlap{)}  & 1.266            \\
\hline
Aug'93           & 7.205(3\rlap{)}  & 6.944(3\rlap{)}  & 6.582(3\rlap{)}  &
6.137(3\rlap{)}  & 1.228(2\rlap{0)} \\
\hline
\end{tabular}
\end{table}

Our $uvby$ lightcurves of HD153919/4U1700$-$37 agree with the lightcurves
obtained by Hammerschlag-Hensberge \& Zuiderwijk (1977). They noticed an
orbital modulation of the colour $m_1$, with a minimum at $\phi\sim0.2$, and
suggested that this might be caused by variability in the emission lines of
\Heii\ 4686, \Ciii\ 4650 and \Niii\ 4634-4641 that are included in the
Str\"{o}mgren b-filter. Our observations indeed confirm a minimum in $m_1$
between $\phi=0.1$ and 0.2. Krzemi\'{n}ski (1976) did not find any orbital
modulation of the \Heii\ 4686 line. Kaper et al.\ (1994) found variability in
the \Heii\ 4686 line profile, but this alone cannot cause observable
variability in the Str\"{o}mgren b-filter. The other two mentioned emission
lines may still be variable.

There is a weak indication for a gradual decrease in the H$\beta$ absorption
(or an increase in emission) towards $\phi=1$, possibly with a secondary weak
minimum around $\phi=0.3$ (Fig.\ C1). This is exactly what Kaper et al.\
(1994) found in their study of the variability in the H$\beta$ P-Cygni line
profile. They attributed this to the presence of a photo-ionization wake.

\section{Covariances method for error and variability analysis}

The degree of variability in a particular spectral point over a series of
spectra can be expressed by the variance in that spectral point. The variances
are often used to derive error estimates on the values of spectral points, and
to detect intrinsic spectral variability. Usually two adjacent spectral points
do not behave independently from eachother. Ignoring the covariability of
adjacent spectral points can lead to serious under-estimation of the error on
a quantity derived by integration along the spectrum. A correct error analysis
therefore involves the calculation of covariances. Better than variances,
covariances are powerful in detecting variability on an intermediate spectral
scale --- i.e.\ exceeding the instrumental profile but considerably smaller
than the total spectral range. Covariances have also been used successfully in
proving the reality of spectral features with low signal-to-noise (van Loon et
al.\ 1996).

For a set of $N$ spectra, each consisting of a number of spectral values $x$,
the covariance of points $a$ \& $b$ is:
\begin{equation}
\sigma_{ab}^2 = \frac{ N \sum_{k=1}^N x_{ka}x_{kb} - (\sum_{k=1}^N x_{ka} )
(\sum_{k=1}^N x_{kb} ) }{ N (N-1) }
\end{equation}
It measures how much and how coherently the values of the two points vary from
one spectrum to another. For $a=b$ the covariance reduces to the variance.
Integrating the spectral values $x$ over a spectral range $(A,B)$ in spectrum
$k$ yields:
\begin{equation}
I_k = \int_A^B x_{ki} {\rm d}i
\end{equation}
with an error estimate $\sigma_{I_k}$ given by:
\begin{equation}
\sigma_{I_k}^2 = \int_A^B \int_A^B \sigma_{ij}^2 {\rm d}j {\rm d}i
\end{equation} 
Only if all points within $(A,B)$ behave statistically independently this
reduces to the commonly used variance-deduced error estimate, because then:
\begin{equation}
\sigma_{ij} = \sigma_{ii} \delta_{ij}
\end{equation}
with $\delta_{ij}$ the Kronecker delta function. In the opposite extreme when
all points within $(A,B)$ behave in phase:
\begin{equation}
\sigma_{ij} = \sigma_{ii}
\end{equation}
and consequently no increase in signal-to-noise can be obtained by integrating
along the spectrum. The integrated covariances $\int_A^B \sigma_{ij}^2 {\rm
d}j$ are calculated in spectral regions without intrinsic spectral variability
(Table D1) and tabulated in flux-wavelength space. Errors can then be assigned
according to flux and wavelength. This assumes that the integrated covariances
are a smooth function of flux and wavelength, which was proven to be true for
IUE spectra (Howarth \& Smith 1995).

%
%
\begin{table}[]
\caption[]{Intervals $(A,B)$, in \AA, used for error estimation.}
\begin{tabular}{lllll}
\hline\hline
Cyg X-1            &
LMC X-4            &
SMC X-1            &
Vela X-1           &
4U1700$-$3\rlap{7} \\
\hline
1250,137\rlap{0}   &
1170,120\rlap{0}   &
1250,138\rlap{0}   &
1153,120\rlap{0}   &
1183,1225          \\
1420,151\rlap{0}   &
1280,138\rlap{0}   &
1415,153\rlap{0}   &
1253,128\rlap{0}   &
1420,1517          \\
1565,192\rlap{6}   &
1430,153\rlap{0}   &
1560,192\rlap{6}   &
1315,138\rlap{0}   &
1730,1947          \\
                   &
1565,192\rlap{6}   &
                   &
1415,153\rlap{0}   &
                   \\
                   &
                   &
                   &
1565,184\rlap{0}   &
                   \\
\hline
\end{tabular}
\end{table}

%
%
\begin{figure*}[]
\centerline{\hbox{
\psfig{figure=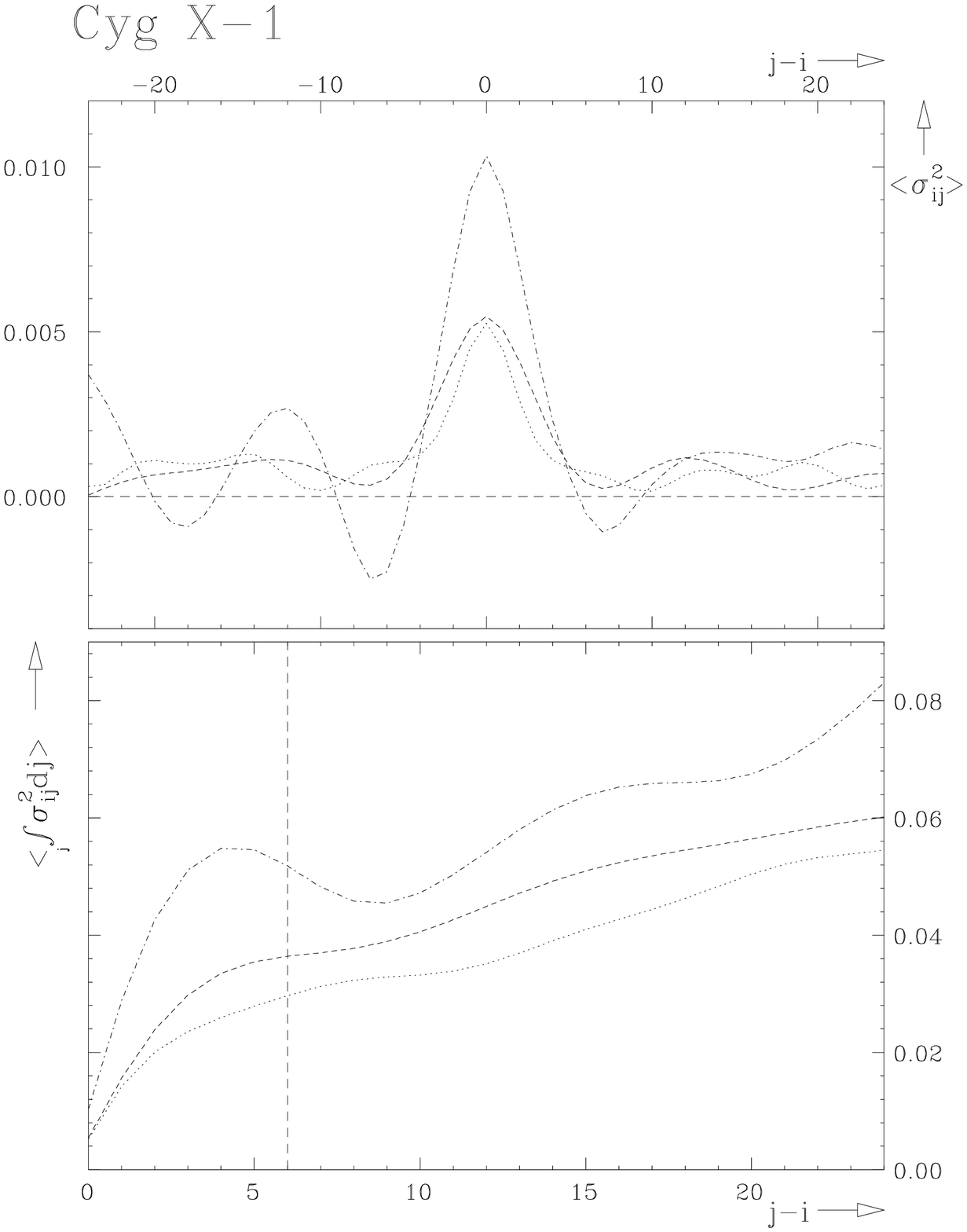,width=88mm}
\psfig{figure=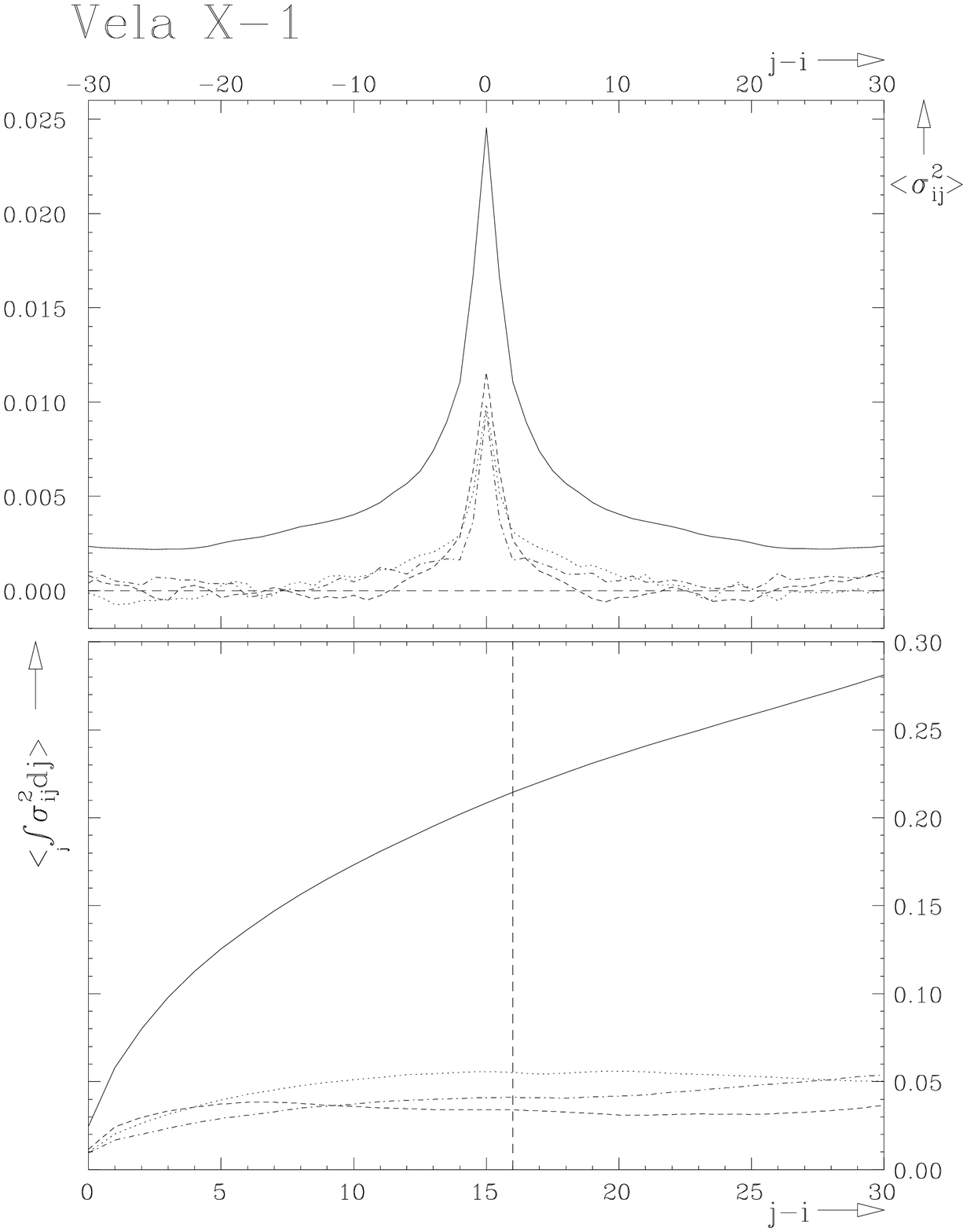,width=88mm}
}}
\caption[]{Covariance (top) and cumulative covariance (bottom) for Cyg X-1
(left) and Vela X-1 (right) as a function of distance along the spectrum with
respect to the spectral point $i$. The drawn, dashed, dotted and dash-dotted
lines represent the averages for the entire spectrum and the three spectral
regions in which the continuum calibration factors were determined,
respectively, with subsequently larger mean wavelength. The vertical
long-dashed line indicates the choice of the integration boundary.}
\end{figure*}

%
%
\begin{figure*}[]
\centerline{\vbox{
\hbox{
\psfig{figure=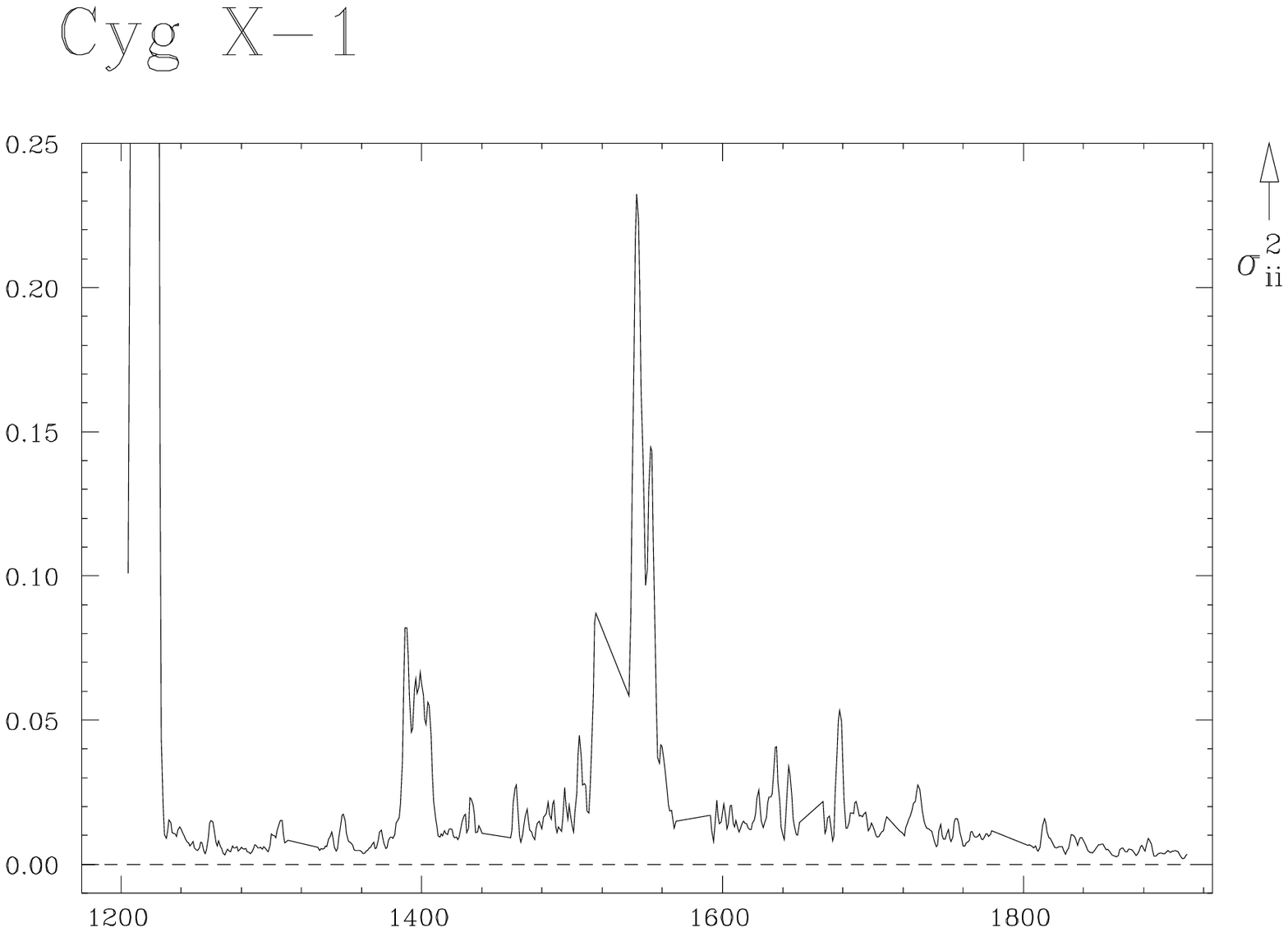,width=94.3mm}
\psfig{figure=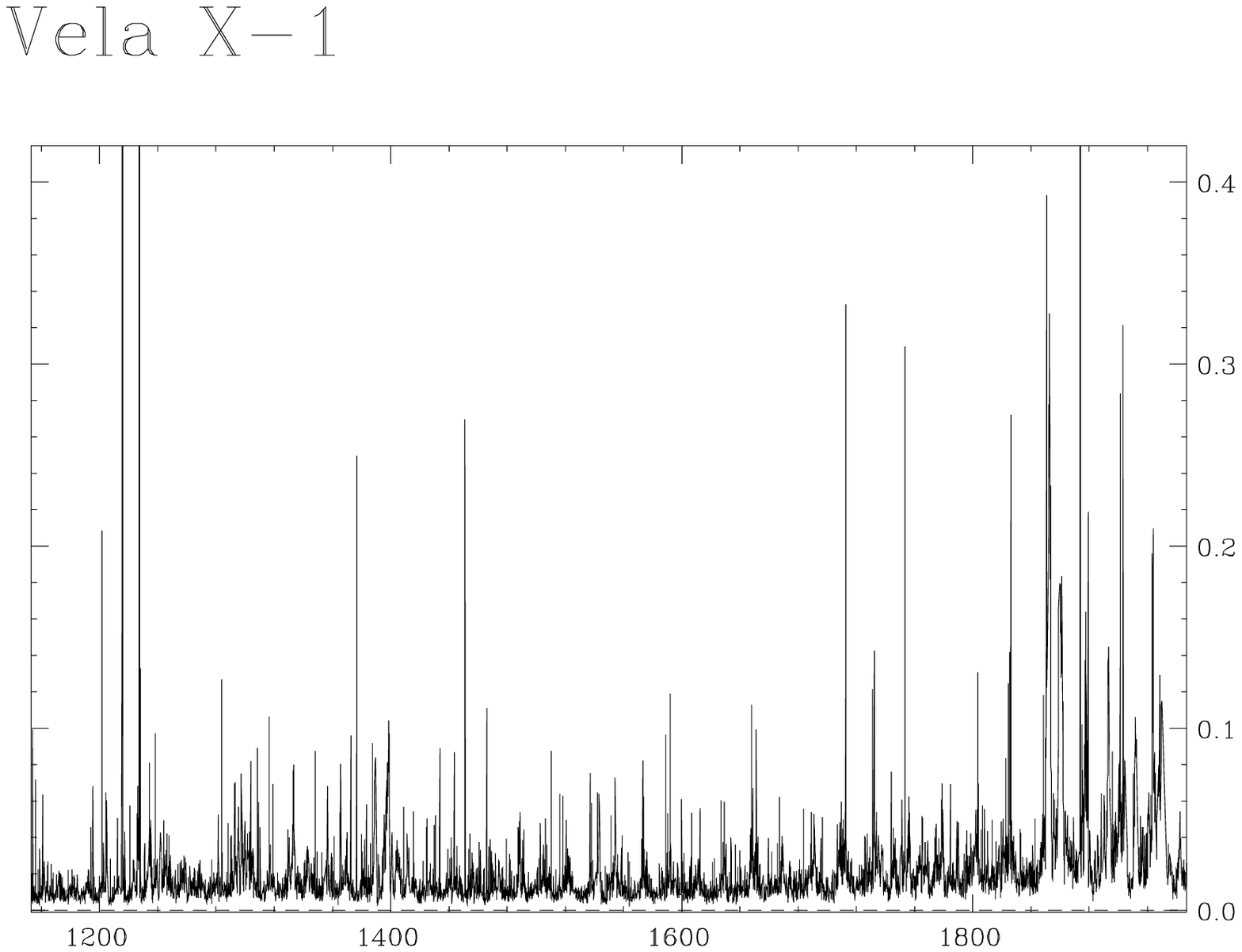,width=85.7mm}}
\hbox{
\psfig{figure=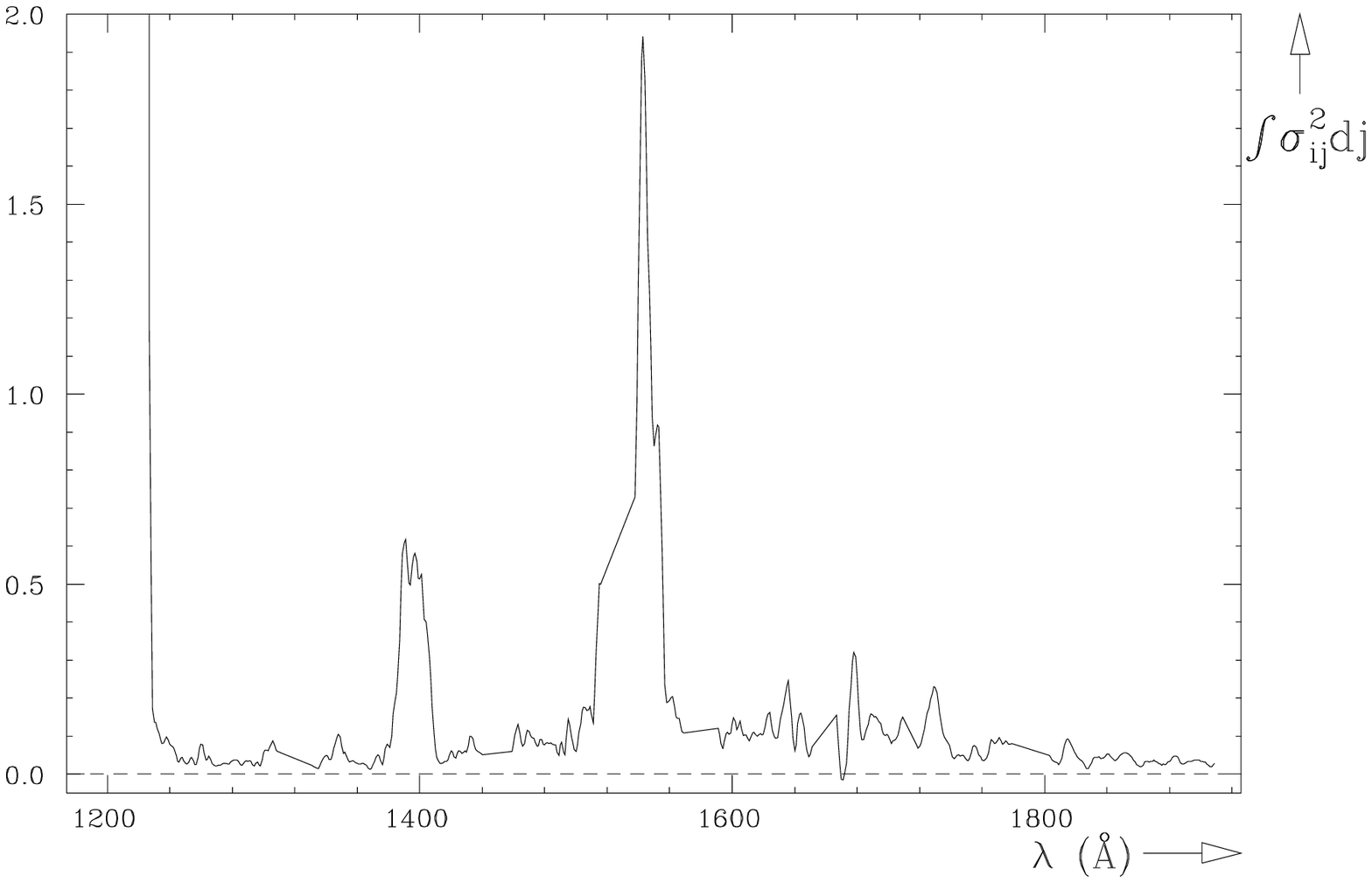,width=94.3mm}
\psfig{figure=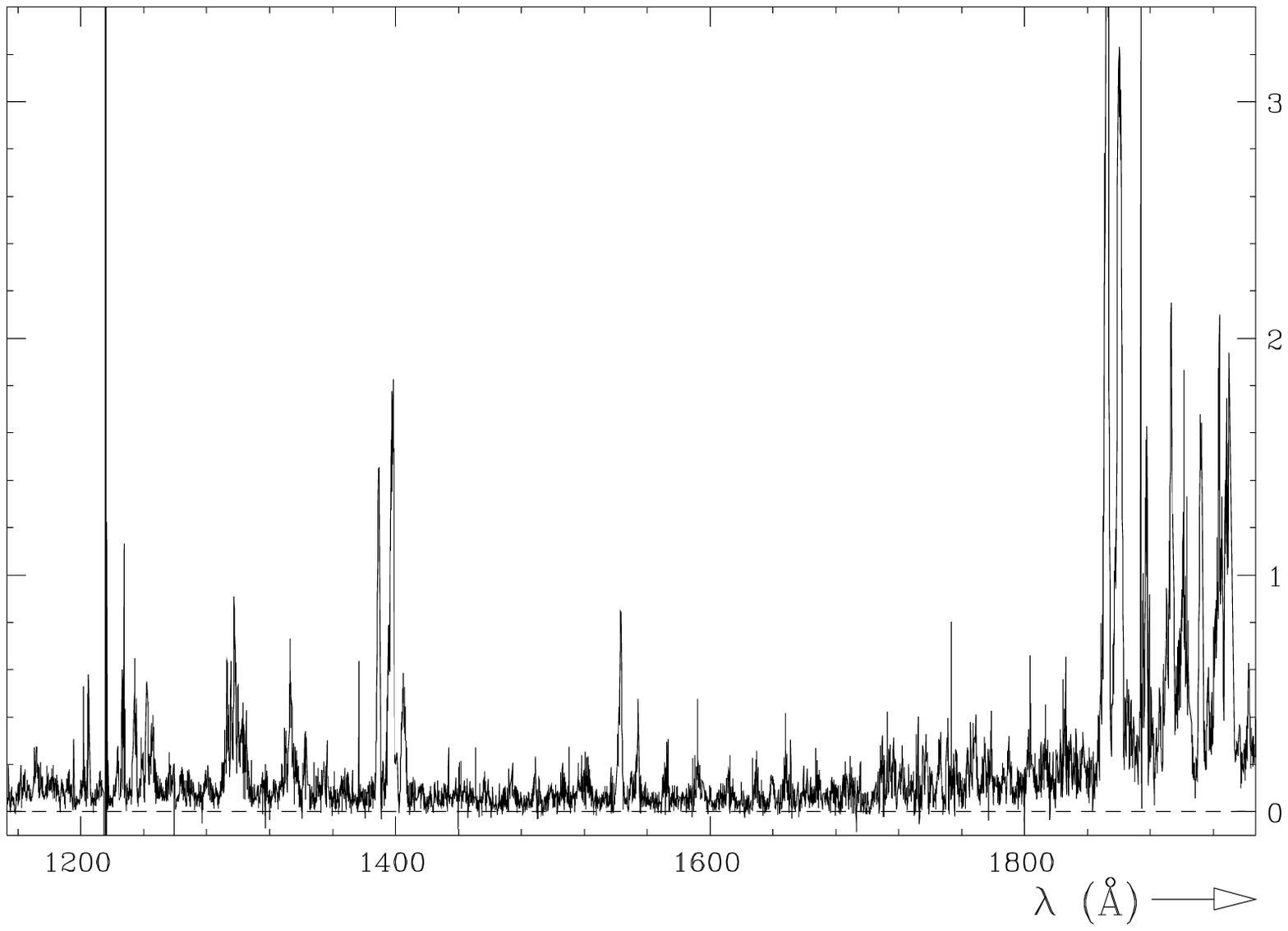,width=85.7mm}}
}}
\caption[]{Variances (top) and integrated covariances (bottom) in the spectra
of Cyg X-1 (left) and Vela X-1 (right), as a function of wavelength of the
spectral points $i$.}
\end{figure*}

The covariances $\sigma_{ij}^2$ decrease with increasing distance $|j-i|$ to
the point $i$. If the integral of the covariances for the point $i$ converges
sufficiently fast, then the integration interval $(A,B)$ in Eq.\ (D3) may be
replaced by a smaller interval such that the integral of the covariances just
reaches convergence. This interval is determined from a covariance profile of
$\sigma_{ij}^2$ versus $j-i$, which represents the spectral shape of the
instrumental profile. As an example the covariance profile and the cumulative
covariance are shown for Cyg X-1 and Vela X-1 (Fig.\ D1). The dashed, dotted
and dash-dotted lines represent the mean curves within the three regions in
which the continuum calibration factors were determined (from shortest to
longest wavelength, respectively), whilst the solid line represents the entire
spectral range --- i.e.\ including wildly variable spectral features. Both the
covariance and the cumulative covariance at zero distance from the point $i$
reduce to the variance --- i.e.\ the covariance of point $i$ with itself. At
larger distances the covariance diminishes to a small oscillation around zero,
whilst the initially rapidly growing cumulative covariance approximates a
constant level.

For Cyg X-1 the solid line is too high to be captured within the frame of the
plots, indicating that some parts of the spectrum are strongly variable. The
width of the covariance profile corresponds to the specified spectral
resolving power of about 250. The cumulative covariance indicates that a
reasonable choice for the integration interval is $\pm6$ \AA. The sharp peak
of the covariance profile of Vela X-1 indicates a spectral resolution in
accordance with the specified spectral resolving power of about $10^{4}$.
Its broad wings, however, prevent the cumulative covariance from converging
within $\sim16$ spectral points ($\equiv1.6$ \AA) --- several times the
specified spectral resolution.

Spectra of the variances and integrated covariances are shown for Cyg X-1 and
Vela X-1 (Fig.\ D2) as an example of the detection of intrinsic spectral
variability. In the low-resolution spectra of Cyg X-1 the spectral features
(notably the \Siiv\ and \Civ\ resonance lines near 1400 and 1550 \AA,
respectively) and their variability are unresolved. The calculation of
covariances does not improve the detection sensitivity much. In the
high-resolution spectra of Vela X-1, however, the integrated covariance
spectrum is much more sensitive than the variance spectrum in detecting
intrinsic variability. Events that are limited to only one or very few
adjacent pixels (like cosmics or dead pixels) cause a ``forest'' of sharp
peaks in the variance spectrum, nearly completely obscuring the resolved but
still narrow peaks of the intrinsic variability --- the "trees" of interest.
By calculating the integrated covariance spectrum the forest is suppressed,
and at the same time the features indicating intrinsic variability are
enhanced. The variable spectral features can be identified easily.

\end{document}